\documentclass[%
showpacs,
amsmath,amssymb,amsfonts,
aps,prx
]{revtex4-1}
\usepackage{braket}
\usepackage{float}
\usepackage{graphicx}
\usepackage{dcolumn}
\usepackage{bm}
\usepackage{subfigure}
\usepackage{color}
\usepackage[normalem]{ulem}
\usepackage{mathtools}
\usepackage[unicode=true,pdfusetitle,bookmarks=true,
bookmarksnumbered=false,bookmarksopen=false,breaklinks=false,
pdfborder={0 0 0},backref=false,colorlinks=true,citecolor=red]
{hyperref}
\usepackage{cleveref}
\usepackage{mathrsfs}
\usepackage{esint}

\renewcommand{\ss}[1]{{#1}}
\newcommand\del{\nabla}
\newcommand\e{\epsilon}
\renewcommand\phi{\varphi}
\newcommand\dd{\mathrm{d}}
\renewcommand\b[1]{{\bf  #1}}
\renewcommand\vec[1]{\boldsymbol{#1}}

\newcommand{\vs}[1]{\rule{0pt}{#1ex}}

\begin{document}
\hyphenation{eq-ua-tions diff-er-ent only sce-nario also however equi-librium fila-ment results works re-mains two still account University} 

\title{Spatial population genetics with fluid flow}

\author{Roberto Benzi$^{1}$}
\author{David R.~Nelson$^{2}$}
\author{Suraj Shankar$^{2}$}
\author{Federico Toschi$^{3}$}
\author{Xiaojue Zhu$^{4}$}
\affiliation{
$^1$Department of Physics and INFN, University of Rome Tor Vergata, I-00133 Rome, Italy\\
$^2$Department of Physics, Harvard University, Cambridge, Massachusetts 02138, USA\\
$^3$Department of Applied Physics, Department of Mathematics and Computer Science, Eindhoven University of Technology, 5600 MB Eindhoven, The Netherlands and CNR-IAC, I-00185 Rome, Italy
$^4$Center of Mathematical Sciences and Applications, and School of Engineering and Applied Sciences, Harvard University, Cambridge, Massachusetts 02138, USA\\
}

\date{\today}

\begin{abstract}
	The growth and evolution of microbial populations is often subjected to advection by fluid flows in spatially extended environments, with immediate consequences for questions of spatial population genetics in marine ecology, planktonic diversity and origin of life scenarios. Here, we review recent progress made in understanding this rich problem in the simplified setting of two competing genetic microbial strains subjected to fluid flows. As a pedagogical example we focus on antagonsim, i.e., two killer microorganism strains, each secreting toxins that impede the growth of their competitors (competitive exclusion), in the presence of stationary fluid flows. By solving two coupled reaction-diffusion equations that include advection by simple steady cellular flows composed of characteristic flow motifs in two dimensions ($2d$), we show how local flow shear and compressibility effects can interact with selective advantage to have a dramatic influence on genetic competition and fixation in spatially distributed populations. We analyze several $1d$ and $2d$ flow geometries including sources, sinks, vortices and saddles, and show how simple analytical models of the dynamics of the genetic interface can be used to shed light on the nucleation, coexistence and flow-driven instabilities of genetic drops. By exploiting an analogy with phase separation with nonconserved order parameters, we uncover how these \emph{genetic} drops harness fluid flows for novel evolutionary strategies, even in the presence of number fluctuations, as confirmed by agent-based simulations as well.

\end{abstract}


\maketitle

\section{Introduction}
The advection of reproducing microbial populations by fluid flows is ubiquitous in the natural world. Hydrodynamic transport continually shapes and reorganizes competing populations across virtually all length and time scales \cite{tel_2005_biological_flows}, and can either mix populations to uniformity or lead to the formation of spatial structures. Turbulent mixing of reproducing phytoplankton near the surface of oceans and lakes provides an example of the latter phenomenon. Provided characteristic eddy turnover times are long compared to microorganism doubling times, effectively compressible two-dimensional fluid flows near the surface can cluster phytoplankton blooms into patchy, fractal-like spatial structures \cite{d_ovido_2010_fluid_niches} that lead to spatial ecological niches and genetic heterogeneity \cite{pigolotti_pop_bio_flow_2013}.

\begin{figure}[htpb!]
	\includegraphics[width=0.6\textwidth]{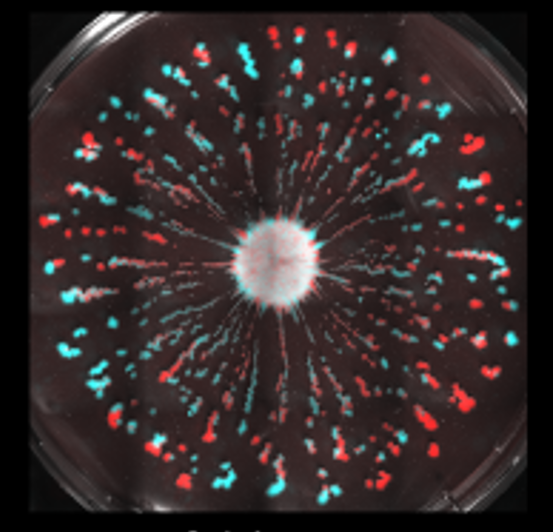}
	\caption{Yeast colony (\textit{Saccharomyces cerevisiae}) growing and advecting at the surface of a very viscous ($\eta=450~\textrm{Pa-s}\approx4.5\times 10^6\eta_{\vs{1.8}{\rm H_2O}}$) nutrient-rich liquid after $84~$hours of growth, subject to a radially symmetric outward velocity field that increases approximately linearly with distance $r$ from the center. An initially circular inoculant was deposited with a $50-50$ concentration of randomly mixed fluorescently labelled genetically neutral cyan and red strains that pass on their genetic labels to their offspring. Reproduced with permission from Ref.~\cite{atis2019microbial}.
	}
	\label{fig:exp}
\end{figure}

Novel laboratory systems for genetically labeled microorganisms subject advection are now becoming available. For example, one recent study uncovered a coupling between the metabolism of cell colonies and self-generated fluid flows \cite{atis2019microbial,kanso2019microbial}. Non-motile microbial colonies trapped by capillary forces on a viscous, nutrient-rich liquid substrate can create a submerged vortex ring in a Petri dish by locally reducing the fluid density. These flows are driven by a baroclinic instability triggered by the \textit{active} metabolism of the organisms, which can lead to remarkable finger-like protrusions at the colony frontier (Fig.~\ref{fig:exp}) as well as entirely fragmented colonies that spread radially many times faster than their expansion speed in the absence of a flow. Although these experiments focused on fluorescently tagged colonies of \textit{Saccharomyces cerevisiae} (baker's yeast), similar phenomena can be engineered for bacterial colonies at interfaces, such as \textit{Escherichia coli}, with flows provided externally by pumps and syringes \cite{atis2019microbial}. In both cases, the characteristic eddy turnover times of the fluid motion is of the order of hours, corresponding to many cell division times, allowing cell reproduction within the colonies to respond to and in turn sculpt the spatial structures embodied in the fluid dynamics. Such experiments provide a rich framework to examine the delicate interplay between fluid flow and spatial population dynamics.

As an illustrative example, see Fig.~\ref{fig:exp} which shows a yeast colony growing and advecting itself for a span of $84~$hours, at the surface of a very viscous ($\eta=450~$Pa-s) nutrient-rich liquid \cite{atis2019microbial}. Cell-proliferation and nutrient consumption spontaneously induce an outward flow field with a radially symmetric velocity field that increases approximately linearly with distance from the center. The original inoculant, containing a well-mixed $50-50$ concentration of fluorescently labeled, genetically neutral cyan and red strains, eventually forms numerous smaller microbial assemblies at its periphery on the surface of the nutrient-rich liquid. The front of the initially circular colony quickly becomes unstable and develops fingerlike structures within the first $24~$hours of growth. As Fig.~\ref{fig:exp} makes clear, such instabilities can have a profound effect on the spatial population genetics. These fingers, formed after the frontier undergoes genetic demixing \cite{hallatschek2007genetic}, typically lead to monoclonal genetic aggregates that grow and break up into small clusters, somewhat reminiscent of line-tension-driven Plateau-Rayleigh instabilities \cite{eggers1997nonlinear}. A striking difference, however, is that the instability is driven by active cell divisions and a colony-generated radial velocity field \cite{atis2019microbial}. One of many interesting challenges, which motivates the simplified reaction-diffusion models studied in this paper, is to understand what happens when the two reproducing strains are \emph{not} genetically neutral, as in Fig.~\ref{fig:exp}, but instead exhibit \textit{antagonism}, i.e., each organism produces secretions that poison the competing strain. Antagonistic interactions can lead to a form of \textit{genetic} phase separation \cite{lavrentovich2014asymmetric,lavrentovich2019_antagonism} with an emergent biological line tension between competing species, another phenomenon quite common in the natural world \cite{barton1985hybrid}. This genetic line tension is in addition to the conventional line tension present between organisms and uncolonized space that plays a prominent role in Fig.~\ref{fig:exp}. Other biological interactions such as \emph{mutualism}, i.e., the two strains benefit from the presence of each other, are also possible, but a complete understanding of how genetic interactions alter experiments like the above would go beyond the scope of this paper.

\begin{figure}[]
	\includegraphics[width=0.6\textwidth]{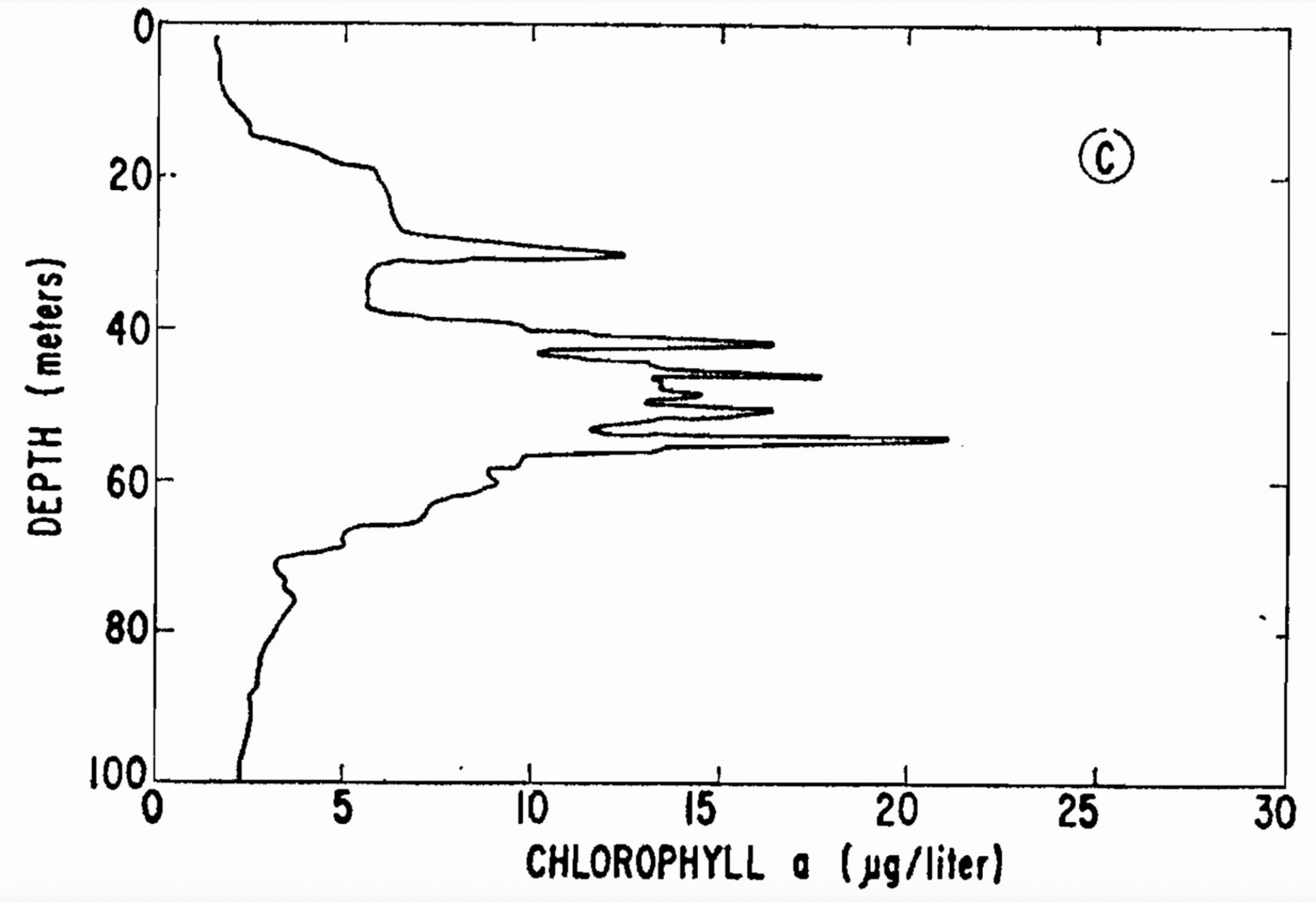}
	\caption{Experimental measurements of the concentration of phytoplankton as a function of depth close to La Jolla. We can clearly see thin layered structures that mark the presence of temporally coherent phytoplankton layers in the ocean. Figure reproduced from  Ref.~\cite{strickland1970ecology}.
	}
	\label{fig:layer}
\end{figure}

To put this review in context, it is worth reminding ourselves that, about $2-3$ billion years ago, the growth and evolution of microorganisms took place in oceans. Oxygen-producing cyanobacteria, which have been dated back to $\sim2.8-3.5$ billions of years ago, transformed the atmosphere via oxygen generating photosynthesis and may have been the ancestors of chloroplasts in plants and eukaryotic algae. Spatial growth, competition and fixation amongst different photosynthetic bacterial variants presumably took place at high Reynolds number in upper layers of the oceans. Such competition persists with organisms swimming and controlling their buoyancy to resist down-welling currents and stay close to the ocean surface.
For instance, planktonic organisms are known to concentrate in so-called thin phytoplankton layers \cite{durham2012thin} which are temporally coherent structures ranging in thickness from several centimeters to a few meters (see Fig.~\ref{fig:layer}), and extending laterally over kilometers in width.
Several mechanisms have been proposed for the formation of such temporally coherent thin layers \cite{durham2012thin} but we do not discuss them further.

Our goal here is to instead ask how fluid flows in such quasi-$2d$ layers affects the evolution of spatially distributed microorganisms. \ss{Early pioneering work by Allan Robinson and others~\cite{robinson1997theory,mcgillicuddy1995coupled,mcgillicuddy1998influence,oschlies1998eddy} highlighted the importance of eddies and turbulent advection in the ocean to sustain nutrient supply and phytoplankton primary production in the euphotic zone. Here, we will assume the organisms are not resource or nutrient limited, and consider the impact of advective effects on spatial population genetics alone.} To do so, we focus on an instructive example of antagonistic organisms that allows us to explore the interplay between biological line tensions and simple steady fluid flows (including radial ones) in two dimensions, using a simple reaction-diffusion model. As we shall see, even this simplified problem is highly informative, with many intriguing aspects. To make this paper relatively self-contained and pedagogical, we first review the essentials of population genetics in the absence of flow and then move on to describe a more general reaction-diffusion model capable of dealing with simple cellular fluid flows and various types of line tensions, both physical and biological \cite{pigolotti_pop_bio_flow_2013}. We build on an analogy with classical phase separation in binary mixtures to analyze the invasion, survival and coexistence of competing species as a droplet nucleation problem, with a nonconserved order parameter (related to the Model A universality class \cite{hohenberg1977critical_dynamics}), additionally aided or impeded by fluid flows, and validate the results by comparing against agent-based simulations.

In Sec.~\ref{sec:popgen}, we review spatial population genetics ideas that underlie competition and interactions between two genotypes in the absence of fluid flows. After a brief discussion of selective advantages (a spatial version of ``survival of the fittest''), and mutualism, we focus down on descriptions of competitions between antagonistic genotypes, for which both genetic line tensions and number fluctuations at interfaces can be important. Here, the overall composite population density at every point in space is held constant, an approximation which breaks down at antagonistic frontiers, when genotypes invade new territory or populations are subject to advection by compressible velocity fields. Then, in Sec.~\ref{sec:flow} we introduce a more general reaction-diffusion model capable of dealing with both fluid flows and invasion into new territory. In a certain regime, but only in this limited region of parameter space, we find agreement with conventional spatial population genetics. However, our model also allows us to determine corrections to simple nucleation theory, due to the depletion of the overall population at antagonistic frontiers. Within this model, we can easily subject two competing, antagonistic genotypes to a simple, steady flow with periodic boundary conditions in two dimensions, one that displays important fluid mechanical building blocks such as vortices and saddle points, as well as point sources and sinks. At a special parameter value, our cellular flows can also exhibit a line source leading to a line sink. Sec.~\ref{sec:drop} focuses on the fate of genetically favored droplets born on isolated sources, sinks, vortices, or saddles, for both steady compressible (Sec.~\ref{subsec:comp}) and steady incompressible flows (Sec.~\ref{subsec:incomp}). \ss{Following the pedagogical approach, we present new results for the growth and form of antagonistic genetic drops advected by fluid flow.} Stationary genetic configurations, where neither genotype takes over at long times, are possible when droplets are born on a sink, while saddle flows tend to stretch and sometimes even tear genetically preferred droplets apart, contributing to their faster demise. Sec.~\ref{subsec:linear} concludes with a discussion of the fate of a strip of the favorable genotype, centered on a line source or sink. Advection by a line source, not surprisingly, helps the favored genotype take over. A stationary genetic configuration, where neither genotype dominates, again arises when the initial configuration is a stripe formed by a ``razor blade inoculation'' \cite{hallatschek2010life} of the favored genotype, centered and parallel to a line sink. However, we show that flat stripes are unstable to undulations, where for this simple problem we can account for number fluctuations explicitly. Finally, in Sec.~\ref{sec:sim}, we compare our understanding of the deterministic reaction-diffusion model with agent-based simulations, and are able to demonstrate that the continuous organism density predictions are approached smoothly when the genotypes are dense, provided number fluctuations due to the discreteness of the competing organisms are taken into account. We conclude in Sec.~\ref{sec:disc} with a brief discussion of open problems and future directions.

\section{Competition and interactions without fluid flows}
\label{sec:popgen}
Biological interactions between distinct spatially distributed genotypes or entirely different organisms are quite common in the world around us. Consider, for simplicity, two competing strains of reproducing microorganisms such as asexual bacteria or yeast in a well-stirred nutrient-rich broth. Such organisms could be selectively neutral, in which case the long time behavior in a well-shaken flask maintained at a constant population size will be determined by the initial proportions of the two species as well as by random fluctuations associated with genetic drift \cite{crow_kimura1970}. Another possibility is that these organisms have differing doubling times, characterized by a simple Darwinian selective advantage $s$. Let $f(t)$ represent the fraction of a genotype or species $A$ with, say, a heritable green fluorescent marker, so that $1-f(t)$ is the fraction of a competing species $B$ with, say, a red heritable genetic marker. Then, if $w_A$ is the growth rate of the green organism ($A$) in a simple logistic growth model and $w_B$ the corresponding growth rate of the red organisms ($B$), a simple selective advantage of $A$ over $B$ is set by $w_A=w_B(1 + s)$, where typically $0<s\ll1$. Provided that number fluctuations associated with small population sizes can be neglected, the dynamics of the green fraction in this simple selective dominance scenario is given by \cite{crow_kimura1970,korolev2010RMP}
\begin{equation}
	\dfrac{\dd f}{\dd t } = \dfrac{s}{\tau_g}f(1-f)\;,\label{eq:sel_dom_ODE}
\end{equation}
where $\tau_g$ is the generation time.
Thus, any nonzero fraction $f$ of the green species ($A$) with a positive selective advantage $s>0$ will take over the population at long times. Upon generalizing to a \textit{spatially} varying green organism fraction $f(\b{r},t)$ in two dimensions ($2d$), with a motility or response to Brownian motion that endows it with an effective diffusion constant $D$, Eq.~\ref{eq:sel_dom_ODE} becomes the famous nonlinear partial differential equation studied by Fisher, Kolmorgorov, Petrovskii and Piskunov (FKPP) \cite{murray2007math_bio}
\begin{equation}
\partial_t f=D\del^2f+ \dfrac{s}{\tau_g}f(1 - f)\;.
\label{eq:FKPP_equation}
\end{equation}
Here again, any small, localized inoculation of the selectively favored green organism that escapes genetic drift will take over the population \cite{korolev2010RMP,murray2007math_bio}. It now does so in a spatial context via a pulled genetic wave \cite{van_saarloos2003front} that is exceeding difficult to stop \cite{tanaka2017spatial_gene_drive}.

A very different scenario, in both well-mixed and spatial situations, arises when the two competing species are mutualists \cite{frey2010evolutionary}. Mutualism arises, for example, when each species exports a public good such as an amino acid that it has in abundance, but which is needed by the other. In well mixed systems, an evolutionarily stable strategy emerges \cite{nowak2004ESS}, such that the population fraction $f(t)$ approaches a stable intermediate fixed point $f^*$ ($0<f^*<1$) for any initial condition that does not consist of $100\%$ red or green organisms. The resulting spatial genetic dynamics has been probed experimentally in two-dimensional range expansions of Baker's yeast \textit{S.~cerevisea} \cite{muller2014mutualism}, where the wavelike establishment of an interpenetrating mutualistic state with an intermediate local $f(\b{r},t)=f^*$, can be studied with both theory and simulations \cite{korolev2011competition,lavrentovich2014asymmetric}.

\begin{figure}[htpb!]
	{\includegraphics[width=0.6\textwidth]{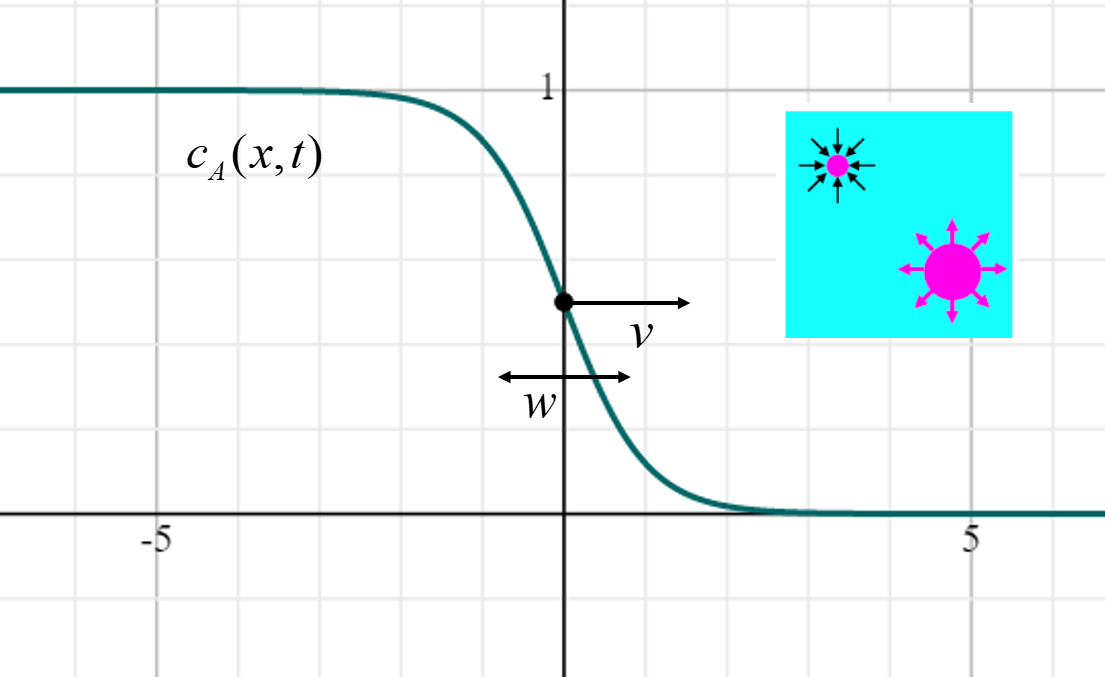}}
	\caption{One-dimensional cross section of the concentration profile $f(x,t)$ of species $A$. The summed concentrations of the two species are assumed to be at their carrying capacity, so the competing species $B$ has a profile $1-f(x,t)$. Inset: A top view of a selectively advantageous circular droplet of species $A$ (\ss{magenta}), growing or shrinking in a background of species $B$ (cyan). Even if species $A$ has a selective advantage, its reciprocal antagonism with species $B$ means the droplet radius $R$ must exceed a critical value before species $A$ can spread to take over the two-dimensional domain. A supercritical droplet is shown in the lower right corner, with a subcritical droplet in the upper left corner.
			}
	\label{fig:wave}
\end{figure}

Another common biological interaction, which will be the main focus of this paper, involves antagonism, such as when two microorganism species $A$ and $B$ each secrete proteins or chemical toxins that actively reduce the growth of the other, either when well mixed in a nutrient-rich liquid \cite{nowak2004ESS}, in close spatial proximity on a hard agar plate or in a natural environment \cite{yanni2019antagonism,frey2010evolutionary,yunker2019_review}. McNally et al.~\cite{yunker2017TypeVI_secretion} have studied antagonism in two distinct, genetically labeled antagonistic bacterial strains, each possessing toxin-antitoxin cassettes. Here, gram negative prey bacteria cells are killed upon contact with their predator bacterial counterparts that secrete specialized proteins that can rupture lipid membranes and cell walls \cite{brunet2013imaging_Type_VI}. With \textit{reciprocal} antagonism, each strain produces a distinct poison that is detrimental to the other, while protecting itself from its own toxin by also producing the corresponding antitoxin. In simple, repeatable experiments on a petri dish, these authors observe coarsening in approximately $50-50$ mixtures of antagonists \cite{yunker2017TypeVI_secretion}, similar, but not identical, to the growth of up/down spin domains in a two-dimensional Ising model quenched below its critical temperature. As will be discussed further below, the mechanism by which one selectively advantageous species $A$ takes over another species $B$ with reciprocal antagonism in $2d$ is via a pushed genetic wave, shown schematically in Fig.~\ref{fig:wave}.

Another important example of antagonistic spatial interactions is provided by hybrid zones of sexually reproducing organisms such as grasshoppers, butterflies or toads \cite{barton1985hybrid,barton1989adaptation}. Hybrid zones are relatively narrow interfaces where genetically distinct populations meet, mate and produce fertile offspring. In natural environments, these two dimensional interfaces can be a few hundred meters thick and hundreds of kilometers long!  Here antagonism arises if both species have been spatially isolated from each other long enough to develop optimal suites of genes corresponding to two distinct fitness maxima, one for each of the initial species. First or second generation hybrids at the interface, formed from parents with two distinct gene complexions, can then be less fit compared to either species $A$ or $B$. Antagonistic interactions and pushed genetic waves also arise over a range of selective \textit{disadvantages} when spatial gene drives (which convert heterozygous chromosome pairs in fertilized eggs with single copy of the drive into chromosomes homozygous for the drive, thus compensating for a selective disadvantage when the fertilized egg matures) are introduced into natural environments \cite{tanaka2017spatial_gene_drive}. Although these examples typically involve competitions on solid surfaces, similar phenomena can arise near the surfaces of lakes and oceans, involving antagonistic microorganisms, confined to a finite layer of liquid in the euphotic zone near the surface by their need to carry out photosynthesis.

A simple model of spatial antagonism between green ($A$) and red ($B$) organisms arises if we again denote by $w_A$ and $w_B$ the corresponding growth rates, but now imagine that these rates depend on local organism fractions $f_A(\b{r},t) \equiv f(\b{r},t)$ and $f_B(\b{r},t) = 1-f(\b{r},t)$ as
\begin{align}
	w_A(\b{r},t)&=g+\e_A \left[1 - f(\b{r},t)\right]\;,\\
	w_B(\b{r},t)&=g+\e_B f(\b{r},t)\;,
\end{align}
where $g$ is the background growth rate of either strain in the \textit{absence} of its antagonist (assumed to be identical for simplicity) and the constants $\e_A<0$ and $\e_B<0$, chosen to be negative, parameterize the degree of antagonism (positive $\e_A>0$ and $\e_B>0$ would describe mutualists \cite{korolev2011competition}, while simple Darwinian selective dominance, as in Eq.~\ref{eq:sel_dom_ODE}, corresponds to $\e_A=-\e_B$ \cite{tanaka2017spatial_gene_drive}). If we allow for local number fluctuations with an effective population size $N_0$ for organisms at their carrying capacity, standard methods of stochastic population dynamics then show that the FKPP equation Eq.~\ref{eq:FKPP_equation} becomes a more complex stochastic nonlinear partial differential equation, namely \cite{korolev2010RMP,lavrentovich2019_antagonism}
\begin{gather}
\label{Model A'}
	\partial _tf = D\nabla ^2f + \dfrac{f(1 - f)}{\tau _g}\left[\sigma (2f - 1) + \dfrac{\delta }{2}\right] + \sqrt{\dfrac{2f(1 - f)}{N_0\tau _g}}\;\xi\;,\\
	\delta =\e_A-\e_B\;,\quad\sigma = -\dfrac{1}{2}(\e_A +\e_B)\;.
\end{gather}
Here, $\delta$ describes the selective advantage associated with asymmetric antagonism, while $\sigma>0$ measures the overall degree of antagonism, as discussed below. The final stochastic term includes number fluctuations via a Gaussian white noise $\xi(\b{r},t)$ with short range correlations in space and time
\begin{eqnarray}
	\left\langle\xi (\b{r},t)\xi (\b{r}',t')\right\rangle  = \delta (\b{r}-\b{r}')\delta (t - t')\;,\label{eq:noise}
\end{eqnarray}
where the multiplicative nonlinearity in the noise (Eq.~\ref{Model A'}) should be interpreted using the Ito calculus \cite{korolev2010RMP}.
Note that the strength of this nonlinear noise vanishes when $f=0$ or $f=1$.

\begin{figure}[]
	{\includegraphics[width=0.6\textwidth]{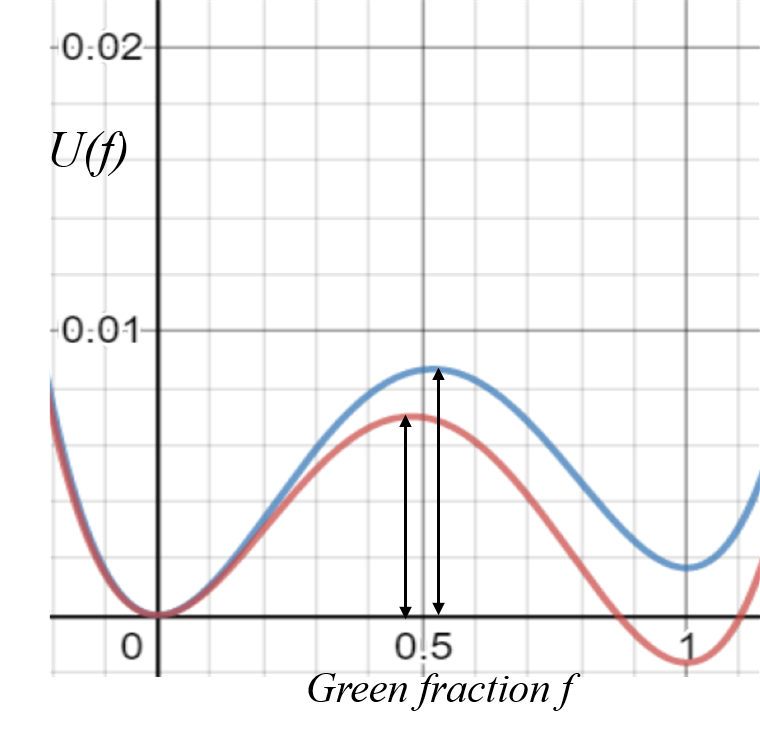}}
	\caption{Plot of the potential function $U(f)$ (Eq.~\ref{eq:pot}) that governs the mean-field dynamics of the green fraction $f$ of antagonistic organisms in Model A$'$ (Eq.~\ref{Model A'}). The two curves are plotted with fixed antagonism ($\sigma = 0.25$) and different selective advantage: $\delta = 0.02$ (red curve) and $\delta = -0.02$ (blue curve). The vertical black arrows indicate the barrier heights $U(f_{\rm max})-U(0)$. The depths of the potential well at $f = 1$ is \ss{controlled} by the selective advantage via $U(1)=U(0)-\delta/12$.
	}
	\label{fig:pot}
\end{figure}
As discussed in Refs.~\cite{tanaka2017spatial_gene_drive,lavrentovich2019_antagonism}, it can be helpful to rewrite Eq.~\eqref{Model A'} as gradient dynamics ($\tau_g\partial_tf=-\delta E[f]/\delta f+\sqrt{2\tau_gf(1-f)/N_0}\;\xi$) in an effective energy landscape given by
\begin{equation}
	E[f]=\int\dd\b{r}\;\left[\dfrac{D\tau_g}{2}|\vec{\del}f|^2+U(f)\right]\;,\quad U(f)=\dfrac{f^2}{12}\left[6\sigma(1-f)^2+\delta(2f-3)\right]\;.\label{eq:pot}
\end{equation}
The quartic potential $U(f)$, plotted in Fig.~\ref{fig:pot}, has minima at $f=0$ (red fraction survives) and $f=1$ (green fraction survives), separated by a single barrier at $f_{\rm max}=(1/2)[1+(\delta/2\sigma)]$. The difference in well-depths is controlled by the selective advantage $\delta$ ($U(1)-U(0)=-\delta/12$), while the barrier height is primarily determined by antagonism ($U(f_{\rm max})\simeq\sigma/32$, for $|\delta|\ll\sigma\ll 1$) \cite{lavrentovich2019_antagonism}. Note that the deterministic part of the dynamics is purely relaxational and equilibrium-like, resembling the time evolution of the well-known Model A of Halperin and Hohenberg \cite{hohenberg1977critical_dynamics} for Ising-like spins below their phase-separation temperature with two coexisting phases. Nonetheless, number fluctuations give rise to an unusual multiplicative noise (Eq.~\ref{Model A'}) that violates detailed balance and breaks the equilibrium analogy. We shall therefore refer to Eq.~\ref{Model A'} as Model A$'$. In the absence of mutations between green and red organisms, the stochastic dynamics arrests entirely upon entering the two pure states ($f=0,1$), which correspond to ``absorbing states'' \cite{hinrichsen2000absorbing_states}. The population then ``fixes'' with either $f(\b{r},t) = 0$ or $f(\b{r},t) = 1$ everywhere in space, at which point the system stops evolving as even the noise terms vanish. While nonequilibrium fluctuations can have important consequences on microbial organization \cite{cates2012diffusive}, much insight can already be gleaned by just considering the deterministic dynamics that admits an equilibrium mapping. Except near the end of this paper, we shall neglect number fluctuations in this paper. More systematic attempts to deal with this unusual noise can be found in Refs.~\cite{barton_rouhani_nucleation_1987} and~\cite{lavrentovich2019_antagonism}.


\section{Reaction-diffusion model of population genetics with advection}
\label{sec:flow}
While the above approach is sufficient and well-suited to describe spatial population genetics in colonies close to their carrying capacity \cite{korolev2010RMP}, its generalization to include arbitrary fluid flows presents interesting challenges \cite{pigolotti2012population}. In particular, independent fluctuations of species numbers is not captured by Model A$'$ (Eq.~\ref{Model A'}), which only tracks the \textit{relative} proportion of green organisms ($f$), with the remaining fraction ($1-f$) corresponding to the red species. When the overall population census is allowed to change as well, it can strongly influence the evolutionary dynamics, especially when mutalistic interactions are present \cite{chotibut2015evolutionary,chotibut2017population}. 

Furthermore, when modeling competition in the ocean, it is often appropriate to consider advection by a \textit{compressible} velocity field, both because of inertial effects \cite{bec2003fractal} associated with fairly large microorganisms (diameter $5-500~\mu$m), and because photosynthetic bacteria and plankton often actively control their buoyancy to stay close to the ocean surface \cite{martin2003phytoplankton}. In the latter case, the coarse-grained velocity field advecting the microorganisms will effectively contain a compressible component to account for upwellings and downwellings that allow both fluid mass and momentum to explore the third dimension \cite{boffetta2004lagrangian}.

\begin{figure}[]
	\includegraphics[width=\textwidth]{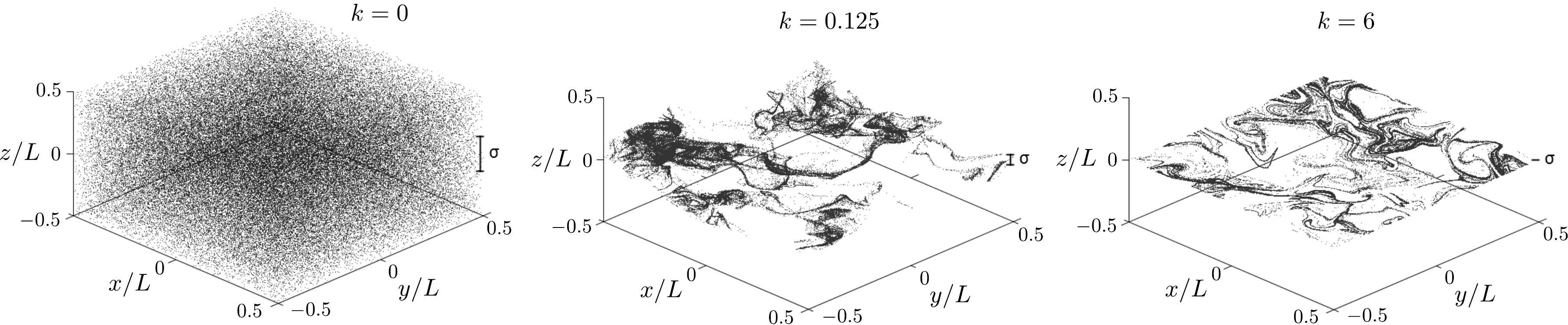}
	\caption{Snapshots from simulations of the evolution of passively advected particles evolving according to the flow in Eq.~\ref{springmodel}. From left to right the plots show the spatial distribution of particles upon varying the magnitude of the force constant, $k$. The magnitude of the slab depth ($\sigma$) is also shown on the right. Left panel: The unconfined case corresponding to $k = 0$. Middle panel: an intermediate confinement case, $k = 0.125$. Right panel: a strong confinement case, $k = 6$. In this last case particles are almost perfectly confined on to a bidimensional plane. It is clearly visible that the presence of increasing vertical confinement induces strong preferential concentration in the plane. \ss{Values of $k$ are reported in frequency units such that the inverse of the large-scale eddy turnover time $1/T_L=L/v_{\rm rms}=0.212$.} Reproduced from Ref.~\cite{depietro2015}. 
	}
	\label{fig:localization-angled}
\end{figure}

Such effects are of importance when considering plankton that tends to form thin layers at a distance below the ocean surface (Fig.~\ref{fig:layer}). Irrespective of the actual mechanisms responsible for the formation of these thin layers, their existence has dramatic implications on both population dynamics and genetics. As argued in Ref.~\cite{pigolotti2012population}, when the generation time ($\tau_g$) of reproducing organisms such as plankton is longer than the time it takes for the flow to transport them over a distance of the order of the thickness of the thin layer, then  microorganisms will experience an effectively compressible, almost two-dimensional velocity field. Importantly, the fact that the size of the organism itself is much smaller than the thickness of the layer is not relevant in this regard \cite{depietro2015}. A simplified model demonstrating that vertical confinement induces an effective compressibility is based on the idea of an organism with a given population density that is transported in a stratified fluid. Given a (turbulent) three-dimensional ($3d$) velocity field $\b{v}(\b{x},t)$ that models oceanic currents, the dynamics of plankton can be described in terms of an effective velocity field $\b{u}(\b{x},t)$ via the relation

\begin{equation}
	\b{u}(\b{x},t)= \b{v}(\b{x},t)-k(z-z_0)\hat{\b{z}}\;,
\label{springmodel}
\end{equation}

where $z_0$ is the position of the thin plankton layer and $k$, with dimension of a frequency, controls the thickness of the plankton layer. For $k=0$ the dynamics is that of a tracer particle transported by the $3d$ incompressible velocity field while, for $k\to\infty$ the dynamics is completely concentrated in the $z=z_0$ plane, \ss{therefore obviously experiencing a compressible two-dimensional velocity field, $(v_x,v_y,0)$ while moving in the $z=z_0$ plane.} Fig.~\ref{fig:localization-angled} shows how upon increasing $k$ the dynamics gets confined to thinner and thinner layers centred around $z=z_0$. From Fig.~\ref{fig:localization-angled} we can further appreciate how the enhanced confinement forces the quasi-$2d$ dynamics to become more spatially heterogeneous and patchy, as a direct consequence of the effective compressibility of the flow. Simple models such as Eq.~\ref{springmodel} can easily be generalised to allow more complex situations such as to describe nonplanar isopycnal (i.e., constant density) surfaces that evolve dynamically \cite{sozza2016}. That confinement induces some degree of compressibility in thin mixing layers is extremely relevant to population dynamics, particularly because high compressibility can effectively reduce the carrying capacity of the ecosystem \cite{perlekar2010}, while even small amounts of compressibility can change the effective selective advantage of a growing population \cite{Plummer2019}.

\begin{figure}[]
	\includegraphics[width=0.6\textwidth]{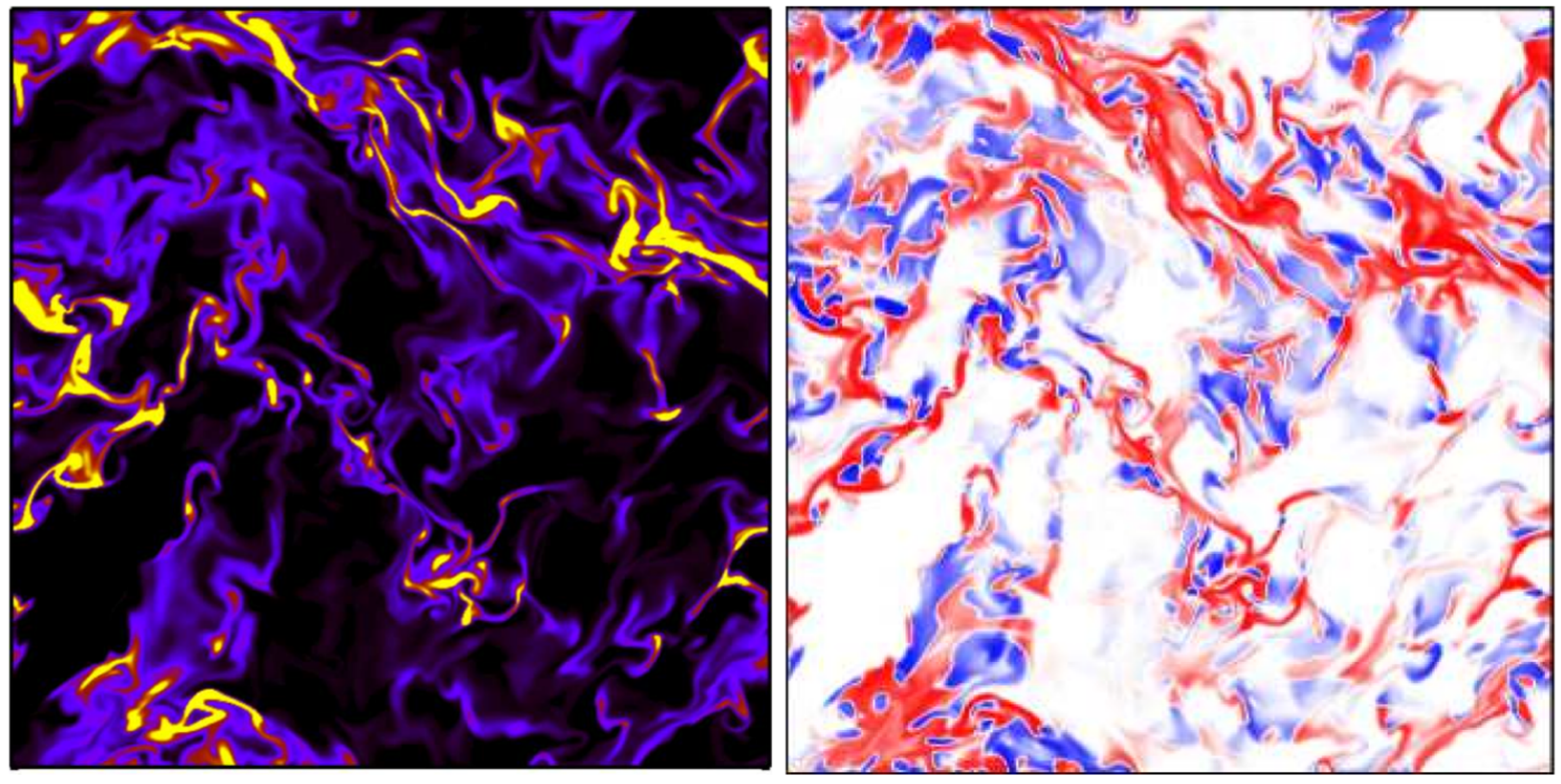}
	\caption{(Left panel) A plot of the concentration field, $c(\b{r},t)$ in pseudocolor where the bright (yellow) color indicates regions of high concentration ($c>0.1$) and the dark black regions indicate low concentration regions. (Right panel) A pseudocolor plot of $c(\b{r})/[0.1+c(\b{r})] \tanh{(-\vec{\nabla}\cdot\b{u})}$ with the red regions indicating negative divergence (and large concentration) whereas the blue regions indicate positive divergence (and large concentration). The plots are made the same time (after the steady state has been reached) on a slice obtained from a $512^2$ numerical simulation of Eq.~\ref{fkpp} for $\tau_{\eta}/\tau_g=0.0045$ and Schmidt number ${\rm Sc}=5.12$. As expected, microorganisms tend to cluster near regions of compression ($\vec{\nabla}\cdot\b{u} < 0$). Reproduced from Ref.~\cite{perlekar2010}.}
	\label{fig:popcomp}
\end{figure}

As compressibility also allows the total concentration of the species to change, it is instructive to instead use a more realistic reaction-diffusion model for advected interacting microorganisms that tracks each species fraction separately \cite{pigolotti_pop_bio_flow_2013}. For a single growing population invading empty space in the presence of fluid flow, we can generalize the FKPP equation (Eq.~\ref{eq:FKPP_equation}) to
\begin{equation} 
	\partial_tc + \vec{\del}\cdot (\b{u} c)= D\del^2 c + \dfrac{1}{\tau_g}c(1-c)
\label{fkpp}
\end{equation}
where $\b{u}$ is the velocity field, $D$ the diffusivity, $\tau_g$ the generation time and the carrying capacity has been rescaled to unity. The relevant dimensionless numbers characterizing the physics of the problem are the Reynolds number of the flow, ${\rm Re}= u_{\rm rms} L/\nu$, the Schmidt number ${\rm Sc}=\nu/D$ and the Damk{\"o}hler number ${\rm Da}=\tau_\eta/\tau_g$, where $L$ is a typical length scale, $\tau_\eta$ is the eddy turnover time and $u_{\rm rms}$ is the root mean squared velocity in a fluid with kinematic viscosity $\nu$. Here the fluctuating random velocity field $\b{u}(\b{x},t)$ is input from, say a simulation of homogeneous isotropic turbulence projected onto a biologically relevant $2d$ plane; see Fig.~\ref{fig:popcomp} for an illustration of simulations of Eq.~\ref{fkpp} based on this model.
Following Ref.~\cite{pigolotti_pop_bio_flow_2013}, we can quantify the degree of compressibility via the spatially averaged dimensionless quantity \ss{${\cal C}=\langle(\vec{\del}\cdot\b{u})^2\rangle/\langle(\partial_iu_j)^2\rangle$}.
Ref.~\cite{perlekar2010} showed that the carrying capacity measured as the spatially averaged concentration
\begin{equation}
	Z(t)=\dfrac{1}{L^2}\int\dd^2x\;c({\bm x},t)\;,
\end{equation}
depends strongly on the degree of compressibility \ss{${\cal C}$}, with $Z(t)\approx 1$ when $\tau_\eta\gg \tau_g$, while $Z(t)\ll 1$ decreasing by an order of magnitude for $\tau_\eta\ll \tau_g$, a condition relevant for marine environments \cite{perlekar2010}.

\ss{From a mathematical point of view, by substituting Eq.~\ref{springmodel} into Eq.~\ref{fkpp}, one obtains a term involving the $3d$ divergence of $\b{u}$, i.e., $\vec{\nabla}\cdot\b{u} c=-kc$, where we use the fact that $\b{v}$ is $3d$ incompressible ($\vec{\nabla}\cdot\b{v}=0$). By comparing this term with the logistic growth term, one easily realizes that compressibility effects are negligible when $k\ll 1/\tau_g$, i.e., when the timescale associated with compressibility is much longer than the generation time of the biological processes.}

For the reasons sketched above, it is natural to consider the effect of a compressible $2d$ velocity field on spatial population genetics.
For two genetically interacting species advected by a (possibly compressible) flow field, we study here two coupled reaction-diffusion equations that track each species fraction separately \cite{pigolotti_pop_bio_flow_2013},
\begin{align}
	\dfrac{\partial c_A}{\partial t}+\vec{\del} \cdot (\b{u}c_A)&=D\nabla^2 c_A+ \dfrac{1}{\tau_g}\;c_A(1-c_A-c_B+\epsilon_Ac_B)\;,\label{eq:cA}\\
	\dfrac{\partial c_B}{\partial t}+\vec{\del}\cdot (\b{u}c_B)&=D\nabla^2 c_B+ \dfrac{1}{\tau_g}\;c_B(1-c_B-c_A+\epsilon_Bc_A)\;,\label{eq:cB}
\end{align}
where $c_A$ is the concentration of the green organism ($A$), $c_B$ the concentration of red organism ($B$), and the two strains are subject to both diffusive ($D$ is the diffusion constant) and advective ($\b{u}$ is the flow velocity) transport.
As before, $\epsilon_A$ and $\epsilon_B$ control genetic competition between the two strains, with $\e_A,\e_B\leq 0$ corresponding to reciprocal antagonism that causes each species to suffer a growth penalty when the competing strain grows near it. For simplicity, we have neglected number fluctuations in the above, but these can be easily incorporated as in Eq.~\ref{Model A'} \cite{pigolotti_pop_bio_flow_2013}.

\begin{figure}[]
	{\includegraphics[width=0.9\textwidth]{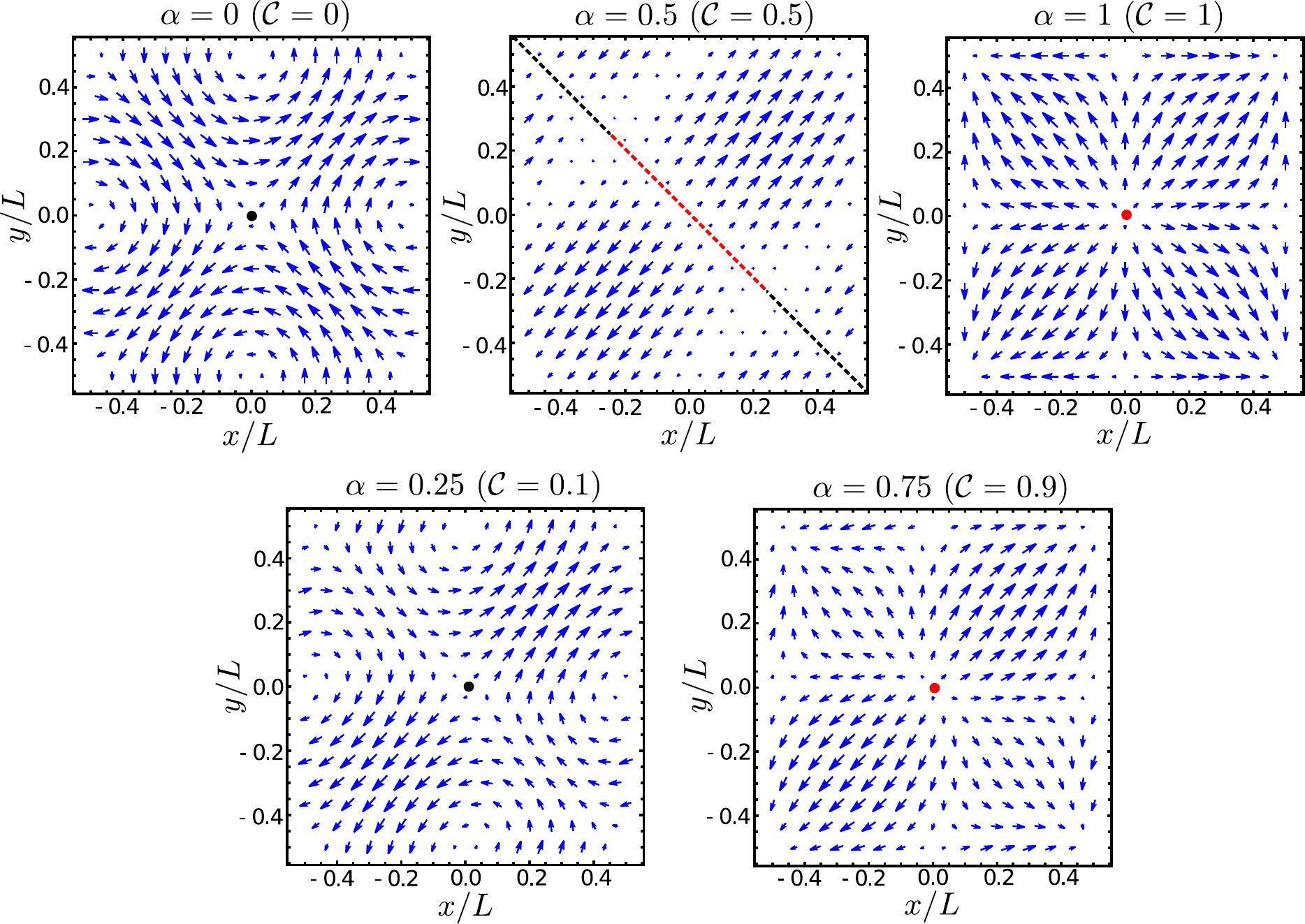}}
	\caption{Steady flows generated by Eq.~\ref{eq:u} with $F=1$ and periodic boundary conditions, for various values of $\alpha$, ranging from the incompressible limit $\alpha=0$ (with two saddle points and both clockwise and counterclockwise vortices) to the potential flow limit $\alpha=1$ (one source, one sink and two saddles). The normalized compressibility parameter \ss{$\cal C$} is also provided alongside $\alpha$. Although the $\alpha=0.25$ and $\alpha=0.75$ (see lower two figures) resemble their counterparts at $\alpha=0$ and $1$, $\alpha=0.5$ produces a variable line source and sink along the diagonal (red and \ss{black} dashed line respectively).
	}
	\label{fig:flow}
\end{figure}

For simplicity, we fix the flow to be stationary and externally imposed and leave the consideration of dynamic flows that respond to the growing colony for future work.
In order to elucidate the primary consequences of fluid motion on evolutionary strategies, we choose characteristic motifs that encode common topological singularities found in both steady and chaotic fluid flows. The following cellular flow model,
\begin{subequations}
\begin{align}
	u_x&=F[\alpha\sin(2\pi x/L)+(1-\alpha)\sin(2\pi y/L)]\;,\\
	u_y&=F[\alpha\sin(2\pi y/L)+(1-\alpha)\sin(2\pi x/L)]\;,
\end{align}\label{eq:u}
\end{subequations}
captures these features on a square periodic domain of size $L$. The strength of the flow is controlled by $F$, while $0\leq\alpha\leq 1$ tunes the compressibility, with $\alpha=0$ corresponding to incompressible flow ($\vec{\del}\cdot\b{u}=0$), while $\alpha\neq 0$ corresponds to compressible flows ($\vec{\del}\cdot\b{u}\neq 0$).
A useful dimensionless measure of the degree of compressibility is the spatial average \ss{${\cal C}=\langle(\vec{\del}\cdot\b{u})^2\rangle/\langle(\partial_iu_j)^2\rangle=\alpha^2/[\alpha^2+(1-\alpha)^2]$} (Fig.~\ref{fig:flow}) \cite{pigolotti_pop_bio_flow_2013}.
Incompressible flows can only host vortices and hyperbolic fixed points (saddles) as characteristic singularities, but compressible flows also admit sinks and sources that can have striking effects on the genetic dynamics, as we shall see below. Fig.~\ref{fig:flow} shows the flow field associated with Eq.~\ref{eq:u} for a variety of $\alpha$ values. Note the line sink and source for $\alpha=0.5$. We will study more robust line sinks and sources below in Sec.~\ref{subsec:linear}.

Let $c_T=c_A+c_B$ be the total concentration and $f={c_A}/{(c_A+c_B)}=c_A/c_T$ as before be the fraction of organism $A$ (correspondingly, $c_B/c_T=1-f$). Upon combining the equations for $c_A$ and $c_B$ (Eqs.~\ref{eq:cA},~\ref{eq:cB}), we have
\begin{align}
	\partial_tc_T+\vec{\del}\cdot(c_T\b{u})&=D\del^2c_T+\dfrac{1}{\tau_g}c_T\left(1-c_T\right)-\dfrac{2\sigma}{\tau_g}c_T^2 f(1-f)\;,\label{eq:cT}\\
	\partial_tf+\b{u}\cdot\vec{\del}f&=D\del^2f+2\dfrac{D}{c_T}\vec{\del}f\cdot\vec{\del}c_T+\dfrac{c_T}{\tau_g}f(1-f)\left[\dfrac{\delta}{2}+\sigma(2f-1)\right]\;\label{eq:dtf}.
\end{align}
In the limit when $|\e_A|,|\e_B| \ll 1$ (so $0<\sigma\ll 1$), we can neglect any variations in the total concentration and simply set $c_T\approx 1$, in which case Eq.~\ref{eq:dtf} reduces to
\begin{eqnarray}
\label{eqn6}
	\partial_t f + \b{u}\cdot\vec{\del}f= D\nabla^2f+\dfrac{1}{\tau_g}f(1-f)\left [ \dfrac{\delta}{2}+\sigma(2f-1) \right]\;.
\end{eqnarray}
Eq.~\ref{eqn6} generalizes Eq.~\ref{Model A'} (albeit without noise) to incorporate fluid flow, so that antagonistic genetic competitions take place in a fluid parcel advected by the velocity field of Eq.~\ref{eq:u}.

\begin{figure}[t!]
\begin{center}
\includegraphics[width=0.80\textwidth]{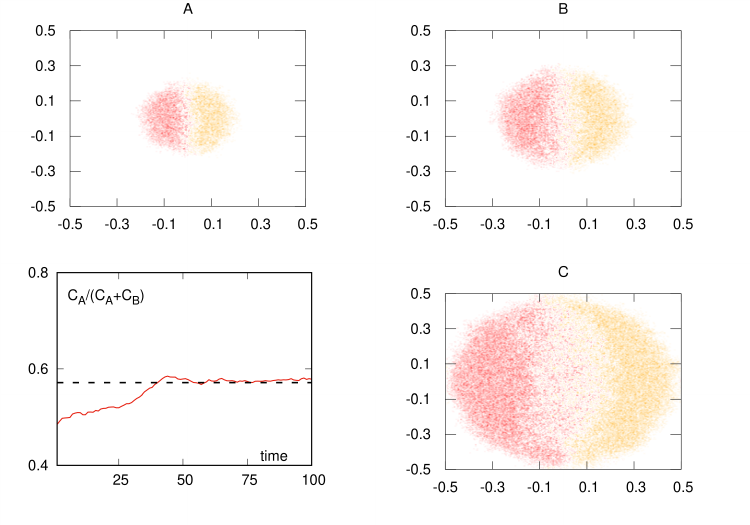}
\end{center}
	\caption{\ss{Agent based simulations of mutualistic organisms in the absence of any flow, with $\e_A=0.4$, $\e_B=0.3$ and effective population size $N_0=2$ (other parameters are fixed as in Sec.~\ref{sec:sim}). The three panels (A-C) show the populations $c_A$ and $c_B$ at time $t=10\tau_g$ (A), $t=20\tau_g$ (B) and $t=45\tau_g$ (C). In the bottom left panel we plot the \ss{fraction} of species A $f(t) = c_A(t)/(c_A(t)+c_B(t))$ as a function of time and show that $f(t)$ approaches the stable fixed point at long times $f^* =\e_A/(\epsilon_A+\epsilon_B )$ (dashed line). While the two populations expand in the plane, a mutualistic region grows outwards from the initial contact line. At long time, the mixed region of ``mutualistic smoke'' dominates.
	}}
\label{fig:mut1}
\end{figure}
It is reassuring to readily recover Eq.~\ref{Model A'} from Eq.~\ref{eqn6} by setting $\b{u}=\b{0}$, a result from spatial population genetics with a rigidly fixed overall population density. However, the reaction-diffusion model embodied in Eqs.~\ref{eq:cA} and~\ref{eq:cB} is much more general (see Ref.~\cite{pigolotti_pop_bio_flow_2013} and Sec.~\ref{subsec:linear} for the generalization that includes number fluctuations). These equations could also be used to study situations where species $A$ and $B$ compete not only with each other, but also with the ``empty set'', i.e., as they invade uncolonized territory. In the latter case, one typically finds \emph{pulled} FKKY population waves~\cite{murray2007math_bio,van_saarloos2003front} invading the frontier, rather than the pushed genetic waves with antagonism \cite{tanaka2017spatial_gene_drive} that are the focus the rest of this paper. For a closely related reaction diffusion model with two colonizing frontier species, one with a selective advantage, see Ref.~\cite{korolev2012selective}. \ss{Note, in this paper, we restrict ourselves to the case of weak antagonism and selective advantage ($\delta<\sigma\ll 1$), for which it is sufficient to work with the effective dynamics (Eq.~\ref{eqn6}) of the relative fraction $f$ of organism A and neglect large scale spatiotemporal variations in the total concentartion ($c_T$). Strong flow gradients and genetic competition that cause large fluctuations in $c_T$ generally necessitate the full model in Eqs.~\ref{eq:cA},~\ref{eq:cB} though.}

It is interesting to use our more general reaction-diffusion model to estimate corrections to the approximate result Eq.~\ref{eqn6}, which assumes that the two populations have grown up such that $c_T(\b{r})\approx 1$ everywhere. For any finite antagonism parameter $\sigma>0$, inserting $c_T=1$ into Eq.~\ref{eq:cT} leads to $\partial_tc_T\approx -(2\sigma/\tau_g)f(1-f)$. Although this term vanishes away from genetic interfaces, when $f=0$ or $f=1$, it shows that the antagonism parameter $\sigma>0$ depletes the overall population at interfaces, where $f(1-f)\approx 1/4$. Indeed, Eq.~\ref{eq:cT} shows that a better approximation to $c_T$ is
\begin{equation}
	c_T(\b{r},t)\approx\dfrac{1}{1+2\sigma\; f(\b{r},t)[1-f(\b{r},t)]}\;.\label{eq:cT1}
\end{equation}
For typical parameters used in our simulations, antagonism leads to $\sim10\%$ dip in the total population near interfaces. This dip modifies the prefactor in the final term on the right-hand of Eq.~\ref{eq:dtf}. Eq.~\ref{eq:cT1} can also be used to estimate the size of the second term on the RHS of Eq.~\ref{eq:dtf}, an effect we also neglect in most of this paper,
\begin{equation}
	2D\vec{\del}f\cdot\vec{\del}\ln(c_T)\approx-2D|\vec{\del}f|^2\dfrac{2\sigma(1-2f)}{1+2\sigma f(1-f)}\;.\label{eq:Dcorr}
\end{equation}
Because $|\vec{\del}f|^2$ is small away from genetic interfaces, and $f=1/2$ at the interface center, we expect this correction to Eq.~\ref{eqn6} to be small everywhere when $\sigma\ll 1$. However, as shown in Appendix~\ref{app:antag}, this term can be viewed as generating a weak but effective velocity sink. See Appendix~\ref{app:antag} for a more detailed exploration of the correction terms above (which can shift the precise values of quantities such critical droplet sizes with and without background flow fields). For most of our analytic calculations, we will work with Eq.~\ref{eqn6} for simplicity, indicating corrections to the basic results caused by dips in the local overall population density when appropriate.

{Before we proceed to discuss the case of antagonistic organisms in detail, we first briefly consider the case of mutualists ($\e_A>0,\,\e_B>0$). The effective potential $U(f)$ (Eq.~\ref{eq:pot}) now instead has a stable minimum at a finite fraction $f^*=\e_A/(\e_A+\e_B)$. In the absence of fluid flows, two neighbouring populations of mutualistic strains initialized as contacting half-discs, will merge and dissolve into each other forming a kind of fully mixed ``mutualistic smoke'' characterized by the stable fixed point $f=f^*$ of the dynamics; see Fig.~\ref{fig:mut1} for a simulation of this scenario using an agent based model that includes demographic noise (details in Sec.~\ref{sec:sim}). Note that the two strains continue to grow larger and invade empty space at their respective genetic Fisher wave speeds, while the interface gets more diffuse and mixed through a slower propagation of a front of ``mutualistic smoke'' into the pure strain populations on either side. Unlike the antagonistic case discussed below, the lack of a potential barrier results in immediate mixing of the two strains at the interface, without any nucleation threshold or an effective line tension.

If we now start with the same initial innoculation geometry, but turn on a converging flow sink centered at the origin ($F<0$ and $\alpha=1$ in Eq.~\ref{eq:u}, so that the flow is compressible, see Fig.~\ref{fig:flow}), we see a strikingly different outcome (Fig.~\ref{fig:mut2}). As explained further in detail for antagonistic organisms in Sec.~\ref{subsec:comp}, the circular drop stops growing and arrests in size at a finite radius when its outward Fisher wave speed exactly matches the inward flow speed of the fluid. Unlike organisms with antagonism, the interface again dissolves into ``mutualistic smoke'' with $f=f^*$ as the stable fixed point, and at long times, the whole stationary drop ends up in this stable mixed state (Fig.~\ref{fig:mut2}). This is evident in both the trace of the species fraction $f(t)$ and the local heterozygosity given by $f(t)[1-f(t)]$. This simple situation with a single stable fixed point already demonstrates some of the key consequences of flow on spatial population genetics. Below we will use the more challenging example of antogonistic interactions in the presence of fluid flows to highlight the key role of biological line tension and explain how fluid advection can dramatically modify our na{\"i}ve expectations of genetic competion and evolution.

\begin{figure}[t!]
\begin{center}
\includegraphics[width=0.80\textwidth]{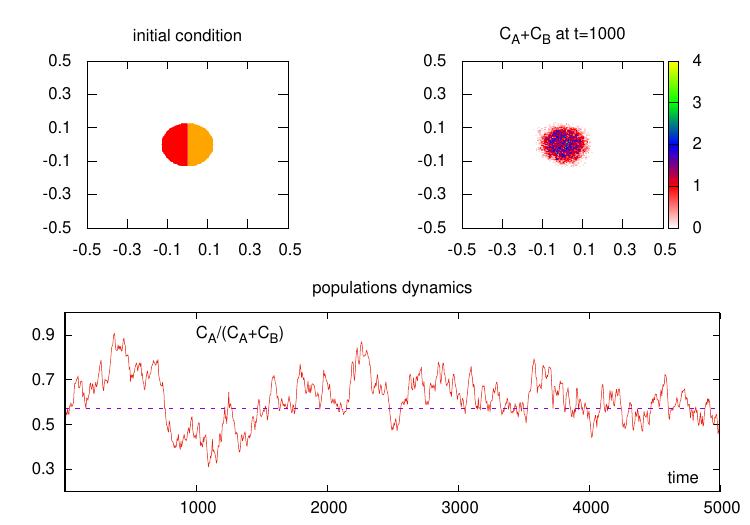}
\end{center}
	\caption{
	\ss{Agent-based numerical simulations of two mutualistic strains in a box of size $L=1$ with periodic boundary conditions, in the presence of a flow sink with $\alpha=1$ and $F=-0.025$ in Eq.~\ref{eq:u} (other parameters as in Fig.~\ref{fig:mut1}). The initial condition is shown in the top left panel: the two populations $c_A$ (red) and $c_B$ (yellow) are set equal to $N=2$ in a disk of radius $L/8$ and zero elsewhere. As time progresses, the drop arrests in size when the sink velocity balances the outward expansion of the two populations. In the meantime, the interface broadens into ``mutualistic smoke'' which eventually takes over the entire stationary drop. In the top right panel we show $c_T=c_A+c_B$ at $t=1000\tau_g$: the two populations are strongly mixed as predicted by the mutualistic {\it stable} solution $f^*=\epsilon_A/(\epsilon_A+\epsilon_B)$. In the bottom panel we show the time behaviour of $f(t)$ (red continuous line) with the dashed horizontal line indicating $f^*$. The population in the droplet fluctuates about $f^*$ at long times.}
	} 
	\label{fig:mut2}
\end{figure}
\section{Nucleation and growth of genetic drops}
\label{sec:drop}

We now return to the case of antagonistic organisms and take $\e_A,\,\e_B<0$.
In the absence of both noise and fluid flows, the simple analogy with equilibrium phase separation (Eq.~\ref{eq:pot}) offers a fruitful way to think about evolutionary dynamics in a spatial setting. As the effective potential $U(f)$ induced by antagonism (see Eq.~\ref{eq:pot}) renders the selectively advantaged strain (say, green for $\delta>0$) as the global minimizer and hence the dominant survivor, it is tempting to conclude that a large circular droplet of green organisms will always grow and invade the surrounding sea of red organisms. After all, this tendency embodies the classical Darwinian paradigm of survival of the fittest! However, note that the ``energy'' functional in Eq.~\ref{eq:pot}, in addition to possessing an asymmetric double well potential, also \textit{penalizes} interfaces between genotypes due to spatial gradients arising from diffusion. As discussed, in, e.g., Refs.~\cite{lavrentovich2019_antagonism} and~\cite{tanaka2017spatial_gene_drive}, and originally explored for spatial population genetics by Barton and collaborators \cite{barton1985hybrid,barton1989adaptation}, the question of how and whether a two-dimensional droplet of green organisms with a selective advantage $\delta>0$ and antagonism $\sigma>0$ takes over in a background of red organisms can be phrased as a nucleation problem \cite{langer1969_nucleation,coleman_nucleation1977,coleman_summer_school_lectures}.

We shall work, for simplicity, in the limit where antagonism dominates the selective advantage ($0<\delta\ll\sigma\ll 1$) and consider fluid flow to be a small perturbation. The fate of a droplet of size $R$ (Fig.~\ref{fig:geom}a) can then be studied by simply looking for stationary saddle point configurations associated with the energy functional in Eq.~\ref{eq:pot}, which in the absence of both stochastic number fluctuations reduces to $\del^2f=-(\sigma/D\tau_g)f(1-f)(2f-1)$. As discussed in, e.g., Refs.~\cite{coleman_summer_school_lectures} and~\cite{bray2002phase_ordering}, droplet solutions spatially interpolating between locally stable minima at $f=0,1$ can be poised on the edge of either expanding or shrinking, depending on the balance between the negative ``energy'' difference between genetically pure states and the positive cost of an interface. In $2d$, an isotropic solution for a large, stationary circular drop (with size $R\gg w$, the interfacial width; see Fig.~\ref{fig:geom}a) separating two competing genotypes in the absence of flow is
\begin{equation}
\label{droplet solution}
	f(r)=\dfrac{1}{2}-\dfrac{1}{2}\tanh \left[\dfrac{r - R}{2w}\right]+\mathcal{O}\left(\dfrac{w}{R}\right)\;,
\end{equation}
where the width of the interface $w=\sqrt{D\tau_g/\sigma}$ is controlled primarily by spatial diffusion ($D$) and antagonism ($\sigma$), the latter of which determines the barrier height. As expected, the effective energy of this droplet solution evaluates to
\begin{equation}
\label{droplet energy}
	E(R) = 2\pi \gamma R - \pi c_0R^2\;,
\end{equation}
and simply compares an interface cost with an effective ``line tension'' $\gamma=(2/3)\sqrt{D\tau_g\sigma}$ (in $2d$) arising from antagonism between the two genotypes, against a bulk ``condensation energy density'' $c_0=\delta/3$, which captures the asymmetric well-depths induced by the selective advantage. This simple description illustrates how antagonism can impede the growth of selectively favored droplets, through an emergent line tension, which sets a critical size $R_c^0=\gamma/c_0=2\sqrt{D\tau_g\sigma}/\delta$ (in $2d$, see Fig.~\ref{fig:geom}c) above which drops will grow and below which they will shrink and die out, just as in cassical nucleation theory.
Precisely this behaviour was recently observed in Petri dish experiments on antagonistic yeast microorganisms by Giometto et al.~\cite{giometto2020antagonism}. Although we foucs on $2d$ drops, a similar analysis can be performed for linear strip-like geometries as well (Fig.~\ref{fig:geom}b). Due to the absence of curvature in $1d$, the nucleation threshold is instead simply of the same order as the interfacial width $w$, and is independent of the selective advantage $\delta$ (Fig.~\ref{fig:geom}d). We shall address this case further in Sec.~\ref{subsec:linear} in the presence of line sinks and sources.

\begin{figure}[]
	\includegraphics[width=0.9\textwidth]{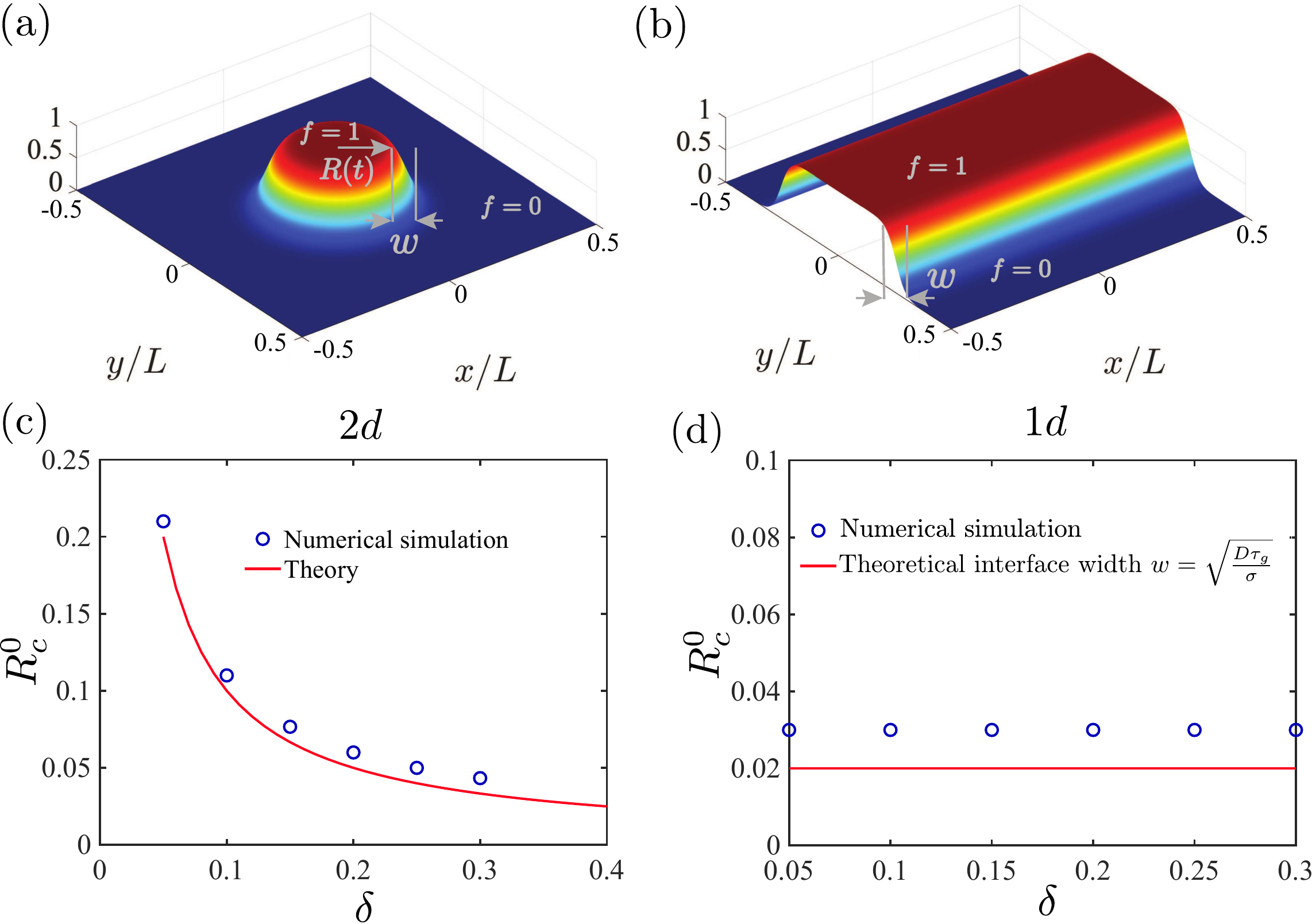}
	\caption{Circular drop (a) and linear strip (b) inoculation geometries along with the selectively-favored organism fraction $f$ displayed with $f=1$ in red and $f=0$ in blue. The width of the interface at the transition from $f=0$ to $f=1$ is $w$ and the radius of the circular drop is $R(t)$. The corresponding critical nucleation thresholds $R_c^0$ for a $2d$ circular drop (c) and a $1d$ linear strip (d), in the absence of flow ($F=0$). The blue circles are obtained from numerical simulations of the full reaction-diffusion model (Eqs.~\ref{eq:cA},~\ref{eq:cB}) and the red line is the theoretical prediction. In $1d$, the critical inoculant size is independent of the selective advantage $\delta$ and is of order the interfacial width $w=\sqrt{D\tau_g/\sigma}$. In $2d$, the critical size is set by a balance between diffusion and the pushed genetic wave, with $R_c^0=2\sqrt{D\tau_g\sigma}/\delta$ decreasing as a function of the selective advantage $\delta$. Parameter values used are $\sigma=0.25$, $D=10^{-4}$ and $\tau_g=1$. \ss{Note all parameters here are nondimensionalized using the generation time $\tau_g=1$ and box size $L=1$.}}
	\label{fig:geom}
\end{figure}

But this simple energetic argument does not capture dynamics of the metastable state! Droplet dynamics is all the more important in the presence of fluid flow. In order to elucidate the nucleation and growth dynamics of a selectively favored drop (say, of green organisms), we derive an effective dynamical equation for the drop radius, in the spirit of similar reduced formulations of coarsening dynamics at thermal equilibrium \cite{bray2002phase_ordering}. As advection by flow is only appreciable when $\vec{\del}f\neq \b{0}$, i.e., in the vicinity of the interface, the fraction of green organisms ($f$) is still smoothly approaches either $f=1$ (within the drop) or $f=0$ (outside the drop). In the simplest case that we first consider below, we assume the drop remains isotropic and circular with $f(r,t)\equiv f(r-R(t))$ (Fig.~\ref{fig:geom}a), with a radius that evolves slowly with a velocity $\dd R/\dd t$. In Secs.~\ref{subsec:incomp},~\ref{subsec:linear}, we generalize the analysis to account for anisotropy and shape deformations as well.

\subsection{Compressible flows: Sinks and sources}
\label{subsec:comp}
\begin{figure}[]
	\includegraphics[width=0.8\textwidth]{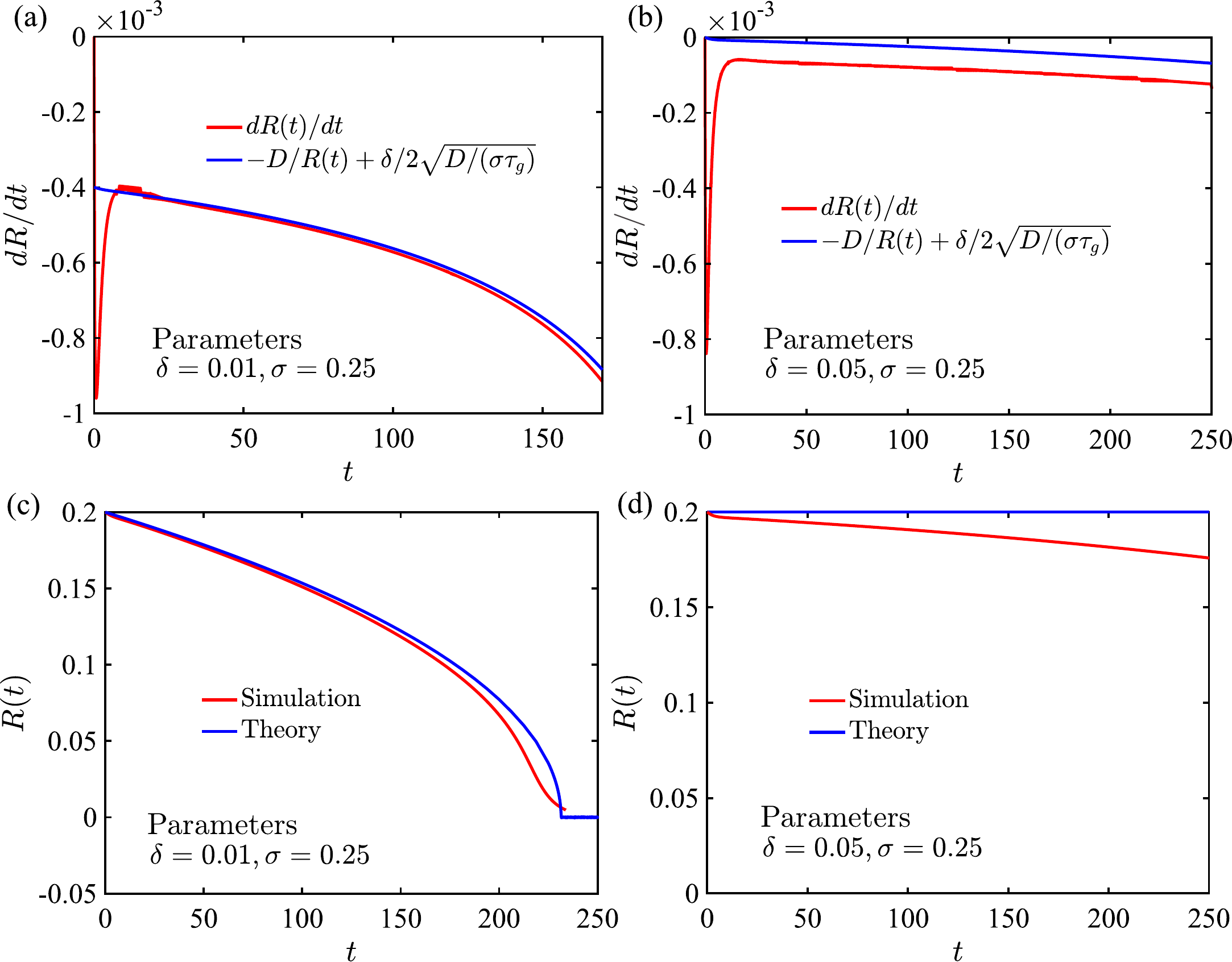}
	\caption{Comparisons between numerical simulations (red curves) employing the full reaction diffusion model (Eqs.~\ref{eq:cA},~\ref{eq:cB}) and the simplified dynamics (blue curves) of the drop interface ($R$), in the absence of flow ($\b{u}=\b{0}$). The initial size of the circular drop $R(0)=0.2$. Both the interfacial velocity and the integrated size evolution match well with the numerical results when the selective advantage is small ($\delta=0.01$; a-c), but display systematic deviations for larger selective advantage ($\delta=0.05$; b-d).}
	\label{fig:noflow}
\end{figure}
Here, we set $\alpha=1$ in the velocity profile (Eq.~\ref{eq:u}) to obtain a fully compressible flow field with either a sink ($F<0$) or a source ($F>0$) located at the origin (compensating defects are also present at the edges of the box due to periodic boundary conditions, but we can neglect their effects for small droplets). For a drop of green organisms with a well localized interface ($w/R\ll 1$), born centered on the origin with a flow source or sink, we assume azimuthal symmetry to write $f(r,t)=f(r-R(t))$, with a comoving profile still given by Eq.~\ref{droplet solution} and $w=\sqrt{D\tau_g/\sigma}$ to lowest order. In the absence of noise, we switch to a frame comoving with the interface to obtain
\begin{equation}
	\left(u_0-\dfrac{\dd R}{\dd t}\right)f'=Df''+\dfrac{D}{r}f'-\dfrac{1}{\tau_g}\dfrac{\dd U}{\dd f}\;,\label{eq:fv}
\end{equation}
where $u_0(r)=(1/2\pi)\int\dd\phi\;\b{u}\cdot\hat{\b{e}}_r$ is the azimuthally averaged local radial velocity, $U(f)$ is given by Eq.~\ref{eq:pot} and the prime denotes $\dd/\dd r$.
Upon multiplying Eq.~\ref{eq:fv} by $f'$ and integrating over all space, we note that $f'$ is sharply peaked at the interface, and vanishingly small elsewhere, and then determine the drop size evolution
\begin{equation}
	\dfrac{\dd R}{\dd t}=-\dfrac{D}{R}+u_0(R)+\dfrac{\delta}{2}\sqrt{\dfrac{D}{\tau_g\sigma}}\;,\quad u_0(R)=2FJ_1\left(\dfrac{2\pi R}{L}\right)\;,\label{eq:compRdot}
\end{equation}
where we have computed the average radial velocity using Eq.~\ref{eq:u} ($J_n(z)$ is a Bessel function of the first kind) and used the fact that $\int\dd r\;f'(r)^2\simeq 1/6w$ for a well localized interface.  As is immediately evident from Eq.~\ref{eq:compRdot}, there are three distinct mechanisms that drive the coarsening dynamics of the drop. The first is simply the pushed wave velocity $v_0=(\delta/2)\sqrt{D/\sigma\tau_g}$ due to the selective advantage of the green organisms (in the drop) over the surrounding red genotype. The second is due to spatial diffusion and interfacial curvature, that induces the Model A variant of a Laplace pressure gradient driving droplet ripening \cite{bray2002phase_ordering}; this effect vanishes for a flat interface ($R\to\infty$). The third mechanism, which is new here, is the advecting flow velocity $u_0(R)$ evaluated at the interface, which depends on the size of the drop itself.

When fluid flow is absent ($F=0$), we recover previous results for a growing genetic droplet of antagonistic organisms \cite{lavrentovich2019_antagonism}. In this case, droplets of the selectively favored organism need to exceed the critical size $R_c^0=D/v_0=2\sqrt{D\tau_g\sigma}/\delta$ ($\delta>0$) to overcome the genotypic line tension from antagonism ($\sigma>0$) to survive and invade the competing strain. Large drops then expand with a nearly constant speed ($R(t)\simeq v_0 t$) set by the pushed genetic wave while small droplets with initial size $R(0)<R_c^0$ shrink and vanish over a typical diffusive time scale $\tau_D\sim R(0)^2/2D$ with a characteristic square-root scaling in time ($R(t)^2=R(0)^2-2Dt$). This behaviour agrees with the simple energetic argument (Eq.~\ref{droplet energy}) and matches numerical simulations of the full reaction diffusion model (Fig.~\ref{fig:noflow}). However, nonzero flow from a source or sink modifies this scenario in a striking way. When $F\delta>0$, the radial fluid velocity $u_r(R)$ aids the pushed wave speed $v_0$ to overcome the diffusive flux, thereby effectively enhancing the selective advantage of the green organisms in the droplet and suppressing the critical nucleation size threshold. We shall work with $\delta>0$ and $R\ll L$, and hence linearize the flow field near the origin as $u_0(R)\approx 2\pi F(R/L)$ to obtain
\begin{equation}
	\dfrac{R_c(F)}{R_c^0}=\dfrac{\sqrt{1+4{\rm Pe}}-1}{2{\rm Pe}}\quad ({\rm Pe}>0)\;,\label{eq:Rc}
\end{equation}
where we have defined a flow-dependent \emph{P{\'e}clet number} ${\rm Pe}=u_0(R_c^0)R_c^0/D\simeq 2\pi F (R_c^0)^2/DL$ (for $R/L\ll 1$) which measures the relative importance of fluid advection to diffusion (alternately, ${\rm Pe}=u_0(R_c^0)/v_0$ is also the ratio of the fluid speed to the genetic wave speed). Note that ${\rm Pe}$ can be of either sign depending on $F$. As expected, $R_c(F)\to R_c^0$ when ${\rm Pe}\to 0$ and decreases monotonically for increasing ${\rm Pe}>0$ (Fig.~\ref{fig:comp}). Hence drops with a selective advantage $\delta>0$ benefit from being born on a source ($F>0$). Conversely, for a given initial size $R(0)$, as the outward flow confers an enhanced selective advantage to the drop, the critical $\delta_c$ required for the drop to grow and take over the entire domain is lowered as
\begin{equation}
	\delta_c(F)=\delta_c(0)-2u_0(R(0))\sqrt{\dfrac{\sigma\tau_g}{D}}\;,\label{eq:deltac}
\end{equation}
where $\delta_c(0)=[2D/R(0)]\sqrt{\sigma\tau_g/D}$ is the critical selective advantage necessary in the absence of flow.
\begin{figure}[]
	\includegraphics[width=0.5\textwidth]{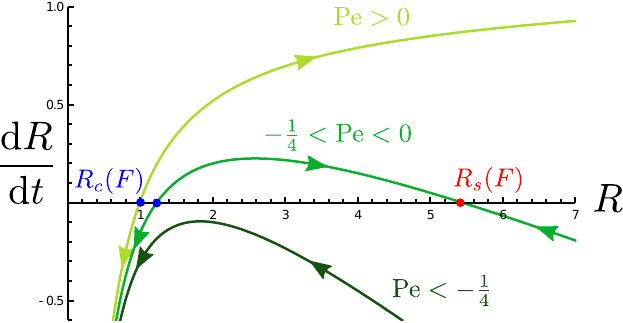}
	\caption{The radial dynamics of a circular drop \ss{(Eq.~\ref{eq:compRdot} with $R/L\ll 1$)} with selective advantage $\delta>0$ born either at a sink (${\rm Pe}<0$) or a source (${\rm Pe}>0$). For ${\rm Pe}>0$, there is only one unstable fixed point, $R_c(F)$ (Eq.~\ref{eq:Rc}), below which drops shrink and die, and above which drops grow indefinitely. Intermediate flows into a sink ($-1/4<{\rm Pe}<0$) arrests the growth of large enough droplets, leading to stable coexistence at size $R_s(F)$ (Eq.~\ref{eq:Rs}) where the outward wave speed balances the inward fluid flow. Stronger sink flows (${\rm Pe}<-1/4$) cannot be balanced by the expanding pushed wave and all drops collapse to extinction, no matter the selective advantage.}
	\label{fig:comp}
\end{figure}

In the opposite limit of a sink ($F\delta<0$, ${\rm Pe}<0$), the dynamics is rather different. For very strong flows ($|{\rm Pe}|\gg 1$), the inward fluid advection out-competes the outward pushed wave and causes all droplets to collapse, irrespective of their selective advantage ($\delta>0$)! Moderate flow strengths in a sink reveal yet another striking scenario. Upon once again linearizing the flow around the origin ($R/L\ll 1$), for $-1/4<{\rm Pe}<0$, a new stable fixed point emerges (Fig.~\ref{fig:comp}). The continuation of Eq.~\ref{eq:Rc} for ${\rm Pe}<0$ (i.e., $R_c(F)=(R_c^0/2|{\rm Pe}|)[1-\sqrt{1-4|{\rm Pe}|}]$) remains an unstable fixed point demarcating the critical size below which drops evaporate and above which they grow. An important new stable fixed point appears at a larger drop size and is given by
\begin{equation}
	\dfrac{R_s(F)}{R_c^0}=\dfrac{1+\sqrt{1-4|{\rm Pe}|}}{2|{\rm Pe}|}\quad \left(-\dfrac{1}{4}<{\rm Pe}<0\right)\;.\label{eq:Rs}
\end{equation}
Remarkably, large enough drops do not coarsen and grow indefinitely, but instead, they now approach a stable finite size given by $R_s(F)$ that diverges as $|{\rm Pe}|\to 0$ (Fig.~\ref{fig:comp}). The flow into the sink balances the outward pushed wave and allows the two antagonistic strains to achieve stable coexistence, in marked contrast to the na{\"i}ve binary outcome predicted on the basis of competitive exclusion and the nonzero selective advantage! Spatial segregation and advection are key to this evolutionary strategy. See Fig.~\ref{fig:cAsim}a for a check on these predictions, using the full reaction-diffusion model.
As a point of comparison, it is useful to note that, similar reaction-diffusion models used to model biomolecular condensates within living cells \cite{Berry_2018,weber2019physics} also exhibit finite size droplets stabilized solely by active chemical fluxes. However, this mechanism is distinct from the hydrodynamic selection mechanism uncovered here, as the stabilization of coexistence in a finite genetic droplet crucially relies on \emph{compressibility} of the fluid flow.

\begin{figure}[]
	\includegraphics[width=0.9\textwidth]{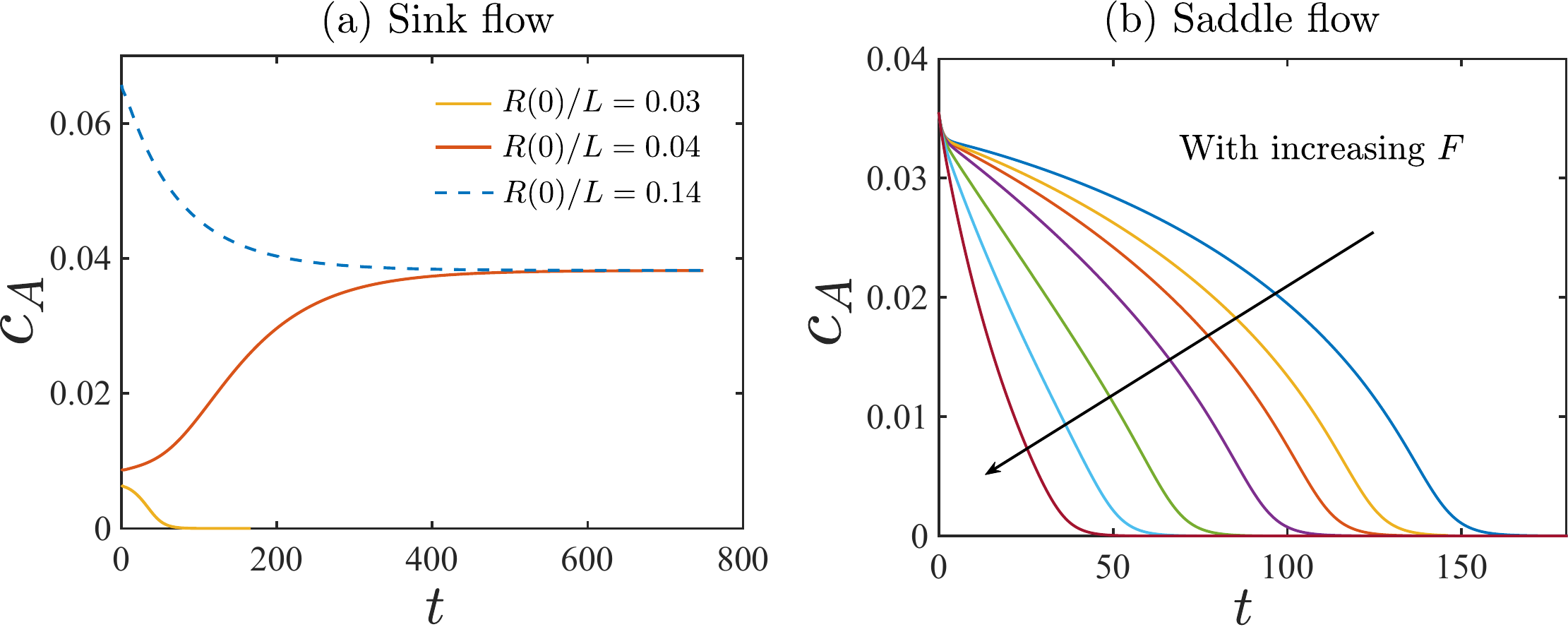}
	\caption{Numerical simulations of the reaction-diffusion model in the presence of fluid flow. (a) Dynamics of the total concentration of genotype $A$ ($c_A$) in a circular drop born centered on a sink ($F<0$). Very small droplets (e.g.~$R(0)/L=0.03$) vanish more rapidly for a finite sink strength $F$, while larger ones (e.g., $R(0)/L=0.04$ and $0.14$) approach a stable asymptotic size, consistent with the theoretical predictions. (b) Extinction dynamics of a drop born on a saddle flow. Total amount of the $A$ genotype for a droplet with $\delta=0.1$, $\sigma=0.25$ and $R(0)/L=0.1$ as a function of time with increasing saddle strength $F$. The blue curve is for $F=0$ and increasing $F$ shortens the droplet lifetime by stretching the drop in one direction and shrinking it in another.}
	\label{fig:cAsim}
\end{figure}

In Appendix~\ref{app:antag}, we show that, in a more accurate description that takes into account the dip in the overall population density at antagonistic interfaces, Eq.~\ref{eq:compRdot} is replaced by
\begin{equation}
	\dfrac{\dd R}{\dd t}=-\dfrac{D}{R}+u_0(R)+\dfrac{\delta}{2}\sqrt{\dfrac{D}{\tau_g\sigma}}\dfrac{1}{(1+\sigma/2)}\;.\label{eq:Rdotimprov}
\end{equation}
In other words, the pushed wave speed $v_0=[\delta/(2+\sigma)]\sqrt{D/\tau_g\sigma}$ gets further suppressed by antagonism. The shift in quantities such as the critical droplet size (or, equivalently, the critical selective advantage required for a droplet of fixed size to grow) is in good agreement with simulations of the full deterministic Eqs.~\ref{eq:cT},~\ref{eq:dtf} (see Fig.~\ref{fig:cAsim}a).

\subsection{Incompressible flows: Saddles and vortices}
\label{subsec:incomp}
We now consider the opposite limit of incompressible flows ($\alpha=0$ in Eq.~\ref{eq:u}). As shown in Fig.~\ref{fig:flow}, the elementary topological singularities are now simple vortices or hyperbolic fixed points, i.e., saddles. A drop of green organisms born on a vortex simply performs rigid body rotation at an angular velocity determined by \ss{(half)} the finite vorticity $\omega=\vec{\del}\times\b{u}$ in the vicinity of the vortex center, i.e., a rotation induced Coriolis force. As the flow itself is primarily tangential, the drop continues to grow isotropically and radially, largely unaffected by the vorticial motion. Stronger inertial flow velocities could potentially also generate an appreciable centrifugal force ($\propto F^2$) which would drive motion in the radial direction, but we neglect these effects here. A more interesting case is that of a saddle flow, where the flow is convergent along one axis and extends outwards along an orthogonal axis, as on the left side of Fig.~\ref{fig:flow}.

\begin{figure}[]
	\includegraphics[width=\textwidth]{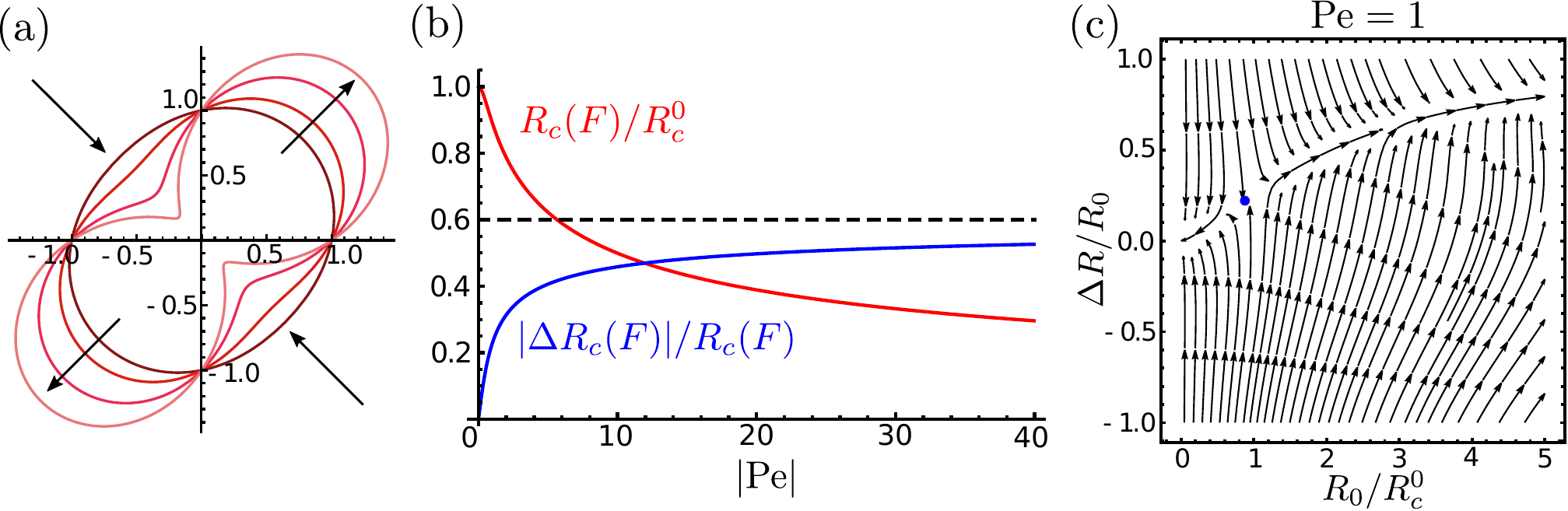}
	\caption{The dynamics of an initially circular drop with selective advantage $\delta>0$ born centered on a saddle flow. (a) Plot of the parametrized drop interface showing a deforming droplet with increasing $\Delta R/R_0$ (dark to light red) as it gets stretched and sheared by a saddle flow. (b) The critical size threshold ($R_c(F)/R_c^0$, red) and corresponding shape anisotropy ($|\Delta R_c(F)|/R_c(F)$, blue) as a function of $|{\rm Pe}|$. Note, while $R_c(F)$ is an even function of Pe, $\Delta R_c(F)$ is an odd function of Pe. For $|{\rm Pe}|\to\infty$, the critical nucleus size decreases with increasing flow as $R_c(F)\propto |{\rm Pe}|^{-1/2}$ and the shape anisotropy asymptotes to a constant $\Delta R_c(F)/R_c(F)\to 0.6$. (c) Phase portrait of the fully nonlinear dynamics (Eq.~\ref{eq:nonlin}, see Appendix~\ref{app:nonlin}) for ${\rm Pe}=1$, with the unstable fixed point marked in blue. At the boundary, when $|\Delta R|/R_0=1$, we have drop splitting and our single drop analysis breaks down.
	}
	\label{fig:incomp}
\end{figure}

Fig.~\ref{fig:cAsim}b shows qualitatively the fate of a small genetically favored droplet, initially centered on a saddle point. The favored genotype does not exceed the critical droplet size even when $F=0$, but dies out all the more rapidly when stretched by a saddle flow.
By virtue of the geometry of the saddle flow streamlines, an initially circular drop will inevitably develop anisotropy, morphing into an oval shape that elongates along the extensional shear axis and shrinks along the convergent axis of the saddle (see Fig.~\ref{fig:incomp}a and Fig.~\ref{fig:incompsim}c-d). A better understanding generalizes our previous analysis to anisotropic drop shapes. By allowing angular variations of the drop shape, we now parametrize the interface of a droplet centered on the origin by a function $R(\phi,t)$, so that the drop solution now becomes $f(\b{r},t)=f(r-R(\phi,t))$. The interfacial velocity $\partial_tR$ is then given by
\begin{equation}
	\partial_tR=u_r-\dfrac{1}{R}u_\phi\partial_\phi R-\dfrac{D}{R}+\dfrac{D}{R^2}\partial^2_\phi R+v_0\;,\label{eq:saddle1}
\end{equation}
where $v_0=(\delta/2)\sqrt{D/\sigma\tau_g}$ is the speed of a flat pushed wave as before, and the radial ($u_r=\hat{\b{e}}_r\cdot\b{u}(R,\phi)$) and tangential ($u_\phi=\hat{\b{e}}_\phi\cdot\b{u}(R,\phi)$) flow velocities depend explicitly on the polar angle $\phi$ and the interface position $R(\phi,t)$. We now decompose Eq.~\ref{eq:saddle1} in angular Fourier harmonics by writing $R(\phi,t)=R_0(t)+\Delta R(t)\sin(2\phi)$, thus retaining only the lowest order modes characterizing shear driven size ($R_0$) and elliptic shape ($\Delta R$) change of the drop (Fig.~\ref{fig:incomp}a). A similar Fourier decomposition of the flow-field in Eq.~\ref{eq:u} (with $\alpha=0$) gives $u_r(R,\phi)=u_0(R_0)(\Delta R/R_0)+\Delta u_r(R_0)\sin(2\phi)$ and $u_{\phi}(R,\phi)=\Delta u_\phi(R_0)\cos(2\phi)$, where
\begin{gather}
	u_0(R_0)=2F\left[\dfrac{L}{\pi R_0}J_2\left(\dfrac{2\pi R_0}{L}\right)-2J_3\left(\dfrac{2\pi R_0}{L}\right)\right]\approx \pi F\dfrac{R_0}{L}+\mathcal{O}\left(\dfrac{R_0}{L}\right)^3\;,\\
	\Delta u_r(R_0)=\dfrac{4FL}{\pi R_0}J_2\left(\dfrac{2\pi R_0}{L}\right)\approx 2\pi F\dfrac{R_0}{L}+\mathcal{O}\left(\dfrac{R_0}{L}\right)^3\;,\\
	\Delta u_\phi(R_0)=4F\left[J_1\left(\dfrac{2\pi R_0}{L}\right)-\dfrac{L}{\pi R_0}J_2\left(\dfrac{2\pi R_0}{L}\right)\right]\approx 2\pi F\dfrac{R_0}{L}+\mathcal{O}\left(\dfrac{R_0}{L}\right)^3\;,
\end{gather}
are the only nonvanishing components to leading order in $\Delta R/R_0$. Here, the $J_n(z)$ are Bessel functions and the final approximate equalities above further assume a small droplet size, $R_0/L\ll 1$. Upon projecting Eq.~\ref{eq:saddle1} onto these low order Fourier modes and expanding to first order in $\Delta R(t)/R_0(t)$, we obtain the effective shape dynamics of the interface, valid for times such that a $2$-mode Fourier truncation is a good approximation,
\begin{figure}[]
	\includegraphics[width=0.8\textwidth]{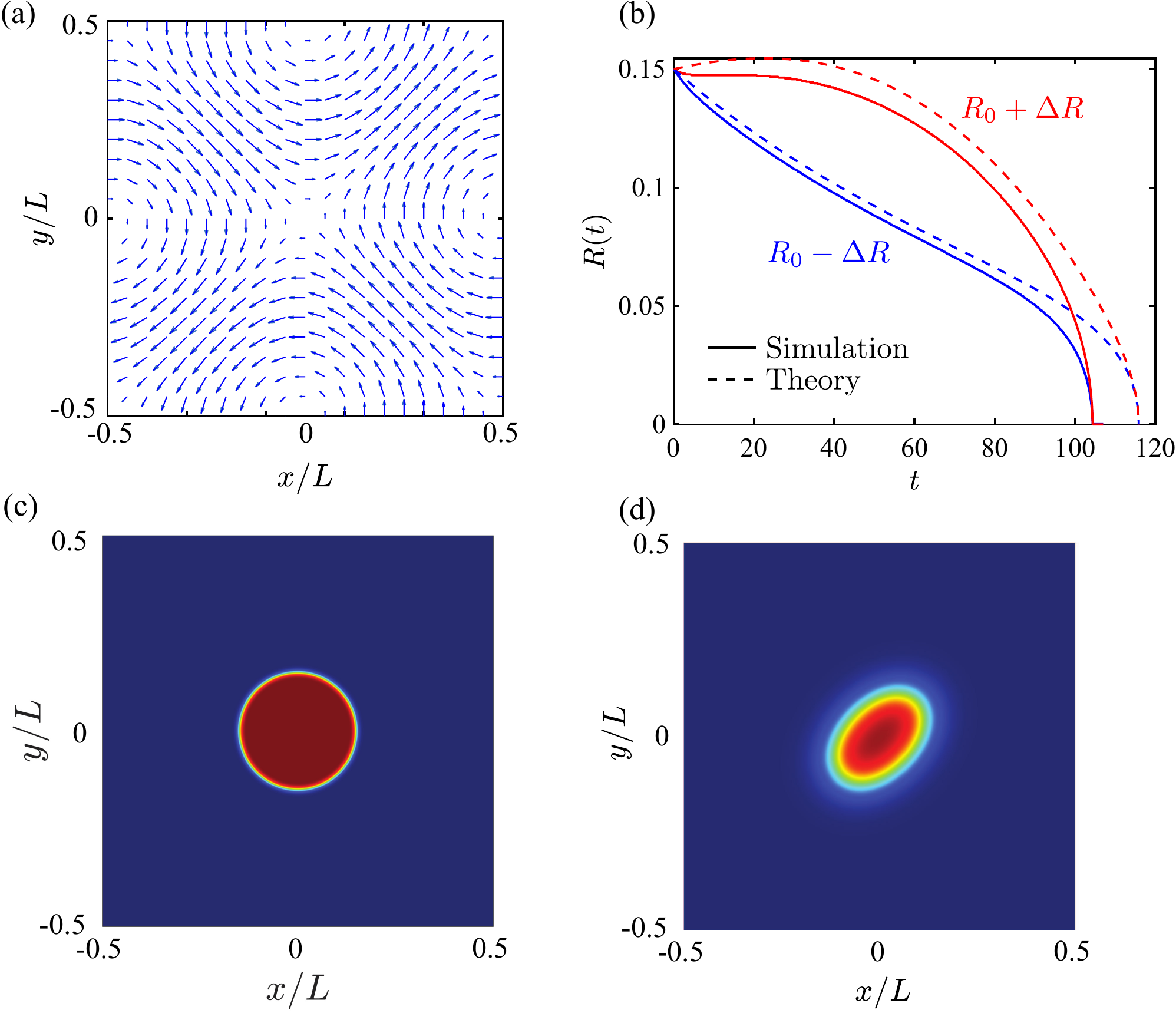}
	\caption{Simulated dynamics of a circular drop born centered on a saddle flow. (a) The flow field is given by Eq.~\ref{eq:u} with $\alpha=0$ (incompressible) and $F=0.0025$. (b) The simplified description of the drop interface (Eqs.~\ref{eq:dR},~\ref{eq:ddR}) approximately captures the dynamics of a drop deformed into an oval shape by the shearing flows, except close to the time the droplet radii vanish or the assumption of sharp interfaces breaks down. The time evolution of the major ($R_0(t)+\Delta R(t)$, red) and minor ($R_0(t)-\Delta R(t)$, blue) axis of the drop is compared between the theoretical prediction (dashed line) from Eqs.~\ref{eq:dR},~\ref{eq:ddR} and the full numerical simulations (solid line) using Eqs.~\ref{eq:cA},~\ref{eq:cB}. (c-d) Snapshots of the selectively favored fraction $f$ ($f=0$: blue, $f=1$: red) showing an initially circular drop getting deformed into an oval shape by the anisotropic fow field, for short times. The initial radius is $R(0)/L=0.11$ with $\delta=0.1$ and $\sigma=0.25$.}
	\label{fig:incompsim}
\end{figure}

\begin{align}
	\dfrac{\dd R_0}{\dd t}&=-\dfrac{D}{R_0}+\left[u_0(R_0)-\Delta u_\phi(R_0)\right]\dfrac{\Delta R}{R_0}+\dfrac{\delta}{2}\sqrt{\dfrac{D}{\sigma\tau_g}}\;,\label{eq:dR}\\
	\dfrac{\dd \Delta R}{\dd t}&=-3D\dfrac{\Delta R}{R_0^2}+\Delta u_r(R_0)\;.\label{eq:ddR}
\end{align}
Provided the principle droplet radii ($R_0\pm\Delta R$) are not close to vanishing, this simple dynamics of the drop size and shape matches well with numerical simulations of the full reaction-diffusion model (Eqs.~\ref{eq:cA},~\ref{eq:cB}), see Fig.~\ref{fig:incompsim}b.
As before, we can search for steady solutions of these equations. Upon linearizing the flow around the origin (again with $R_0/L\ll 1$), we once again define an effective flow P{\'e}clet number ${\rm Pe}=\Delta u(R^0_c)R^0_c/D\simeq 2\pi F(R^0_c)^2/DL$. Here, $\Delta u=(\Delta u_r+\Delta u_\phi)/2$ measures the average anisotropic flow strength in the saddle. The generalization of Eqs.~\ref{eq:dR},~\ref{eq:ddR} to fully account for nonlinear shape deformations to all orders in $\Delta R/R_0$ is given in Appendix~\ref{app:nonlin}. Upon analyzing these equations, we find that for all values of Pe, there is only one unstable fixed point present (see Fig.~\ref{fig:incomp}c for a representative phase portrait of the dynamics). For small $|{\rm Pe}|\ll 1$, the fixed point values as a function of the flow strength $F$ (as embodied in the P{\'e}clet number Pe) are given by
\begin{equation}
	\dfrac{R_c(F)}{R_c^0}=1-\dfrac{2}{9}{\rm Pe}^2+\mathcal{O}({\rm Pe}^4)\;,\quad\dfrac{\Delta R_c(F)}{R_c(F)}=\dfrac{{\rm Pe}}{3}+\mathcal{O}({\rm Pe}^3)\;.\label{eq:Rc1}
\end{equation}
Eq.~\ref{eq:Rc1} gives the critical nucleation shape parameters below which drops succumb to interfacial tension. Shear flow around the saddle point deforms the drop ($\Delta R_c(F)\propto {\rm Pe}$) into an oval shape and hence suppresses the nucleation threshold ($R_c(F)<R_c^0$), irrespective of the sign of $F$. The full dependence of the fixed point values on Pe is shown in Fig.~\ref{fig:incomp}b. For large values of ${\rm Pe}\to\infty$, we find
\begin{equation}
	\dfrac{R_c(F)}{R_c^0}=\left(\dfrac{125}{24|{\rm Pe}|}\right)^{1/2}+\mathcal{O}\left(\dfrac{1}{|{\rm Pe}|}\right)\;,\quad\dfrac{|\Delta R_c(F)|}{R_c(F)}=\dfrac{3}{5}+\mathcal{O}\left(\dfrac{1}{|{\rm Pe}|^{1/2}}\right)\;.\label{eq:Rc2}
\end{equation}
\begin{figure}[]
	\includegraphics[width=\textwidth]{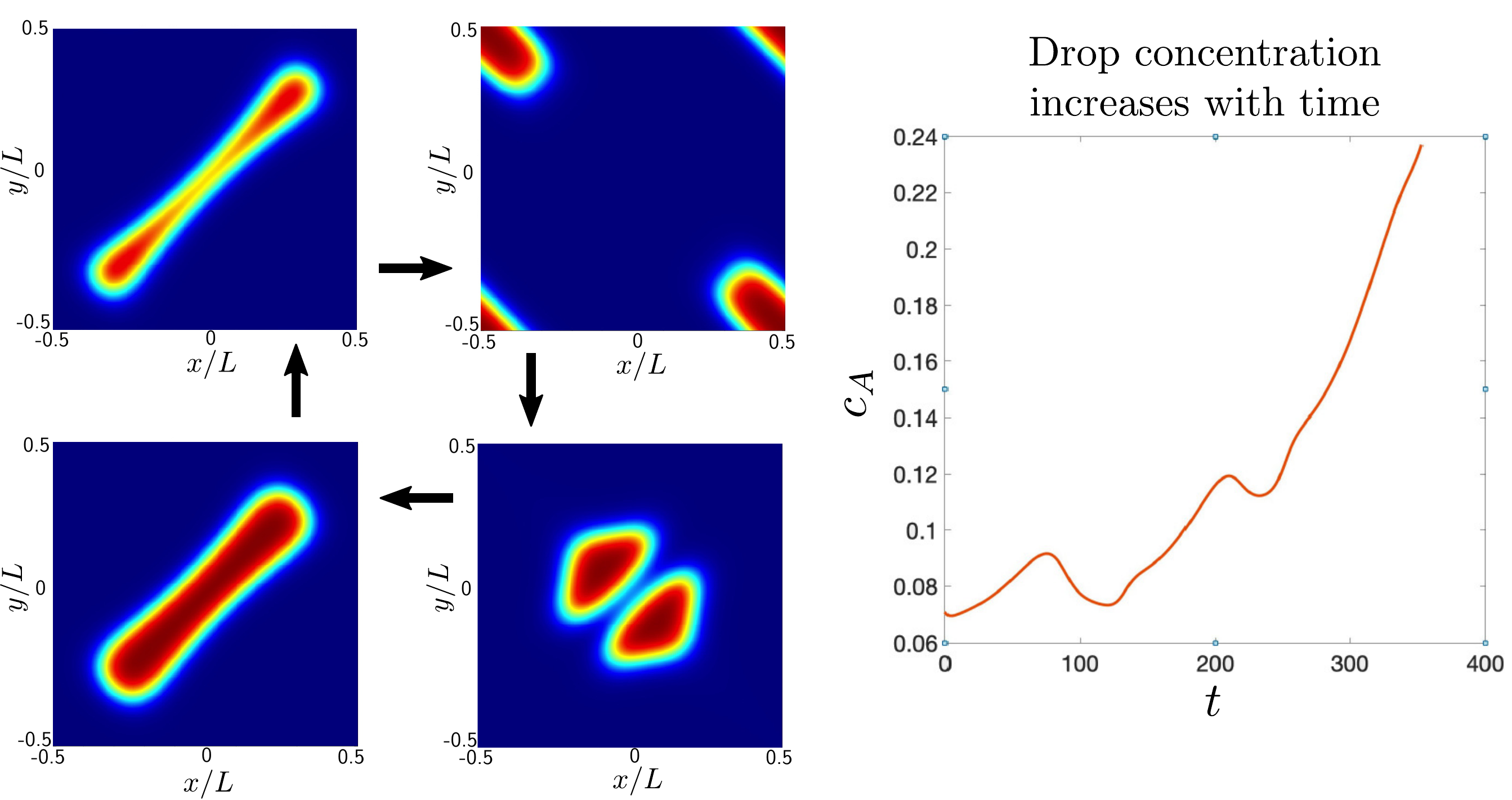}
	\caption{Evolution of a large drop born on a periodic saddle flow (Eq.~\ref{eq:u} with $\alpha=0$). The time evolution of a \emph{large} selectively favored drop born initially on a weakly shearing saddle point that eventually takes over the entire periodic domain. This happens via an oscillatory process, with splitting and recombination events, as the drop is exchanged between two orthogonally oriented saddle points, one at the origin and the other at the corners of the periodic box (see Fig.~\ref{fig:incompsim}a). The total concentration $c_A$ of this genotype in the periodic box grows nonmonotonically (right), with the drop exploiting the periodic boundary conditions to survive and grow. In these simulations, the initial radius $R(0)/L=0.15$ and the flow strength $F=0.0025$.}
	\label{fig:split}
\end{figure}
\ss{Note, the $|{\rm Pe}|^{-1/2}$ scaling for the critical size can be simply understood as the size $R$ at which the (inverse) diffusion time on the scale of the drop ($\sim D/R^2$) is comparable to the flow shear rate ($\sim F/L$).}
Strong shearing by the saddle flow thus produces an asymptotically constant shape at the nucleation threshold, beyond which the drop continues to grow, albeit anisotropically. However, unlike the previous cases of no-flow and compressible flows with sources and sinks that allow the drop to grow and expand eternally beyond the nucleation threshold, saddle flows reveal a distinct phenomenon. As the drop grows, drops get sheared and progressively stretched out by the saddle flow, and the $2$-mode approximation above breaks down. Eventually the drop develops a neck when $|\Delta R|/R_0=1$, and the drop then pinches off and splits into two daughter droplets. Once the drop divides into two, more complex phenomena can occur. For instance, the resulting daughter drops may each have a size that is smaller than the nucleation threshold, thereby causing them to shrink and go extinct. However, in the presence of periodic boundaries and weakly shearing flows, the split drops can get advected away to the corners of the box, only to get reinjected back into the saddle point at the origin, where the colliding droplets coalesce and get sheared and torn apart yet again (Fig.~\ref{fig:split}). This remarkable nonmonotonic dynamics enables drops to grow and eventually colonize the entire system, despite the shear induced deformations generated by the advecting velocity (see right panel of Fig.~\ref{fig:split}).

Unlike mechanisms of shear-driven hydrodynamic scale selection underlying the arrest of phase separation in passive binary mixtures by turbulent advection \cite{perlekar2014spinodal,berti2005turbulence} (described by Model $B$ dynamics \cite{hohenberg1977critical_dynamics} instead of the Model $A$ dynamics explored here), or in self-shearing flows in active fluids \cite{zwicker2017growth,singh2019hydrodynamically}, our continually growing genetic drops are never stabilized into finite size coexistence by shear. Instead, strong hyperbolic flows rip apart drops faster than they can grow, causing rapid extinction of the small daughter droplets, while weakly shearing flows strikingly allow subsequent growth and dominance through cycles of drop splitting and merger events that crucially exploit the periodicity of the cellular flow field.

\subsection{Linear strip: Instabilities, pinch-off and fluctuations}
\label{subsec:linear}
\begin{figure}[]
	\includegraphics[width=0.8\textwidth]{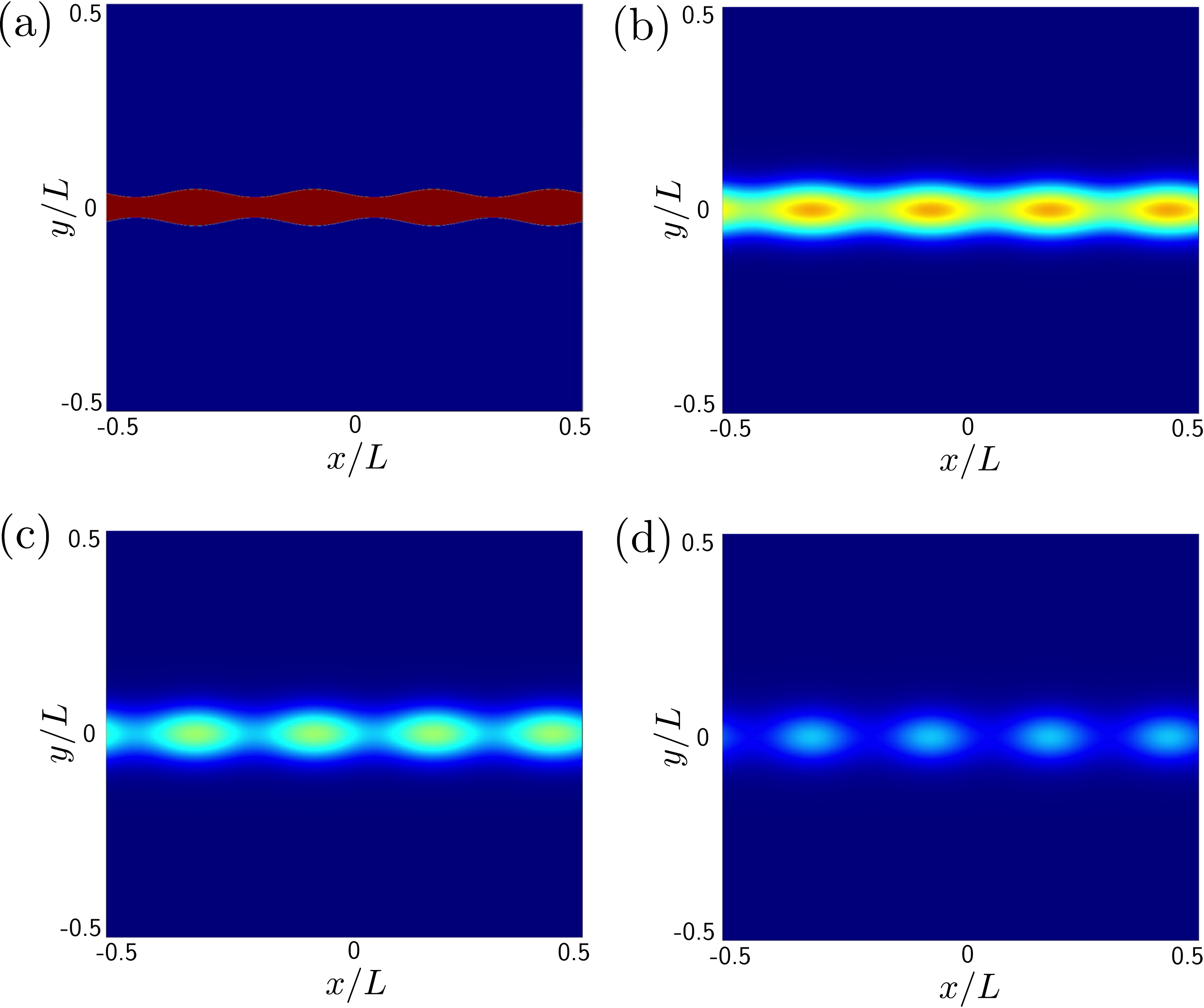}
	\caption{Deterministic dynamics of a linear strip with a spatially modulated initial condition, with periodic boundary conditions along the $x$-axis. A selectively favored red strip of genotype $A$ ($\delta=0.1$ and $\sigma= 0.25$) has been inserted into a blue background genotype, and centered on a line source (of strength $F = 0.0025$).To a strip of initial width $W/L = 0.07$ has been added a modulation of the form $0.01\sin[2\pi(4x/L)]$ (the stochastic number fluctuations discussed in the text would produce similar perturbations at all wavelengths). Instead of expanding indefinitely in the vertical direction, as would be the case in the absence of the sinusoidal modulation, the strip is instead unstable to this particular undulation, with a wavelength $1/4$ of the box size. The strip now breaks up into four droplets, each smaller than the critical radius in two dimensions, which then disappear.}
	\label{fig:linsim}
\end{figure}
We now consider the simpler situation of a one-dimensional ($1d$) strip inoculated on a line sink or source ($\b{u}=F(y/L)\hat{\b{e}}_y$). This problem is simple enough to allow us to reintroduce number fluctuations to study the effect of noise (i.e., ``genetic drift'') on the growth of the interface between the green and red organisms (line sources and sinks also arise, e.g., for the flows in Eq.~\ref{eq:u} with $\alpha=0.5$, see Fig.~\ref{fig:flow}). By parametrizing an interface oriented on average along the $x$-axis as $R(x,t)$ (so $f(\b{r},t)=f(y-R(x,t))$, see Fig.~\ref{fig:geom}b and Fig.~\ref{fig:linsim}), we can perform the same kind of calculation, starting with Eq.~\ref{eqn6}, as was done in Sec.~\ref{subsec:incomp}, to obtain
\begin{equation}
	\partial_tR=D\partial_x^2R+u_y(R)+v_0+\eta\;,\label{eq:dRdtlin}
\end{equation}
where now in addition to the diffusive relaxation of curvature ($D\partial_x^2R$), flow advection up the $y$-axis ($u_y$) and the pushed wave speed ($v_0=(\delta/2)\sqrt{D/\sigma\tau_g}>0$, assuming a positive selective advantage $\delta>0$), we also have stochastic number fluctuations $\eta(x,t)$ that are Gaussian and occur right at the interface (where $f\approx 1/2$ in Eq.~\ref{Model A'}), with zero mean and correlations
\begin{equation}
	\langle\eta(x,t)\eta(x',t')\rangle=\Delta \delta(t-t')\delta(x-x')\;,
\end{equation}
where $\Delta=(2/\tau_gN_0)[\int\dd y\;f(1-f)f'(y)^2]/[\int\dd y\;f'(y)^2]^2=12w/(5\tau_gN_0)$ is the effective (constant) noise strength. Note that, as we are focusing just on the interface, the noise is no longer multiplicative (as in Eq.~\ref{Model A'}), but is simply additive. This approximation is justified as number fluctuations simply vanish away from the interface where $f$ is zero or unity. We now assume $R(x,t)=R_0(t)+\int(\dd q/2\pi)\;e^{-iqx}R_q(t)$ and explore the dynamics of Fourier components $R_q(t)$ with wave-vector $q\neq 0$ and a uniform translation represented by $R_0(t)$. This decomposition decouples the Fourier components to give
\begin{equation}
	\dfrac{\dd R_0(t)}{\dd t}=v_0+F\dfrac{R_0(t)}{L}+\eta_0(t)\;,\quad\dfrac{\dd R_q(t)}{\dd t}=\left(\dfrac{F}{L}-Dq^2\right)R_q(t)+\tilde{\eta}_q(t)\;,
\end{equation}
where $\langle\eta_0(t)\eta_0(t')\rangle=\Delta\delta(t-t')$ and $\langle\tilde{\eta}_q(t)\tilde{\eta}_{q'}(t')\rangle=2\pi\Delta\delta(q+q')\delta(t-t')$. In the absence of flow ($F=0$), the interface simply advances on average with a constant speed ($\langle R_0(t)\rangle=R_0(0)+v_0 t$) and all other finite wavelength perturbations relax exponentially to zero ($\langle R_q(t)\rangle=R_q(0)e^{-Dq^2t}$). Number fluctuations, however, maintain a fluctuating interface with long-ranged correlations as revealed by the equal-time spectrum associated with Eq.~\ref{eq:dRdtlin},
\begin{equation}
	\langle R_q(t)R_{q'}(t)\rangle=\dfrac{\Delta}{2Dq^2}\;2\pi\delta(q+q')\;.
\end{equation}

Classical nucleation theory behaves differently in $1d$ than for $2d$ droplets: The critical size beyond which the strip of, say, green organisms propagates to invade a red genetic background, is of order the interface width ($w$), see Fig.~\ref{fig:geom}d and Ref.~\cite{tanaka2017spatial_gene_drive}.
In the presence of flow ($F\neq 0$), the interface could, in principle, propagate exponentially quickly when $F>0$, i.e., $R_0(t)\sim e^{(F/L)t}$. However, long-wavelength modes are destabilized by the outward advective flow from the line source (short wavelength modes with $q>\sqrt{F/LD}$ remain stable due to diffusion). The unstable $q\to 0$ perturbations eventually lead to pinch-off on a time scale $\sim L/F$, reminiscent of, but distinct from, the Rayleigh-Plateau instability \cite{chandrasekhar1981hydrodynamic} in fluid jets \footnote{Note that the conventional Rayleigh-Plateau instability is driven by surface tension and requires a $3d$ fluid column \cite{chandrasekhar1981hydrodynamic}, unlike the flow driven pinch-off of a $1d$ strip obtained here.}. This phenomenon is illustrated in Fig.~\ref{fig:linsim}, which shows a deterministic simulation of the reaction-diffusion model of Eqs.~\ref{eq:cA},~\ref{eq:cB} with a line source.

In contrast, a line sink ($F<0$) stabilizes and pins the interface at a distance $\langle R_0\rangle=v_0L/|F|$. This is the $1d$ analog of the coexisting drop stabilized on a $2d$ sink, analyzed in Sec.~\ref{subsec:comp}. As expected, the inward fluid flux balances the outward pushed wave leading to a stably positioned interface. Finite wavelength fluctuations are now stable with an exponential relaxation time $\sim|F|/L+Dq^2$ that remains finite as $q\to 0$. As a consequence, the steady state fluctuations of the interface are short-ranged and suppressed by the advective flow,
\begin{equation}
	\langle R_q(t)R_{q'}(t)\rangle=\dfrac{\Delta L}{2|F|\left[1+(q\xi)^2\right]}\;2\pi\delta(q+q')\;,
\end{equation}
with a finite correlation length $\xi=\sqrt{DL/|F|}$ that diverges as $|F|\to 0$.

\section{Simulations with number fluctuations and the $N_0\to\infty$ limit}
\label{sec:sim}
In this section we compare the deterministic theory described in the previous sections against agent-based simulations that include number fluctuations. We use the numerical method illustrated in Ref.~\cite{guccione2019} and consider two specific examples, namely the effect of a source in $2d$ and a sink on a $1d$ line. In both cases, the system size ($L=1$) and the generation time ($\tau_g=1$) are used to fix the units of length and time. The diffusivity and antagonism parameters are fixed to $D=10^{-4}L^2/\tau_G$ and $\sigma = 0.25$ respectively, which gives a genetic interface width of $w=0.02L$. Finally, the {\it deme} size is $a = L/256$ and $N_0$ denotes the number of agents in a deme box of size $a^d$, where $d=1,2$ is the system dimension. \ss{Note the term `deme' refers to a (spatially) local subpopulation of organisms that are subjected to selection as a well-mixed group, rather than tracking each individual.} In the $N_0\to \infty$ limit, number fluctuations become negligible and we expect to recover the deterministic results presented above.

\subsection{Two dimensional source}
We set $\alpha=1$ in Eq.~\ref{eq:u} to obtain a $2d$ source flow at the origin for $F>0$ (Fig.~\ref{fig:flow}). The system is initialized at $t=0$ with a total concentration $c_T=N_0$ everywhere and $c_A=N_0, c_B=0$, corresponding to $f=1$, in a disk of radius $R(0)=1/8$ centred at the origin.
In the absence of flow ($F=0$), we can check that, even for fairly small $N_0$, the interfacial profile of the drop closely matches our deterministic solution (Eq.~\ref{droplet solution}) upto random fluctuations $\sim\mathcal{O}(N_0^{-1/2})$ (see Fig.~\ref{fig:2dsource}a).
Near $f=1$ and $f=0$ we observe fixation into the two absorbing states, while diffusion prevents the local fixation from being stationary. This effect is expected since in each deme of size $a$ with $f\approx 1$ or $0$, the probability to reach the absorbing states is close to $1$ on a time scale $\sim N_0\tau_g$.

\begin{figure}[t!]
\begin{center}
\includegraphics[width=\textwidth]{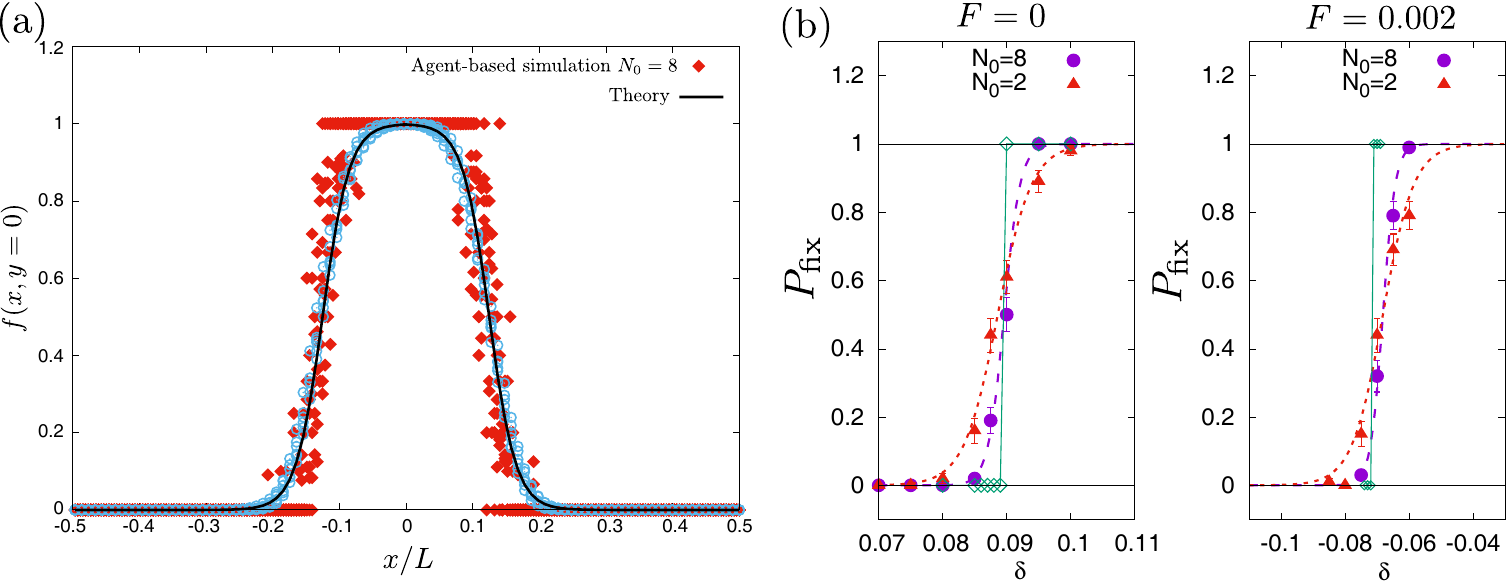}
\end{center}
\label{figure1}
	\caption{Agent-based simulations of a drop born at a point \ss{source} in $2d$. (a) The cross-section of a drop with selective advantage $\delta=0.095$ showing the spatial profile of $f$ at $y=0$. The initial condition is given by $f=1$ on a disk of radius $R(0)=1/8$ centred at the origin. The figure shows the value of $f(x,y=0)$ obtained at $t=10$. The continuous line is given by Eq.~\ref{droplet solution} with $w=0.02$ and $R(0)=1/8$. The open symbols represent a Bezier smoothing procedure applied to the original data set. (b) The fixation probability $P_{\rm fix}(\delta)$ plotted as a function of $\delta$ for $N_0=2$ (triangles) and $N_0=8$ (circles). The continuous line show the deterministic transition at $\delta=\delta_c(F)$, with the left panel for $F=0$ ($\delta_c(0)=0.09$, Eq.~\ref{eq:deltac1}) and the right panel for $F=0.002$ ($\delta_c(F)=-0.07$, Eq.~\ref{eq:deltac1}).}
\label{fig:2dsource}
\end{figure}

From our nucleation analysis in Sec.~\ref{subsec:comp}, we know that a genetic drop born on a source flow acquires an enhanced selective advantage due to fluid advection (Eq.~\ref{eq:deltac}). Upon including the suppression of the total concentration ($c_T$) at the interface due to antagonism (Appendix~\ref{app:antag}), for a given flow strength $F$ and initial drop radius $R(0)$, we find the flow dependent critical selective advantage (from Eq.~\ref{eq:Rdotimprov}) to be
\begin{equation}
	\delta_c(F)=\delta_c(0)-u_0(R(0))\left(2+\sigma\right)\dfrac{\tau_g}{w}\;,\label{eq:deltac1}
\end{equation}
where $\delta_c(0)=(2+\sigma)D\tau_g/[w R(0)]$ is the selective advantage threshold in the absence of flow and $u_0(R)=2FJ_1(2\pi R/L)$ (Eq.~\ref{eq:compRdot}) is the azimuthally averaged radial flow.

In the presence of number fluctuations, the drop will grow (shrink) for $\delta>\delta_c(F)$ ($\delta<\delta_c(F)$), \emph{on average}.
More precisely, one should consider the probability $P_{\rm fix}(\delta)$ to reach fixation at $f=1$ given an initial drop size $R(0)$ and population number $N_0$. The fixation probability smoothly transitions from zero to one upon varying $\delta$ around $\delta_c$, with the transition getting sharper for increasing $N_0$, eventually approaching a step function in the deterministic limit ($N_0\to\infty$).

This expectation is confirmed in Fig.~\ref{fig:2dsource}b which shows the numerically computed $P_{\rm fix}(\delta)$ for $F=0$ (left) and $F=0.002$ (right) corresponding to a point source. Even for small population sizes of $N_0=2$ (triangles) and $N_0=8$ (circles), we see good agreement of the numerical results with the deterministic transition threshold (continuous line) obtained from reaction-diffusion model (Eqs.~\ref{eq:cA},~\ref{eq:cB}). As expected, the agreement improves for larger $N_0$, with the transition becoming sharper.

\begin{figure}[t!]
\begin{center}
\includegraphics[width=\textwidth]{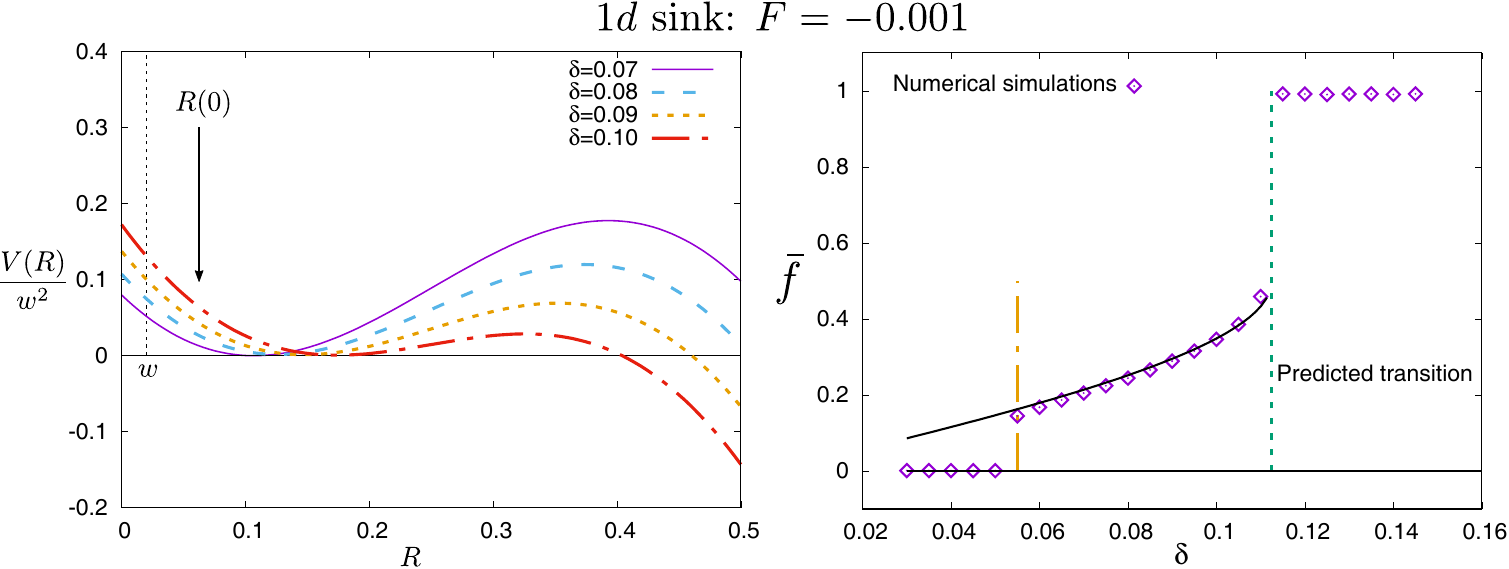}
\end{center}
	\caption{Deterministic dynamics of a drop in $1d$ born on a sink with flow strength $F=-0.001$. (a) The effective potential $V(R)$ (Eq.~\ref{eq:1d}) nondimensionalized by the interface width $w$ and plotted for different values of $\delta$. Note that $V(R)$ is shifted by an unimportant constant to maintain $V(R_{\rm min})=0$, for easy visualization. The dashed vertical line indicates the point where $R=w$ while the vertical arrow indicates the value of the initial drop size, $R(0)=1/16$, used in the numerical simulation. (b) The spatially averaged fraction ($\bar{f}$, open diamonds) obtained from numerical solutions of the reaction-diffusion model (Eqs.~\ref{eq:cA},~\ref{eq:cB}) in $1d$, plotted against varying $\delta$. The simulations are performed using a second order accurate finite difference method on a mesh of $256$ grid points. The continuous line corresponds to the analytical expression $\bar{f}\approx 2R_{\rm min}(\delta)/L$ computed from $V(R)$. Two clear transitions are evident and highlighted with vertical dashed lines. At $\delta=\delta_c$ (blue dashed line) the deterministic solution shows a sudden jump to $\bar{f}=1$ as predicted by our analysis. For small $\delta$, we have a different transition (orange dashed line) to $\bar{f}=0$, which occurs when $R_{\rm min}(\delta)=w$ and corresponds to the $1d$ critical nucleation threshold.}
\label{fig:1dsink}
\end{figure}
\subsection{One dimensional sink}
We now consider the case of selectively favored organisms born at $x=0$ in a $1d$ system with a periodic flow field $u(x)=F\sin(2\pi x/L)$ that hosts a sink at the origin for $F<0$. Just as in Sec.~\ref{subsec:linear}, we can obtain the stochastic time evolution of the interface, which we conveniently write as
\begin{equation}
	\dfrac{\dd R}{\dd t}=-\dfrac{1}{\tau_g}\dfrac{\dd V(R)}{\dd R}+\eta(t)\;,\label{eq:1d}
\end{equation}
where the effective potential $V(R)=\tau_g(FL/2\pi)\cos(2\pi R/L)-w\delta R/(2+\sigma)$ includes the effect of interfacial density suppression from antagonism, on the pushed wave speed. The Gaussian white noise $\eta(t)$ has zero mean and correlation $\langle\eta(t)\eta(t')\rangle=\Delta\delta(t-t')$ with a constant variance $\Delta=12w/(5\tau_gN_0)$. By virtue of being strictly in $1d$, there are no diffusive curvature fluxes present.

Fig.~\ref{fig:1dsink}a shows the effective potential $V(R)$ (offset by a constant) for a sink of strength $F=-0.001$ and different values of $\delta$. Due to the periodicity of the velocity field and the sink flow at the origin, for any $\delta>0$, $V(R)$ displays a minimum at $R=R_{\rm min}(\delta)$ and a maximum at $R=R_{\rm max}(\delta)$. The size $R_{\rm min}(\delta)$ corresponds to a stable local minimum of the deterministic dynamics in Eq.~\ref{eq:1d}, while $R_{\rm max}(\delta)$ corresponds to the distance to the barrier, beyond which the periodicity of the flow becomes important and the dynamics evolves towards large $R$ and complete colonization. Upon increasing $\delta$, both $R_{\rm min}$ and $R_{\rm max}$ eventually merge at a critical $\delta_c$, beyond which no finite size stable solution exists. In the absence of genetic drift ($N_0\to\infty$), when $\delta>\delta_c$ the system reaches fixation at $f=1$ everywhere ($R=L/2$) while for $\delta<\delta_c$ the system settles into a stationary stable configuration of finite size ($R=R_{\rm min}$). As discussed earlier (see for instance, Fig.~\ref{fig:geom}d), the nucleation threshold (without flow) in $1d$ is given by the width of the interface, i.e., $R_c^0=w$. Hence, the above conclusions are correct as long as $R>w$ and the simple description in Eq.~\ref{eq:1d} breaks down for $R<w$.
Fig.~\ref{fig:1dsink}b confirms these deterministic results using the spatially averaged fraction at steady state $\bar{f}=(1/L)\int\dd x\;f(x)$, plotted as a function of $\delta$. As expected, for very small $\delta$, $R_{\rm min}(\delta)<w$ and $\bar{f}=0$ (extinction), while for larger values of $\delta$, we either have finite coexistence with $\bar{f}\approx 2R_{\rm min}/L$ for $\delta<\delta_c$ and complete colonization with $\bar{f}=1$ for $\delta>\delta_c$.
\begin{figure}[t!]
\begin{center}
\includegraphics[width=0.60\textwidth]{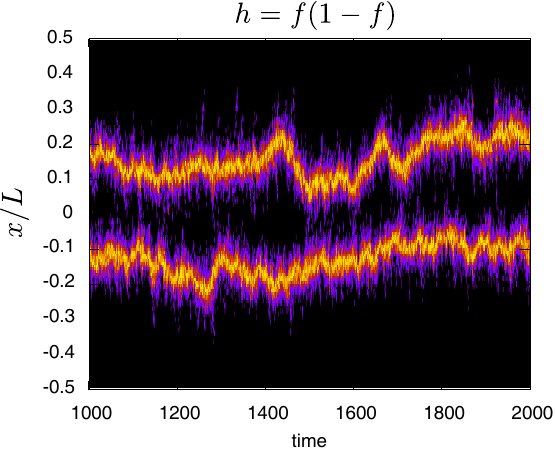}
\end{center}
	\caption{The space-time behavior of $h(x,t)=f(x,t)(1-f(x,t))$ obtained from agent-based simulations with $N_0=32$ and $\delta=0.08$. The quantity $h(x,t)$ is peaked near the interface where $f\approx 1/2$ ($h\approx 1/4$). The yellow color represent the maximum value of $h=1/4$ and black is $h=0$.
} 
\label{fig:h}
\end{figure}

We now compare the above deterministic results with agent-based simulations at finite $N_0$. Number fluctuations cause the stable minimum in $V(R)$ (Fig.~\ref{fig:1dsink}a) to become metastable with a finite lifetime, beyond which the system reaches fixation either at $f=0$ or $f=1$.
Even for a relatively large value of $N_0=32$, number fluctuations have a significant effect on the dynamics. In Fig.~\ref{fig:h}, we show the space-time behavior of $h(x,t)=f(x,t)(1-f(x,t))$ obtained from agent-based simulations with $N_0=32$ and $\delta=0.08$. Note, the quantity $\langle h(x,t)\rangle$ is the local heterozygosity. The interfaces are easily identified by the yellow regions in Fig.~\ref{fig:h} (corresponding to the maximum value of $h=1/4$) and are seen to strongly fluctuate due to stochastic noise. For a more quantitative analysis, we look at the dynamics of the system for different $N_0$ up to a maximum time $T=3L^2/D = 3\times 10^4$ (in units of the generation time $\tau_g$), which allows the agents to sufficiently diffuse across the entire system a few times. We consider two important quantities: $P_{\rm surv}$ which is the probability to survive (not reaching fixation) in $t \in [0:T]$ and $P_{\rm fix}$ which is the probability to reach fixation at $f=1$ in the same time interval. We expect that as $N_0\to\infty$, we approach the limit $P_{\rm surv}\to 1$.

\begin{figure}[t!]
\begin{center}
\includegraphics[width=\textwidth]{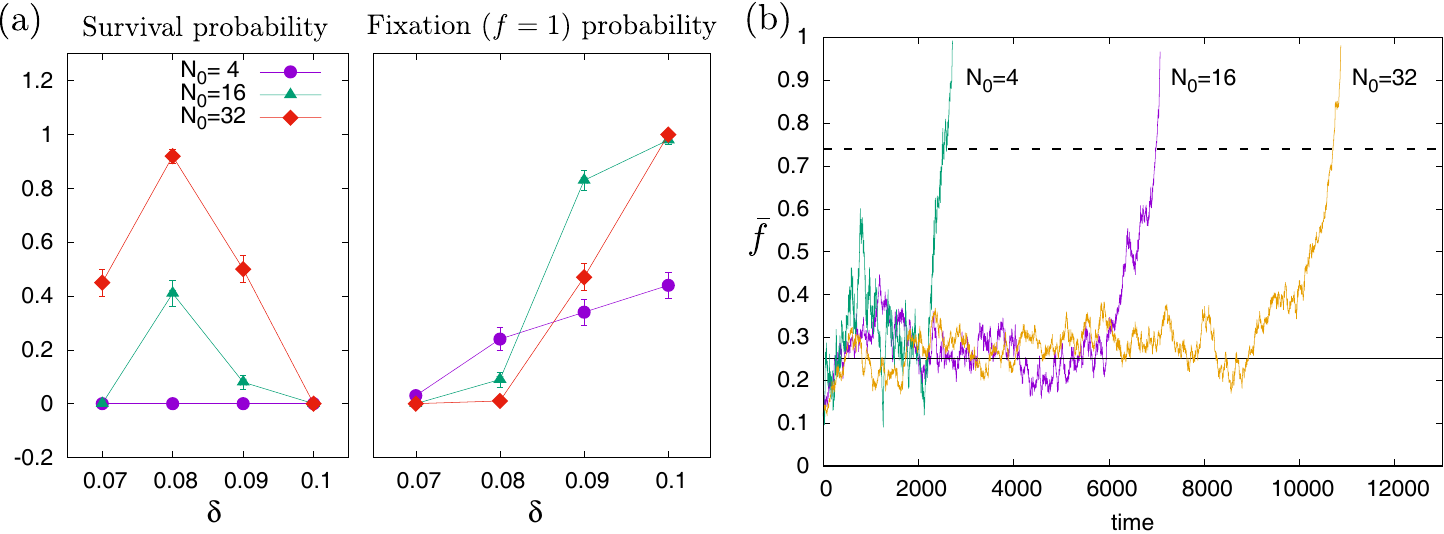}
\end{center}
	\caption{Agent-based simulations of genetic dynamics on a $1d$ sink in the presence of number fluctuations. (a) Survival (left) and fixation probabilities computed for a system with a flow strength of $F=-0.001$ and different values of $\delta$ and $N_0$. All the numerical simulations were performed using a maximum time $T=3L^2/D$, with the statistical quantities evaluated using $100$ independent runs. In the left panel, we plot the probability for the system to survive up to a time $T$, i.e. the system does not reach fixation in $f=1$ or $f=0$ within the time interval $t\in[0:T]$. For increasing $N_0$ the probability to survive increases at $\delta=0.08$ in qualitative agreement with our theoretical considerations (see text). In the right panel, we show the probability to reach fixation at $f=1$ within the same time interval. Upon increasing $N_0$ the transition at $\delta = \delta_c $ (see Fig.~\ref{fig:1dsink}) becomes dominant as expected. Notice that at $\delta = 0.1$ and $N_0=4$ there is still a significant probability to reach the fixation point $f=0$ due to large fluctuations. (b) Three different cases for different $N_0$ illustrating the temporal behaviour of $\bar{f}\simeq 2R(t)$ at $\delta = 0.08$, eventually attaining complete fixation with $\bar{f}=1$. The continuous horizontal line represents $R_{\rm min}(\delta)$ (the deterministic stable stationary state) while the dashed line represents the value of $R_{\rm max}(\delta)$ (size threshold at barrier crossing), see Fig.~\ref{fig:1dsink}.
} 
\label{fig:1dprob}
\end{figure}

By viewing the loss of stability as a Kramer's escape problem, we can estimate the characteristic time taken by the system to exit from the potential well centred at $R=R_{\rm min}(\delta)$ to be proportional to $\exp (c\;N_0\tau_g\Delta V/w)$ where $c$ is an order unity numerical constant and $\Delta V$ is the height of the barrier.
We can then use our deterministic analysis (Fig.~\ref{fig:1dsink}) to qualitatively predict the behaviour of $P_{\rm surv}$ and $P_{\rm fix}$. For small values of $N_0$ (strong noise), we expect $P_{\rm surv}\to0$ for all $\delta$ and $P_{\rm fix}$ to increase with $\delta$, simply because the fluctuations are strong enough to observe large excursions of the interface that allow rapid fixation at $f=0$ or $f=1$. For larger values of $N_0$, we expect $P_{\rm surv}$ and $P_{\rm fix}$ to reflect the deterministic results. For small values of $\delta$, the survival probability remains small, as the system goes extinct below the critical nucleation threshold ($R<w$). Large values of $\delta>\delta_c$ (Fig.~\ref{fig:1dsink}b) allow the system to reach fixation at $f=1$, rapidly colonizing the entire space. As a result, $P_{\rm surv}\approx 0$ in this limit as well. Hence, we expect a nonmonotonic trend in the survival probability, which reaches a peak at intermediate values of $\delta\approx 0.08$ (Fig.~\ref{fig:1dprob}a, left).
In contrast, $P_{\rm fix}$ always increase with the selective advantage $\delta$ (Fig.~\ref{fig:1dprob}a, right). 
We can confirm these expectations by also looking at the temporal evolution of the spatially averaged fraction $\bar{f}\simeq 2R(t)$ for $N_0 = 4,\,16$ and $32$ (Fig.~\ref{fig:1dprob}b). At long time, all three cases achieve fixation at $f=1$, once the drop size escapes from the metastable potential minimum and crosses the barrier (Fig.~\ref{fig:1dsink}a).


\section{Discussion}
\label{sec:disc}
In this paper, we have reviewed some of the dramatic consequences fluid advection has on the spatial population genetics of genetically interacting organisms. Upon using antagonism as an illustrative example, we have \ss{discussed and presented some new results using} a reaction-diffusion model that accurately captures the competitive exclusion of genetic strains in the presence of flow, and characterized the novel evolutionary strategies that emerge through an interplay of selective advantage, local flow motifs and spatial diffusion. Our analysis employs an analogy with phase separation for nonconserved densities \cite{hohenberg1977critical_dynamics} to construct simple models of nucleation and growth dynamics of isolated genetic drops. Importantly, antagonism induces a \emph{genetic} line tension that plays a crucial role in setting a size threshold for clonal dominance in a spatial setting. Motivated by turbulent mixing in stratified layers in the ocean, we have explored the effects of both compressible and incompressible flow patterns on genetic droplet growth. Finally, we have also used agent-based simulations to complement the deterministic results and highlight the crucial role of number fluctuations in noise driven extinction or colonization transitions. While we focused on steady cellular flows as a simple example, dynamic flows generated by turbulent advection or autonomously through metabolic activity and swimming \cite{atis2019microbial} remain an open frontier with regards to spatial population genetics. The complex interplay of taxis, motility and cellular growth offer further directions for exploring the effects of active transport on microbial evolution. We hope the explorations in this review will stimulate a more comprehensive analysis in biologically realistic settings, such as in the exploration of how chaotic flows in an ocean lead to the emergence of spatial ecological niches for competing microorganisms in turbulent marine environments.

\acknowledgments
DRN acknowledges support through NSF grant DMR-1608501 and via the Harvard Materials Science Research and Engineering Center, through NSF grant DMR-2011754. SS acknowedges support from the Harvard Society of Fellows.

\appendix
\section{Antagonistic depletion at genetic interfaces}
\label{app:antag}
In this appendix, we provide more details on how subtle interfacial effects arise in the presence of antagonism that play an important role in quantitative comparisons with the numerical simulations.
The derivation of simplified models of interfacial dynamics (see Eqs.~\ref{eq:compRdot} and~\ref{eq:saddle1} for instance) rely on the following assumptions (in both $1d$ and $2d$):
\begin{enumerate}
\item The interface is sharp and strongly localized, i.e.,  $w/R \to 0$,
\item The total concentration is constant everywhere, $c_T\equiv c_A+c_B = 1$ and
\item Diffusive fluxes arising from gradients of the total concentration ($\sim D\vec{\del}f\cdot\vec{\del}c_T$) are negligible.
\end{enumerate}

In all our numerical simulations we chose $D=10^{-4}$ and $\sigma=0.25$ so that $w = 0.02$ (in units where the system size $L=1$ and the generation time $\tau_g=1$). We can now evaluate how valid the above three assumptions are and ask how deviations from this idealized limit lead to quantitative corrections that may be important in understanding the numerical results. As before, we define $\delta_c$ as the critical value of the selective advantage above which the initial population with radius $R(0)$ starts to grow. Numerically the value of $\delta_c$ may receive corrections that deviate from the simple result obtained in Eq.~\ref{eq:deltac}, for instance.

Let us relax the first assumption of an infinitely sharp interface. For simplicity, we work in the $1d$ limit and set all external flow to zero ($F=0$). Upon setting $c_T=1$ for $\sigma\ll 1$, we obtain the dynamics for the interface at $R$ to be
\begin{equation}
\label{a5}
	-\dfrac{\dd R}{ \dd t}f'=Df'' + \dfrac{1}{\tau_g} f (1-f)\left[\dfrac{\delta}{2}+\sigma(2f-1)\right]\;.
\end{equation}
If $w/R\ll 1$, but finite, then from Eq.~\ref{droplet solution}, we know that the value of $f$ in the interior of the drop ($x=0$) is smaller than $1$, namely
\begin{equation}
	f(0)=\dfrac{1}{2}\left[1+\tanh\left(\dfrac{R}{2w}\right)\right]\;.
\end{equation}
Upon multiplying Eq.~\ref{a5} by $f'$ and integrating across $x$ we then obtain the interface dynamics to be
\begin{equation}
\label{a6}
	\dfrac{1}{6w} \dfrac{\dd R}{\dd t} =- \dfrac{\sigma}{2} f(0)^2 [1-f(0)]^2 + \dfrac{\delta}{2} \left[ \dfrac{f(0)^2}{2}- \dfrac{f(0)^3}{3} \right]\;,
\end{equation}
which now incorporates the small, but nonnegligible size dependent suppression of $f$ in the interior of the drop, due to a finite interface width. As a result, even in $1d$, we then obtain a critical value of $\delta$ for the growth of a nucleated drop, given by
\begin{equation}
\label{a7}
	\delta_c = \dfrac{2\sigma[1-f(0)]^2}{1-2f(0)/3}\;.
\end{equation}
As $w/R\to 0$, $f(0)\to 1$ and the threshold $\delta_c\to 0$ as expected of nucleation in $1d$. This effect is numerically negligible as long as $R(0)\geq 3 w$ (in which case we can simply use the sharp interface approximation), but it can be relevant with $\delta_c\sim 0.1$ when $R(0)=w$.

\begin{figure}[t!]
\begin{center}
\includegraphics[width=\textwidth]{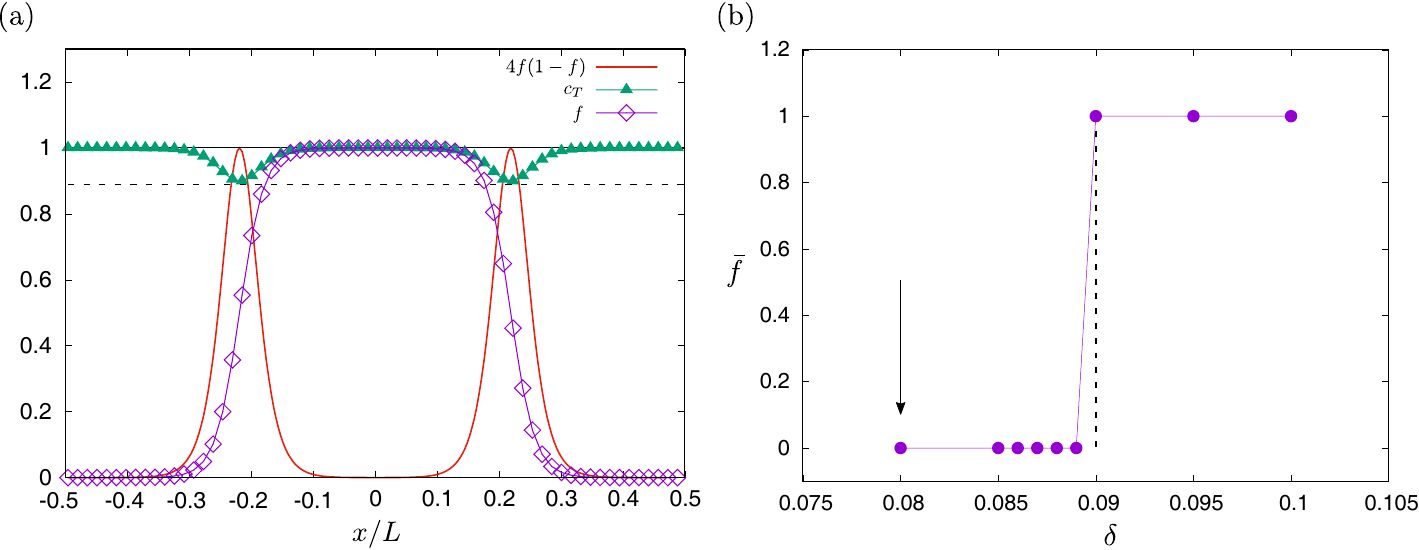}
\end{center}
	\caption{Antagonistic depletion at interfaces. (a) Numerically computed concentration profiles ($c_T$: triangles, $f$: diamonds) at time $t=1200$ for a system in $1d$ initialized with $R(0)=1/16$ and $\delta = 0.05$. The continuous red line shows $4f(1-f)$ which peaks at the interface when $f=1/2$ and the total concentration dips to $c_T= 1/(1+\sigma/2)$, shown as the dashed line. (b) The numerically computed critical $\delta_c$ for a drop in $2d$ with an initial radius of $R(0)=1/8$. The spatially averaged asymptotic fraction $\bar{f}$ transitions sharply from $\bar{f}=0$ to $\bar{f}=1$ upon varying $\delta$. The theoretically predicted value of the threshold selective advantage matches well the numerical result only when the antagonistic suppression of $c_T$ at an interface is included ($\delta_c=0.09$, dashed line). If the total concentration is assumed to be constant everywhere and set to $c_T=1$, then we obtain the erroneous prediction of $\delta_c=0.08$ (arrow), which is significantly different from the numerical result.
}
\label{fig:app}
\end{figure}

The second assumption regarding the constancy of the total concentration ($c_T\approx 1$) is more consequential. Any finite antagonism generates a deviation from the $c_T=1$ limit, with significant dips at genetic interfaces (see Fig.~\ref{fig:app}a). In Eq.~\ref{eq:cT}, if we assume the dynamics of $c_T(\b{r},t)$ to be rapid (compared to that of $f(\b{r},t)$, as is typically the case), we can adiabatically slave the total concentration to the local fraction $f(\b{r},t)$. Upon setting $\partial_tc_T\approx 0$, we then get
\begin{equation}
\label{a8}
	D\del^2c_T+\dfrac{1}{\tau_g}c_T\left [1-c_T -2\sigma\; c_T f (1-f)\right] = 0\;.
\end{equation}
Eq.~\ref{a8} can be solved to lowest order in gradients to get $c_T=1/[1+2\sigma\;f(1-f)]+\mathcal{O}(D\tau_g\del^2f)$ (Eq.~\ref{eq:cT1}). Far from the interface, when $f=0$ or $f=1$, we recover the simple result of $c_T=1$, but at a genetic interface where $f=1/2$, antagonism suppresses the total concentration to $c_T=1/(1+\sigma/2)$. This effect is confirmed in Fig.~\ref{fig:app}a, which shows the concentration profiles at time $t=1200\tau_g$ computed using numerical simulations of the reaction-diffusion model in $1d$ with $\delta = 0.05$ and the initial condition $R(0)=1/16$. The continuous red line displays the spatial profile of the function $4f(1-f)$ which peaks right at the interface where $f=1/2$ and the total concentration correspondingly dips to the value of $c_T= 1/(1+\sigma/2)$ shown as the dashed line. This interfacial decrease in $c_T$ due to antagonism ($\sigma>0$) consequently lowers the pushed genetic wave speed to $v_0=[\delta/(2+\sigma)]\sqrt{D/\tau_g\sigma}$ (Eq.~\ref{eq:Rdotimprov}).

As a result, the threshold selective advantage $\delta_c$ (even in the absence of flow) gets enhanced by a small correction proportional to the antagonsim $\sigma$.
We confirm the validity of this result by computing the critical $\delta_c$ from numerical simulations for a $2d$ drop with an initial radius of $R(0)=1/8$ and comparing it to the predicted $\delta_c$. Along with the parameters $D=10^{-4}$ and $\sigma=0.25$ (so $w=0.02$), we obtain $\delta_c=0.08$ if we set $c_T=1$, while upon including the interfacial suppression of $c_T$, we obtain $\delta_c= 0.09$. The latter prediction is in complete agreement with our numerical result (Fig.~\ref{fig:app}b) which shows the long-time (asymptotic) value of the spatially averaged fraction ($\bar{f}$) for different values of $\delta$. As expected, a sharp transition from $\bar{f}=0$ to $\bar{f}=1$ is observed for $\delta= \delta_c=0.09$, which is signficantly different from the threshold ($\delta_c=0.08$, arrow in Fig.~\ref{fig:app}b) predicted by neglecting the antagonistic depeletion at the interface. A generalization to include advection by fluid flow is straightforward.

Since $c_T$ is not constant, we must take care to also include any effects arising from diffusive fluxes that involve $\vec{\del}c_T$.
From Eq.~\ref{eq:cT}, we find the following term takes the form
\begin{equation}
\label{a10}
	2D \vec{\del} f\cdot\vec{\del}\ln(c_T)= - 2D\dfrac{2\sigma (1-2f)}{1+2\sigma f(1-f)}|\vec{\del} f|^2\;,
\end{equation}
where we have used Eq.~\ref{eq:cT1} to write $c_T$ in terms of $f$. For a well-localized interface, we can show that $|\vec{\del}f|^2\simeq(1/6w)\delta(r-R)$ (Eq.~\ref{droplet solution}), which is sharply peaked right at the interface. At the same time, as $f=1/2$ at $r=R$, the term in Eq.~\ref{a10} vanishes at the interface. Nonetheless, by viewing $-2D\vec{\del}\ln(c_T)\equiv \b{u}_{\rm eff}$ as an effective self-induced flow field, we can see that $\b{u}_{\rm eff}(R)=\b{0}$ at the interface, but $\vec{\del}\cdot\b{u}_{\rm eff}(R)<0$, i.e., the self-induced flow displays a weak sink at the interface.
As a result, $\b{u}_{\rm eff}$ tends to oppose any increase in the initial radius $R(0)$ and leads to a minor increase in the critical $\delta_c$ for the parameter values used here (the overall magnitude of $|\b{u}_{\rm eff}|$ is proportional to $D\sigma/w \sim 10^{-3}$ which only introduces an error of the order of a few precent in $\delta_c$).

\section{Nonlinear dynamics of a drop for arbitrary $\alpha$}
\label{app:nonlin}
Here we generalize the calculation in the main text to consider the fully nonlinear evolution of a deformed droplet interface for arbitrary $\alpha$. Upon once again starting with Eq.~\ref{eq:saddle1}, we parametrize the drop shape via $R(\phi,t)=R_0(t)+\Delta R(t)\sin(2\phi)$, consistent with the symmetries of the flow. We shall continue to assume small drops ($R/L\ll 1$) allowing us to linearize the flow field near the origin, but we retain arbitrary dependence on the shape variation $\Delta R(t)$ of the drop. The flow field (for arbitrary $\alpha$) then takes the form
\begin{equation}
	u_r(r,\phi)=2\pi F\dfrac{r}{L}\left[\alpha+(1-\alpha)\sin(2\phi)\right]\;,\quad u_\phi(r,\phi)=2\pi F\dfrac{r}{L}(1-\alpha)\cos(2\phi)\;.
\end{equation}
We introduce a rescaled time $\tau=tD/(R^0_c)^2$, a nondimensionalized drop size $r_0(t)=R_0(t)/R_c^0$, shape anisotropy parameter $z(t)=\Delta R(t)/R_0(t)$ ($-1\leq z\leq 1$) and the flow P{\'e}clet number ${\rm Pe}=2\pi F (R_c^0)^2/DL$. Recall that $R_c^0=D/v_0$ is the critical size threshold in the absence of flow. Upon projecting Eq.~\ref{eq:saddle1} onto angular Fourier modes, we obtain
\begin{subequations}
\begin{gather}
	\dfrac{\dd r_0}{\dd \tau}=\dfrac{1}{2r_0(1-z^2)}\left[2r_0(1-z^2)+2\dfrac{5z^2-1}{\sqrt{1-z^2}}+{\rm Pe}\;r_0^2(1-z^2)\left(2\alpha-z(1-\alpha)\right)\right]\;,\\
	\dfrac{\dd z}{\dd\tau}=-\dfrac{1}{r_0^2z(1-z^2)}\left[2(1-z^2)(10+z^2 r_0)-\dfrac{2(10-17z^2-5z^4)}{\sqrt{1-z^2}}-{\rm Pe}\;(1-\alpha)r_0^2 z(2-z^2-z^4)\right]\;.
\end{gather}\label{eq:nonlin}
\end{subequations}
After expanding these equations for small $|z|\ll 1$ and setting $\alpha=0$, we recover the nondimensionalized versions of the dynamical equations quoted in the main text (Eqs.~\ref{eq:dR},~\ref{eq:ddR}). Fig.~\ref{fig:incomp}b-c are plotted using the full nonlinear dynamics in Eq.~\ref{eq:nonlin} for $\alpha=0$. One can show that, while saddle-like flows ($0\leq \alpha< 1/2$) do not support any physical stable fixed points, sink-like flows achieved for $0.5<\alpha\leq 1$ and $F<0$ admit stable finite size droplets for a range of ${\rm Pe}<0$. In the $\alpha=1$ (fully compressible, potential flow) limit, we have stable circular drops when born on sinks, as described in Sec.~\ref{subsec:comp} and all anisotropic shape fluctuations decay exponentially rapidly. Upon deviating from the $\alpha=1$ limit, for $\alpha=1-\Delta\alpha$, $\Delta\alpha\ll 1$, we find that the stable drop acquires a small anisotropic deformation proportional to $\Delta\alpha$. This is easily seen by linearizing Eq.~\ref{eq:nonlin} for $|z|\ll 1$ and $\Delta\alpha\ll 1$ which gives for the stable fixed point
\begin{equation}
	r_0^s=\dfrac{1+\sqrt{1-4|{\rm Pe}|}}{2|{\rm Pe}|}+\mathcal{O}(\Delta\alpha)\;,\quad z_s=-\Delta\alpha|{\rm Pe}|\dfrac{(r_0^s)^2}{2+r_0^s}+\mathcal{O}(\Delta\alpha^2)\;.
\end{equation}
This fixed point exists for $-1/4\leq{\rm Pe}\leq 0$ to leading order in $\mathcal{O}(\Delta\alpha)$. As a result, partially compressible flows with $\Delta\alpha=1-\alpha\ll 1$ stabilize shape anisotropy of a finite genetic droplet.




\begin{thebibliography}{60}%
\makeatletter
\providecommand \@ifxundefined [1]{%
 \@ifx{#1\undefined}
}%
\providecommand \@ifnum [1]{%
 \ifnum #1\expandafter \@firstoftwo
 \else \expandafter \@secondoftwo
 \fi
}%
\providecommand \@ifx [1]{%
 \ifx #1\expandafter \@firstoftwo
 \else \expandafter \@secondoftwo
 \fi
}%
\providecommand \natexlab [1]{#1}%
\providecommand \enquote  [1]{``#1''}%
\providecommand \bibnamefont  [1]{#1}%
\providecommand \bibfnamefont [1]{#1}%
\providecommand \citenamefont [1]{#1}%
\providecommand \href@noop [0]{\@secondoftwo}%
\providecommand \href [0]{\begingroup \@sanitize@url \@href}%
\providecommand \@href[1]{\@@startlink{#1}\@@href}%
\providecommand \@@href[1]{\endgroup#1\@@endlink}%
\providecommand \@sanitize@url [0]{\catcode `\\12\catcode `\$12\catcode
  `\&12\catcode `\#12\catcode `\^12\catcode `\_12\catcode `\%12\relax}%
\providecommand \@@startlink[1]{}%
\providecommand \@@endlink[0]{}%
\providecommand \url  [0]{\begingroup\@sanitize@url \@url }%
\providecommand \@url [1]{\endgroup\@href {#1}{\urlprefix }}%
\providecommand \urlprefix  [0]{URL }%
\providecommand \Eprint [0]{\href }%
\providecommand \doibase [0]{http://dx.doi.org/}%
\providecommand \selectlanguage [0]{\@gobble}%
\providecommand \bibinfo  [0]{\@secondoftwo}%
\providecommand \bibfield  [0]{\@secondoftwo}%
\providecommand \translation [1]{[#1]}%
\providecommand \BibitemOpen [0]{}%
\providecommand \bibitemStop [0]{}%
\providecommand \bibitemNoStop [0]{.\EOS\space}%
\providecommand \EOS [0]{\spacefactor3000\relax}%
\providecommand \BibitemShut  [1]{\csname bibitem#1\endcsname}%
\let\auto@bib@innerbib\@empty
\bibitem [{\citenamefont {T{\'e}l}\ \emph {et~al.}(2005)\citenamefont
  {T{\'e}l}, \citenamefont {de~Moura}, \citenamefont {Grebogi},\ and\
  \citenamefont {K{\'a}rolyi}}]{tel_2005_biological_flows}%
  \BibitemOpen
  \bibfield  {author} {\bibinfo {author} {\bibfnamefont {T.}~\bibnamefont
  {T{\'e}l}}, \bibinfo {author} {\bibfnamefont {A.}~\bibnamefont {de~Moura}},
  \bibinfo {author} {\bibfnamefont {C.}~\bibnamefont {Grebogi}}, \ and\
  \bibinfo {author} {\bibfnamefont {G.}~\bibnamefont {K{\'a}rolyi}},\
  }\href@noop {} {\bibfield  {journal} {\bibinfo  {journal} {Physics reports}\
  }\textbf {\bibinfo {volume} {413}},\ \bibinfo {pages} {91} (\bibinfo {year}
  {2005})}\BibitemShut {NoStop}%
\bibitem [{\citenamefont {d’Ovidio}\ \emph {et~al.}(2010)\citenamefont
  {d’Ovidio}, \citenamefont {De~Monte}, \citenamefont {Alvain}, \citenamefont
  {Dandonneau},\ and\ \citenamefont {L{\'e}vy}}]{d_ovido_2010_fluid_niches}%
  \BibitemOpen
  \bibfield  {author} {\bibinfo {author} {\bibfnamefont {F.}~\bibnamefont
  {d’Ovidio}}, \bibinfo {author} {\bibfnamefont {S.}~\bibnamefont
  {De~Monte}}, \bibinfo {author} {\bibfnamefont {S.}~\bibnamefont {Alvain}},
  \bibinfo {author} {\bibfnamefont {Y.}~\bibnamefont {Dandonneau}}, \ and\
  \bibinfo {author} {\bibfnamefont {M.}~\bibnamefont {L{\'e}vy}},\ }\href@noop
  {} {\bibfield  {journal} {\bibinfo  {journal} {Proceedings of the National
  Academy of Sciences}\ }\textbf {\bibinfo {volume} {107}},\ \bibinfo {pages}
  {18366} (\bibinfo {year} {2010})}\BibitemShut {NoStop}%
\bibitem [{\citenamefont {Pigolotti}\ \emph {et~al.}(2013)\citenamefont
  {Pigolotti}, \citenamefont {Benzi}, \citenamefont {Perlekar}, \citenamefont
  {Jensen}, \citenamefont {Toschi},\ and\ \citenamefont
  {Nelson}}]{pigolotti_pop_bio_flow_2013}%
  \BibitemOpen
  \bibfield  {author} {\bibinfo {author} {\bibfnamefont {S.}~\bibnamefont
  {Pigolotti}}, \bibinfo {author} {\bibfnamefont {R.}~\bibnamefont {Benzi}},
  \bibinfo {author} {\bibfnamefont {P.}~\bibnamefont {Perlekar}}, \bibinfo
  {author} {\bibfnamefont {M.~H.}\ \bibnamefont {Jensen}}, \bibinfo {author}
  {\bibfnamefont {F.}~\bibnamefont {Toschi}}, \ and\ \bibinfo {author}
  {\bibfnamefont {D.~R.}\ \bibnamefont {Nelson}},\ }\href@noop {} {\bibfield
  {journal} {\bibinfo  {journal} {Theoretical population biology}\ }\textbf
  {\bibinfo {volume} {84}},\ \bibinfo {pages} {72} (\bibinfo {year}
  {2013})}\BibitemShut {NoStop}%
\bibitem [{\citenamefont {Atis}\ \emph {et~al.}(2019)\citenamefont {Atis},
  \citenamefont {Weinstein}, \citenamefont {Murray},\ and\ \citenamefont
  {Nelson}}]{atis2019microbial}%
  \BibitemOpen
  \bibfield  {author} {\bibinfo {author} {\bibfnamefont {S.}~\bibnamefont
  {Atis}}, \bibinfo {author} {\bibfnamefont {B.~T.}\ \bibnamefont {Weinstein}},
  \bibinfo {author} {\bibfnamefont {A.~W.}\ \bibnamefont {Murray}}, \ and\
  \bibinfo {author} {\bibfnamefont {D.~R.}\ \bibnamefont {Nelson}},\
  }\href@noop {} {\bibfield  {journal} {\bibinfo  {journal} {Physical review
  X}\ }\textbf {\bibinfo {volume} {9}},\ \bibinfo {pages} {021058} (\bibinfo
  {year} {2019})}\BibitemShut {NoStop}%
\bibitem [{\citenamefont {Kanso}\ and\ \citenamefont
  {Septer}(2019)}]{kanso2019microbial}%
  \BibitemOpen
  \bibfield  {author} {\bibinfo {author} {\bibfnamefont {E.}~\bibnamefont
  {Kanso}}\ and\ \bibinfo {author} {\bibfnamefont {A.~N.}\ \bibnamefont
  {Septer}},\ }\href@noop {} {\bibfield  {journal} {\bibinfo  {journal}
  {Physics}\ }\textbf {\bibinfo {volume} {12}},\ \bibinfo {pages} {71}
  (\bibinfo {year} {2019})}\BibitemShut {NoStop}%
\bibitem [{\citenamefont {Hallatschek}\ \emph {et~al.}(2007)\citenamefont
  {Hallatschek}, \citenamefont {Hersen}, \citenamefont {Ramanathan},\ and\
  \citenamefont {Nelson}}]{hallatschek2007genetic}%
  \BibitemOpen
  \bibfield  {author} {\bibinfo {author} {\bibfnamefont {O.}~\bibnamefont
  {Hallatschek}}, \bibinfo {author} {\bibfnamefont {P.}~\bibnamefont {Hersen}},
  \bibinfo {author} {\bibfnamefont {S.}~\bibnamefont {Ramanathan}}, \ and\
  \bibinfo {author} {\bibfnamefont {D.~R.}\ \bibnamefont {Nelson}},\
  }\href@noop {} {\bibfield  {journal} {\bibinfo  {journal} {Proceedings of the
  National Academy of Sciences}\ }\textbf {\bibinfo {volume} {104}},\ \bibinfo
  {pages} {19926} (\bibinfo {year} {2007})}\BibitemShut {NoStop}%
\bibitem [{\citenamefont {Eggers}(1997)}]{eggers1997nonlinear}%
  \BibitemOpen
  \bibfield  {author} {\bibinfo {author} {\bibfnamefont {J.}~\bibnamefont
  {Eggers}},\ }\href@noop {} {\bibfield  {journal} {\bibinfo  {journal}
  {Reviews of modern physics}\ }\textbf {\bibinfo {volume} {69}},\ \bibinfo
  {pages} {865} (\bibinfo {year} {1997})}\BibitemShut {NoStop}%
\bibitem [{\citenamefont {Lavrentovich}\ and\ \citenamefont
  {Nelson}(2014)}]{lavrentovich2014asymmetric}%
  \BibitemOpen
  \bibfield  {author} {\bibinfo {author} {\bibfnamefont {M.~O.}\ \bibnamefont
  {Lavrentovich}}\ and\ \bibinfo {author} {\bibfnamefont {D.~R.}\ \bibnamefont
  {Nelson}},\ }\href@noop {} {\bibfield  {journal} {\bibinfo  {journal}
  {Physical review letters}\ }\textbf {\bibinfo {volume} {112}},\ \bibinfo
  {pages} {138102} (\bibinfo {year} {2014})}\BibitemShut {NoStop}%
\bibitem [{\citenamefont {Lavrentovich}\ and\ \citenamefont
  {Nelson}(2019)}]{lavrentovich2019_antagonism}%
  \BibitemOpen
  \bibfield  {author} {\bibinfo {author} {\bibfnamefont {M.~O.}\ \bibnamefont
  {Lavrentovich}}\ and\ \bibinfo {author} {\bibfnamefont {D.~R.}\ \bibnamefont
  {Nelson}},\ }\href@noop {} {\bibfield  {journal} {\bibinfo  {journal}
  {Physical Review E}\ }\textbf {\bibinfo {volume} {100}},\ \bibinfo {pages}
  {042406} (\bibinfo {year} {2019})}\BibitemShut {NoStop}%
\bibitem [{\citenamefont {Barton}\ and\ \citenamefont
  {Hewitt}(1985)}]{barton1985hybrid}%
  \BibitemOpen
  \bibfield  {author} {\bibinfo {author} {\bibfnamefont {N.~H.}\ \bibnamefont
  {Barton}}\ and\ \bibinfo {author} {\bibfnamefont {G.~M.}\ \bibnamefont
  {Hewitt}},\ }\href@noop {} {\bibfield  {journal} {\bibinfo  {journal} {Annual
  review of Ecology and Systematics}\ }\textbf {\bibinfo {volume} {16}},\
  \bibinfo {pages} {113} (\bibinfo {year} {1985})}\BibitemShut {NoStop}%
\bibitem [{\citenamefont {Durham}\ and\ \citenamefont
  {Stocker}(2012)}]{durham2012thin}%
  \BibitemOpen
  \bibfield  {author} {\bibinfo {author} {\bibfnamefont {W.~M.}\ \bibnamefont
  {Durham}}\ and\ \bibinfo {author} {\bibfnamefont {R.}~\bibnamefont
  {Stocker}},\ }\href@noop {} {\bibfield  {journal} {\bibinfo  {journal}
  {Annual review of marine science}\ }\textbf {\bibinfo {volume} {4}},\
  \bibinfo {pages} {177} (\bibinfo {year} {2012})}\BibitemShut {NoStop}%
\bibitem [{\citenamefont {Strickland}(1970)}]{strickland1970ecology}%
  \BibitemOpen
  \bibfield  {author} {\bibinfo {author} {\bibfnamefont {J.}~\bibnamefont
  {Strickland}},\ }\href@noop {} {\emph {\bibinfo {title} {The Ecology of the
  Plankton Off La Jolla, California: In the Period April Through September,
  1967}}},\ Bulletin of the Scripps Institution of Oceanography, La Jolla,
  California\ (\bibinfo  {publisher} {University of California Press},\
  \bibinfo {year} {1970})\BibitemShut {NoStop}%
\bibitem [{\citenamefont {Robinson}(1997)}]{robinson1997theory}%
  \BibitemOpen
  \bibfield  {author} {\bibinfo {author} {\bibfnamefont {A.~R.}\ \bibnamefont
  {Robinson}},\ }\href@noop {} {\bibfield  {journal} {\bibinfo  {journal}
  {Proceedings of the Royal Society of London. Series A: Mathematical, Physical
  and Engineering Sciences}\ }\textbf {\bibinfo {volume} {453}},\ \bibinfo
  {pages} {2295} (\bibinfo {year} {1997})}\BibitemShut {NoStop}%
\bibitem [{\citenamefont {McGillicuddy~Jr}\ \emph {et~al.}(1995)\citenamefont
  {McGillicuddy~Jr}, \citenamefont {Robinson},\ and\ \citenamefont
  {McCarthy}}]{mcgillicuddy1995coupled}%
  \BibitemOpen
  \bibfield  {author} {\bibinfo {author} {\bibfnamefont {D.}~\bibnamefont
  {McGillicuddy~Jr}}, \bibinfo {author} {\bibfnamefont {A.}~\bibnamefont
  {Robinson}}, \ and\ \bibinfo {author} {\bibfnamefont {J.}~\bibnamefont
  {McCarthy}},\ }\href@noop {} {\bibfield  {journal} {\bibinfo  {journal} {Deep
  Sea Research Part I: Oceanographic Research Papers}\ }\textbf {\bibinfo
  {volume} {42}},\ \bibinfo {pages} {1359} (\bibinfo {year}
  {1995})}\BibitemShut {NoStop}%
\bibitem [{\citenamefont {McGillicuddy}\ \emph {et~al.}(1998)\citenamefont
  {McGillicuddy}, \citenamefont {Robinson}, \citenamefont {Siegel},
  \citenamefont {Jannasch}, \citenamefont {Johnson}, \citenamefont {Dickey},
  \citenamefont {McNeil}, \citenamefont {Michaels},\ and\ \citenamefont
  {Knap}}]{mcgillicuddy1998influence}%
  \BibitemOpen
  \bibfield  {author} {\bibinfo {author} {\bibfnamefont {D.}~\bibnamefont
  {McGillicuddy}}, \bibinfo {author} {\bibfnamefont {A.}~\bibnamefont
  {Robinson}}, \bibinfo {author} {\bibfnamefont {D.}~\bibnamefont {Siegel}},
  \bibinfo {author} {\bibfnamefont {H.}~\bibnamefont {Jannasch}}, \bibinfo
  {author} {\bibfnamefont {R.}~\bibnamefont {Johnson}}, \bibinfo {author}
  {\bibfnamefont {T.}~\bibnamefont {Dickey}}, \bibinfo {author} {\bibfnamefont
  {J.}~\bibnamefont {McNeil}}, \bibinfo {author} {\bibfnamefont
  {A.}~\bibnamefont {Michaels}}, \ and\ \bibinfo {author} {\bibfnamefont
  {A.}~\bibnamefont {Knap}},\ }\href@noop {} {\bibfield  {journal} {\bibinfo
  {journal} {Nature}\ }\textbf {\bibinfo {volume} {394}},\ \bibinfo {pages}
  {263} (\bibinfo {year} {1998})}\BibitemShut {NoStop}%
\bibitem [{\citenamefont {Oschlies}\ and\ \citenamefont
  {Garcon}(1998)}]{oschlies1998eddy}%
  \BibitemOpen
  \bibfield  {author} {\bibinfo {author} {\bibfnamefont {A.}~\bibnamefont
  {Oschlies}}\ and\ \bibinfo {author} {\bibfnamefont {V.}~\bibnamefont
  {Garcon}},\ }\href@noop {} {\bibfield  {journal} {\bibinfo  {journal}
  {Nature}\ }\textbf {\bibinfo {volume} {394}},\ \bibinfo {pages} {266}
  (\bibinfo {year} {1998})}\BibitemShut {NoStop}%
\bibitem [{\citenamefont {Hohenberg}\ and\ \citenamefont
  {Halperin}(1977)}]{hohenberg1977critical_dynamics}%
  \BibitemOpen
  \bibfield  {author} {\bibinfo {author} {\bibfnamefont {P.~C.}\ \bibnamefont
  {Hohenberg}}\ and\ \bibinfo {author} {\bibfnamefont {B.~I.}\ \bibnamefont
  {Halperin}},\ }\href@noop {} {\bibfield  {journal} {\bibinfo  {journal}
  {Reviews of Modern Physics}\ }\textbf {\bibinfo {volume} {49}},\ \bibinfo
  {pages} {435} (\bibinfo {year} {1977})}\BibitemShut {NoStop}%
\bibitem [{\citenamefont {Hallatschek}\ and\ \citenamefont
  {Nelson}(2010)}]{hallatschek2010life}%
  \BibitemOpen
  \bibfield  {author} {\bibinfo {author} {\bibfnamefont {O.}~\bibnamefont
  {Hallatschek}}\ and\ \bibinfo {author} {\bibfnamefont {D.~R.}\ \bibnamefont
  {Nelson}},\ }\href@noop {} {\bibfield  {journal} {\bibinfo  {journal}
  {Evolution: International Journal of Organic Evolution}\ }\textbf {\bibinfo
  {volume} {64}},\ \bibinfo {pages} {193} (\bibinfo {year} {2010})}\BibitemShut
  {NoStop}%
\bibitem [{\citenamefont {Crow}\ and\ \citenamefont
  {Kimura}(1970)}]{crow_kimura1970}%
  \BibitemOpen
  \bibfield  {author} {\bibinfo {author} {\bibfnamefont {J.~F.}\ \bibnamefont
  {Crow}}\ and\ \bibinfo {author} {\bibfnamefont {M.}~\bibnamefont {Kimura}},\
  }\href@noop {} {\emph {\bibinfo {title} {An introduction to population
  genetics theory}}}\ (\bibinfo  {publisher} {New York, Evanston and London:
  Harper \& Row, Publishers},\ \bibinfo {year} {1970})\BibitemShut {NoStop}%
\bibitem [{\citenamefont {Korolev}\ \emph {et~al.}(2010)\citenamefont
  {Korolev}, \citenamefont {Avlund}, \citenamefont {Hallatschek},\ and\
  \citenamefont {Nelson}}]{korolev2010RMP}%
  \BibitemOpen
  \bibfield  {author} {\bibinfo {author} {\bibfnamefont {K.~S.}\ \bibnamefont
  {Korolev}}, \bibinfo {author} {\bibfnamefont {M.}~\bibnamefont {Avlund}},
  \bibinfo {author} {\bibfnamefont {O.}~\bibnamefont {Hallatschek}}, \ and\
  \bibinfo {author} {\bibfnamefont {D.~R.}\ \bibnamefont {Nelson}},\
  }\href@noop {} {\bibfield  {journal} {\bibinfo  {journal} {Reviews of modern
  physics}\ }\textbf {\bibinfo {volume} {82}},\ \bibinfo {pages} {1691}
  (\bibinfo {year} {2010})}\BibitemShut {NoStop}%
\bibitem [{\citenamefont {Murray}(2007)}]{murray2007math_bio}%
  \BibitemOpen
  \bibfield  {author} {\bibinfo {author} {\bibfnamefont {J.~D.}\ \bibnamefont
  {Murray}},\ }\href@noop {} {\emph {\bibinfo {title} {Mathematical biology: I.
  An introduction}}},\ Vol.~\bibinfo {volume} {17}\ (\bibinfo  {publisher}
  {Springer Science \& Business Media},\ \bibinfo {year} {2007})\BibitemShut
  {NoStop}%
\bibitem [{\citenamefont {Van~Saarloos}(2003)}]{van_saarloos2003front}%
  \BibitemOpen
  \bibfield  {author} {\bibinfo {author} {\bibfnamefont {W.}~\bibnamefont
  {Van~Saarloos}},\ }\href@noop {} {\bibfield  {journal} {\bibinfo  {journal}
  {Physics reports}\ }\textbf {\bibinfo {volume} {386}},\ \bibinfo {pages} {29}
  (\bibinfo {year} {2003})}\BibitemShut {NoStop}%
\bibitem [{\citenamefont {Tanaka}\ \emph {et~al.}(2017)\citenamefont {Tanaka},
  \citenamefont {Stone},\ and\ \citenamefont
  {Nelson}}]{tanaka2017spatial_gene_drive}%
  \BibitemOpen
  \bibfield  {author} {\bibinfo {author} {\bibfnamefont {H.}~\bibnamefont
  {Tanaka}}, \bibinfo {author} {\bibfnamefont {H.~A.}\ \bibnamefont {Stone}}, \
  and\ \bibinfo {author} {\bibfnamefont {D.~R.}\ \bibnamefont {Nelson}},\
  }\href@noop {} {\bibfield  {journal} {\bibinfo  {journal} {Proceedings of the
  National Academy of Sciences}\ }\textbf {\bibinfo {volume} {114}},\ \bibinfo
  {pages} {8452} (\bibinfo {year} {2017})}\BibitemShut {NoStop}%
\bibitem [{\citenamefont {Frey}(2010)}]{frey2010evolutionary}%
  \BibitemOpen
  \bibfield  {author} {\bibinfo {author} {\bibfnamefont {E.}~\bibnamefont
  {Frey}},\ }\href@noop {} {\bibfield  {journal} {\bibinfo  {journal} {Physica
  A: Statistical Mechanics and its Applications}\ }\textbf {\bibinfo {volume}
  {389}},\ \bibinfo {pages} {4265} (\bibinfo {year} {2010})}\BibitemShut
  {NoStop}%
\bibitem [{\citenamefont {Nowak}\ \emph {et~al.}(2004)\citenamefont {Nowak},
  \citenamefont {Sasaki}, \citenamefont {Taylor},\ and\ \citenamefont
  {Fudenberg}}]{nowak2004ESS}%
  \BibitemOpen
  \bibfield  {author} {\bibinfo {author} {\bibfnamefont {M.~A.}\ \bibnamefont
  {Nowak}}, \bibinfo {author} {\bibfnamefont {A.}~\bibnamefont {Sasaki}},
  \bibinfo {author} {\bibfnamefont {C.}~\bibnamefont {Taylor}}, \ and\ \bibinfo
  {author} {\bibfnamefont {D.}~\bibnamefont {Fudenberg}},\ }\href@noop {}
  {\bibfield  {journal} {\bibinfo  {journal} {Nature}\ }\textbf {\bibinfo
  {volume} {428}},\ \bibinfo {pages} {646} (\bibinfo {year}
  {2004})}\BibitemShut {NoStop}%
\bibitem [{\citenamefont {M{\"u}ller}\ \emph {et~al.}(2014)\citenamefont
  {M{\"u}ller}, \citenamefont {Neugeboren}, \citenamefont {Nelson},\ and\
  \citenamefont {Murray}}]{muller2014mutualism}%
  \BibitemOpen
  \bibfield  {author} {\bibinfo {author} {\bibfnamefont {M.~J.}\ \bibnamefont
  {M{\"u}ller}}, \bibinfo {author} {\bibfnamefont {B.~I.}\ \bibnamefont
  {Neugeboren}}, \bibinfo {author} {\bibfnamefont {D.~R.}\ \bibnamefont
  {Nelson}}, \ and\ \bibinfo {author} {\bibfnamefont {A.~W.}\ \bibnamefont
  {Murray}},\ }\href@noop {} {\bibfield  {journal} {\bibinfo  {journal}
  {Proceedings of the National Academy of Sciences}\ }\textbf {\bibinfo
  {volume} {111}},\ \bibinfo {pages} {1037} (\bibinfo {year}
  {2014})}\BibitemShut {NoStop}%
\bibitem [{\citenamefont {Korolev}\ and\ \citenamefont
  {Nelson}(2011)}]{korolev2011competition}%
  \BibitemOpen
  \bibfield  {author} {\bibinfo {author} {\bibfnamefont {K.}~\bibnamefont
  {Korolev}}\ and\ \bibinfo {author} {\bibfnamefont {D.~R.}\ \bibnamefont
  {Nelson}},\ }\href@noop {} {\bibfield  {journal} {\bibinfo  {journal}
  {Physical Review Letters}\ }\textbf {\bibinfo {volume} {107}},\ \bibinfo
  {pages} {088103} (\bibinfo {year} {2011})}\BibitemShut {NoStop}%
\bibitem [{\citenamefont {Yanni}\ \emph
  {et~al.}(2019{\natexlab{a}})\citenamefont {Yanni}, \citenamefont
  {M{\'a}rquez-Zacar{\'\i}as}, \citenamefont {Yunker},\ and\ \citenamefont
  {Ratcliff}}]{yanni2019antagonism}%
  \BibitemOpen
  \bibfield  {author} {\bibinfo {author} {\bibfnamefont {D.}~\bibnamefont
  {Yanni}}, \bibinfo {author} {\bibfnamefont {P.}~\bibnamefont
  {M{\'a}rquez-Zacar{\'\i}as}}, \bibinfo {author} {\bibfnamefont {P.~J.}\
  \bibnamefont {Yunker}}, \ and\ \bibinfo {author} {\bibfnamefont {W.~C.}\
  \bibnamefont {Ratcliff}},\ }\href@noop {} {\bibfield  {journal} {\bibinfo
  {journal} {Current biology}\ }\textbf {\bibinfo {volume} {29}},\ \bibinfo
  {pages} {R545} (\bibinfo {year} {2019}{\natexlab{a}})}\BibitemShut {NoStop}%
\bibitem [{\citenamefont {Yanni}\ \emph
  {et~al.}(2019{\natexlab{b}})\citenamefont {Yanni}, \citenamefont
  {M{\'a}rquez-Zacar{\'\i}as}, \citenamefont {Yunker},\ and\ \citenamefont
  {Ratcliff}}]{yunker2019_review}%
  \BibitemOpen
  \bibfield  {author} {\bibinfo {author} {\bibfnamefont {D.}~\bibnamefont
  {Yanni}}, \bibinfo {author} {\bibfnamefont {P.}~\bibnamefont
  {M{\'a}rquez-Zacar{\'\i}as}}, \bibinfo {author} {\bibfnamefont {P.~J.}\
  \bibnamefont {Yunker}}, \ and\ \bibinfo {author} {\bibfnamefont {W.~C.}\
  \bibnamefont {Ratcliff}},\ }\href@noop {} {\bibfield  {journal} {\bibinfo
  {journal} {Current biology}\ }\textbf {\bibinfo {volume} {29}},\ \bibinfo
  {pages} {R545} (\bibinfo {year} {2019}{\natexlab{b}})}\BibitemShut {NoStop}%
\bibitem [{\citenamefont {McNally}\ \emph {et~al.}(2017)\citenamefont
  {McNally}, \citenamefont {Bernardy}, \citenamefont {Thomas}, \citenamefont
  {Kalziqi}, \citenamefont {Pentz}, \citenamefont {Brown}, \citenamefont
  {Hammer}, \citenamefont {Yunker},\ and\ \citenamefont
  {Ratcliff}}]{yunker2017TypeVI_secretion}%
  \BibitemOpen
  \bibfield  {author} {\bibinfo {author} {\bibfnamefont {L.}~\bibnamefont
  {McNally}}, \bibinfo {author} {\bibfnamefont {E.}~\bibnamefont {Bernardy}},
  \bibinfo {author} {\bibfnamefont {J.}~\bibnamefont {Thomas}}, \bibinfo
  {author} {\bibfnamefont {A.}~\bibnamefont {Kalziqi}}, \bibinfo {author}
  {\bibfnamefont {J.}~\bibnamefont {Pentz}}, \bibinfo {author} {\bibfnamefont
  {S.~P.}\ \bibnamefont {Brown}}, \bibinfo {author} {\bibfnamefont {B.~K.}\
  \bibnamefont {Hammer}}, \bibinfo {author} {\bibfnamefont {P.~J.}\
  \bibnamefont {Yunker}}, \ and\ \bibinfo {author} {\bibfnamefont {W.~C.}\
  \bibnamefont {Ratcliff}},\ }\href@noop {} {\bibfield  {journal} {\bibinfo
  {journal} {Nature communications}\ }\textbf {\bibinfo {volume} {8}},\
  \bibinfo {pages} {1} (\bibinfo {year} {2017})}\BibitemShut {NoStop}%
\bibitem [{\citenamefont {Brunet}\ \emph {et~al.}(2013)\citenamefont {Brunet},
  \citenamefont {Espinosa}, \citenamefont {Harchouni}, \citenamefont {Mignot},\
  and\ \citenamefont {Cascales}}]{brunet2013imaging_Type_VI}%
  \BibitemOpen
  \bibfield  {author} {\bibinfo {author} {\bibfnamefont {Y.~R.}\ \bibnamefont
  {Brunet}}, \bibinfo {author} {\bibfnamefont {L.}~\bibnamefont {Espinosa}},
  \bibinfo {author} {\bibfnamefont {S.}~\bibnamefont {Harchouni}}, \bibinfo
  {author} {\bibfnamefont {T.}~\bibnamefont {Mignot}}, \ and\ \bibinfo {author}
  {\bibfnamefont {E.}~\bibnamefont {Cascales}},\ }\href@noop {} {\bibfield
  {journal} {\bibinfo  {journal} {Cell reports}\ }\textbf {\bibinfo {volume}
  {3}},\ \bibinfo {pages} {36} (\bibinfo {year} {2013})}\BibitemShut {NoStop}%
\bibitem [{\citenamefont {Barton}\ and\ \citenamefont
  {Hewitt}(1989)}]{barton1989adaptation}%
  \BibitemOpen
  \bibfield  {author} {\bibinfo {author} {\bibfnamefont {N.~H.}\ \bibnamefont
  {Barton}}\ and\ \bibinfo {author} {\bibfnamefont {G.~M.}\ \bibnamefont
  {Hewitt}},\ }\href@noop {} {\bibfield  {journal} {\bibinfo  {journal}
  {Nature}\ }\textbf {\bibinfo {volume} {341}},\ \bibinfo {pages} {497}
  (\bibinfo {year} {1989})}\BibitemShut {NoStop}%
\bibitem [{\citenamefont {Hinrichsen}(2000)}]{hinrichsen2000absorbing_states}%
  \BibitemOpen
  \bibfield  {author} {\bibinfo {author} {\bibfnamefont {H.}~\bibnamefont
  {Hinrichsen}},\ }\href@noop {} {\bibfield  {journal} {\bibinfo  {journal}
  {Advances in physics}\ }\textbf {\bibinfo {volume} {49}},\ \bibinfo {pages}
  {815} (\bibinfo {year} {2000})}\BibitemShut {NoStop}%
\bibitem [{\citenamefont {Cates}(2012)}]{cates2012diffusive}%
  \BibitemOpen
  \bibfield  {author} {\bibinfo {author} {\bibfnamefont {M.~E.}\ \bibnamefont
  {Cates}},\ }\href@noop {} {\bibfield  {journal} {\bibinfo  {journal} {Reports
  on Progress in Physics}\ }\textbf {\bibinfo {volume} {75}},\ \bibinfo {pages}
  {042601} (\bibinfo {year} {2012})}\BibitemShut {NoStop}%
\bibitem [{\citenamefont {Barton}\ and\ \citenamefont
  {Rouhani}(1987)}]{barton_rouhani_nucleation_1987}%
  \BibitemOpen
  \bibfield  {author} {\bibinfo {author} {\bibfnamefont {N.}~\bibnamefont
  {Barton}}\ and\ \bibinfo {author} {\bibfnamefont {S.}~\bibnamefont
  {Rouhani}},\ }\href@noop {} {\bibfield  {journal} {\bibinfo  {journal}
  {Journal of theoretical biology}\ }\textbf {\bibinfo {volume} {125}},\
  \bibinfo {pages} {397} (\bibinfo {year} {1987})}\BibitemShut {NoStop}%
\bibitem [{\citenamefont {Pigolotti}\ \emph {et~al.}(2012)\citenamefont
  {Pigolotti}, \citenamefont {Benzi}, \citenamefont {Jensen},\ and\
  \citenamefont {Nelson}}]{pigolotti2012population}%
  \BibitemOpen
  \bibfield  {author} {\bibinfo {author} {\bibfnamefont {S.}~\bibnamefont
  {Pigolotti}}, \bibinfo {author} {\bibfnamefont {R.}~\bibnamefont {Benzi}},
  \bibinfo {author} {\bibfnamefont {M.~H.}\ \bibnamefont {Jensen}}, \ and\
  \bibinfo {author} {\bibfnamefont {D.~R.}\ \bibnamefont {Nelson}},\
  }\href@noop {} {\bibfield  {journal} {\bibinfo  {journal} {Physical review
  letters}\ }\textbf {\bibinfo {volume} {108}},\ \bibinfo {pages} {128102}
  (\bibinfo {year} {2012})}\BibitemShut {NoStop}%
\bibitem [{\citenamefont {Chotibut}\ and\ \citenamefont
  {Nelson}(2015)}]{chotibut2015evolutionary}%
  \BibitemOpen
  \bibfield  {author} {\bibinfo {author} {\bibfnamefont {T.}~\bibnamefont
  {Chotibut}}\ and\ \bibinfo {author} {\bibfnamefont {D.~R.}\ \bibnamefont
  {Nelson}},\ }\href@noop {} {\bibfield  {journal} {\bibinfo  {journal}
  {Physical Review E}\ }\textbf {\bibinfo {volume} {92}},\ \bibinfo {pages}
  {022718} (\bibinfo {year} {2015})}\BibitemShut {NoStop}%
\bibitem [{\citenamefont {Chotibut}\ and\ \citenamefont
  {Nelson}(2017)}]{chotibut2017population}%
  \BibitemOpen
  \bibfield  {author} {\bibinfo {author} {\bibfnamefont {T.}~\bibnamefont
  {Chotibut}}\ and\ \bibinfo {author} {\bibfnamefont {D.~R.}\ \bibnamefont
  {Nelson}},\ }\href@noop {} {\bibfield  {journal} {\bibinfo  {journal}
  {Journal of Statistical Physics}\ }\textbf {\bibinfo {volume} {167}},\
  \bibinfo {pages} {777} (\bibinfo {year} {2017})}\BibitemShut {NoStop}%
\bibitem [{\citenamefont {Bec}(2003)}]{bec2003fractal}%
  \BibitemOpen
  \bibfield  {author} {\bibinfo {author} {\bibfnamefont {J.}~\bibnamefont
  {Bec}},\ }\href@noop {} {\bibfield  {journal} {\bibinfo  {journal} {Physics
  of fluids}\ }\textbf {\bibinfo {volume} {15}},\ \bibinfo {pages} {L81}
  (\bibinfo {year} {2003})}\BibitemShut {NoStop}%
\bibitem [{\citenamefont {Martin}(2003)}]{martin2003phytoplankton}%
  \BibitemOpen
  \bibfield  {author} {\bibinfo {author} {\bibfnamefont {A.}~\bibnamefont
  {Martin}},\ }\href@noop {} {\bibfield  {journal} {\bibinfo  {journal}
  {Progress in oceanography}\ }\textbf {\bibinfo {volume} {57}},\ \bibinfo
  {pages} {125} (\bibinfo {year} {2003})}\BibitemShut {NoStop}%
\bibitem [{\citenamefont {Boffetta}\ \emph {et~al.}(2004)\citenamefont
  {Boffetta}, \citenamefont {Davoudi}, \citenamefont {Eckhardt},\ and\
  \citenamefont {Schumacher}}]{boffetta2004lagrangian}%
  \BibitemOpen
  \bibfield  {author} {\bibinfo {author} {\bibfnamefont {G.}~\bibnamefont
  {Boffetta}}, \bibinfo {author} {\bibfnamefont {J.}~\bibnamefont {Davoudi}},
  \bibinfo {author} {\bibfnamefont {B.}~\bibnamefont {Eckhardt}}, \ and\
  \bibinfo {author} {\bibfnamefont {J.}~\bibnamefont {Schumacher}},\
  }\href@noop {} {\bibfield  {journal} {\bibinfo  {journal} {Physical review
  letters}\ }\textbf {\bibinfo {volume} {93}},\ \bibinfo {pages} {134501}
  (\bibinfo {year} {2004})}\BibitemShut {NoStop}%
\bibitem [{\citenamefont {De~Pietro}\ \emph {et~al.}(2015)\citenamefont
  {De~Pietro}, \citenamefont {van Hinsberg}, \citenamefont {Biferale},
  \citenamefont {Clercx}, \citenamefont {Perlekar},\ and\ \citenamefont
  {Toschi}}]{depietro2015}%
  \BibitemOpen
  \bibfield  {author} {\bibinfo {author} {\bibfnamefont {M.}~\bibnamefont
  {De~Pietro}}, \bibinfo {author} {\bibfnamefont {M.~A.~T.}\ \bibnamefont {van
  Hinsberg}}, \bibinfo {author} {\bibfnamefont {L.}~\bibnamefont {Biferale}},
  \bibinfo {author} {\bibfnamefont {H.~J.~H.}\ \bibnamefont {Clercx}}, \bibinfo
  {author} {\bibfnamefont {P.}~\bibnamefont {Perlekar}}, \ and\ \bibinfo
  {author} {\bibfnamefont {F.}~\bibnamefont {Toschi}},\ }\href@noop {}
  {\bibfield  {journal} {\bibinfo  {journal} {Phys. Rev. E}\ }\textbf {\bibinfo
  {volume} {91}},\ \bibinfo {pages} {053002} (\bibinfo {year}
  {2015})}\BibitemShut {NoStop}%
\bibitem [{\citenamefont {Sozza}\ \emph {et~al.}(2016)\citenamefont {Sozza},
  \citenamefont {De~Lillo}, \citenamefont {Musacchio},\ and\ \citenamefont
  {Boffetta}}]{sozza2016}%
  \BibitemOpen
  \bibfield  {author} {\bibinfo {author} {\bibfnamefont {A.}~\bibnamefont
  {Sozza}}, \bibinfo {author} {\bibfnamefont {F.}~\bibnamefont {De~Lillo}},
  \bibinfo {author} {\bibfnamefont {S.}~\bibnamefont {Musacchio}}, \ and\
  \bibinfo {author} {\bibfnamefont {G.}~\bibnamefont {Boffetta}},\ }\href@noop
  {} {\bibfield  {journal} {\bibinfo  {journal} {Phys. Rev. Fluids}\ }\textbf
  {\bibinfo {volume} {1}},\ \bibinfo {pages} {052401} (\bibinfo {year}
  {2016})}\BibitemShut {NoStop}%
\bibitem [{\citenamefont {Perlekar}\ \emph {et~al.}(2010)\citenamefont
  {Perlekar}, \citenamefont {Benzi}, \citenamefont {Nelson},\ and\
  \citenamefont {Toschi}}]{perlekar2010}%
  \BibitemOpen
  \bibfield  {author} {\bibinfo {author} {\bibfnamefont {P.}~\bibnamefont
  {Perlekar}}, \bibinfo {author} {\bibfnamefont {R.}~\bibnamefont {Benzi}},
  \bibinfo {author} {\bibfnamefont {D.~R.}\ \bibnamefont {Nelson}}, \ and\
  \bibinfo {author} {\bibfnamefont {F.}~\bibnamefont {Toschi}},\ }\href@noop {}
  {\bibfield  {journal} {\bibinfo  {journal} {Phys. Rev. Lett.}\ }\textbf
  {\bibinfo {volume} {105}},\ \bibinfo {pages} {144501} (\bibinfo {year}
  {2010})}\BibitemShut {NoStop}%
\bibitem [{\citenamefont {Plummer}\ \emph {et~al.}(2019)\citenamefont
  {Plummer}, \citenamefont {Benzi}, \citenamefont {Nelson},\ and\ \citenamefont
  {Toschi}}]{Plummer2019}%
  \BibitemOpen
  \bibfield  {author} {\bibinfo {author} {\bibfnamefont {A.}~\bibnamefont
  {Plummer}}, \bibinfo {author} {\bibfnamefont {R.}~\bibnamefont {Benzi}},
  \bibinfo {author} {\bibfnamefont {D.~R.}\ \bibnamefont {Nelson}}, \ and\
  \bibinfo {author} {\bibfnamefont {F.}~\bibnamefont {Toschi}},\ }\href@noop {}
  {\bibfield  {journal} {\bibinfo  {journal} {Proceedings of the National
  Academy of Sciences}\ }\textbf {\bibinfo {volume} {116}},\ \bibinfo {pages}
  {373} (\bibinfo {year} {2019})}\BibitemShut {NoStop}%
\bibitem [{\citenamefont {Korolev}\ \emph {et~al.}(2012)\citenamefont
  {Korolev}, \citenamefont {M{\"u}ller}, \citenamefont {Karahan}, \citenamefont
  {Murray}, \citenamefont {Hallatschek},\ and\ \citenamefont
  {Nelson}}]{korolev2012selective}%
  \BibitemOpen
  \bibfield  {author} {\bibinfo {author} {\bibfnamefont {K.~S.}\ \bibnamefont
  {Korolev}}, \bibinfo {author} {\bibfnamefont {M.~J.}\ \bibnamefont
  {M{\"u}ller}}, \bibinfo {author} {\bibfnamefont {N.}~\bibnamefont {Karahan}},
  \bibinfo {author} {\bibfnamefont {A.~W.}\ \bibnamefont {Murray}}, \bibinfo
  {author} {\bibfnamefont {O.}~\bibnamefont {Hallatschek}}, \ and\ \bibinfo
  {author} {\bibfnamefont {D.~R.}\ \bibnamefont {Nelson}},\ }\href@noop {}
  {\bibfield  {journal} {\bibinfo  {journal} {Physical biology}\ }\textbf
  {\bibinfo {volume} {9}},\ \bibinfo {pages} {026008} (\bibinfo {year}
  {2012})}\BibitemShut {NoStop}%
\bibitem [{\citenamefont {Langer}(1969)}]{langer1969_nucleation}%
  \BibitemOpen
  \bibfield  {author} {\bibinfo {author} {\bibfnamefont {J.~S.}\ \bibnamefont
  {Langer}},\ }\href@noop {} {\bibfield  {journal} {\bibinfo  {journal} {Annals
  of Physics}\ }\textbf {\bibinfo {volume} {54}},\ \bibinfo {pages} {258}
  (\bibinfo {year} {1969})}\BibitemShut {NoStop}%
\bibitem [{\citenamefont {Coleman}(1977)}]{coleman_nucleation1977}%
  \BibitemOpen
  \bibfield  {author} {\bibinfo {author} {\bibfnamefont {S.}~\bibnamefont
  {Coleman}},\ }\href@noop {} {\bibfield  {journal} {\bibinfo  {journal}
  {Physical Review D}\ }\textbf {\bibinfo {volume} {15}},\ \bibinfo {pages}
  {2929} (\bibinfo {year} {1977})}\BibitemShut {NoStop}%
\bibitem [{\citenamefont {Kaiser}(2018)}]{coleman_summer_school_lectures}%
  \BibitemOpen
  \bibfield  {author} {\bibinfo {author} {\bibfnamefont {D.}~\bibnamefont
  {Kaiser}},\ }\href@noop {} {\emph {\bibinfo {title} {Lectures Of Sidney
  Coleman On Quantum Field Theory: Foreword By David Kaiser}}}\ (\bibinfo
  {publisher} {World Scientific Publishing},\ \bibinfo {year}
  {2018})\BibitemShut {NoStop}%
\bibitem [{\citenamefont {Bray}(2002)}]{bray2002phase_ordering}%
  \BibitemOpen
  \bibfield  {author} {\bibinfo {author} {\bibfnamefont {A.~J.}\ \bibnamefont
  {Bray}},\ }\href@noop {} {\bibfield  {journal} {\bibinfo  {journal} {Advances
  in Physics}\ }\textbf {\bibinfo {volume} {51}},\ \bibinfo {pages} {481}
  (\bibinfo {year} {2002})}\BibitemShut {NoStop}%
\bibitem [{\citenamefont {Giometto}\ \emph {et~al.}(2021)\citenamefont
  {Giometto}, \citenamefont {Nelson},\ and\ \citenamefont
  {Murray}}]{giometto2020antagonism}%
  \BibitemOpen
  \bibfield  {author} {\bibinfo {author} {\bibfnamefont {A.}~\bibnamefont
  {Giometto}}, \bibinfo {author} {\bibfnamefont {D.~R.}\ \bibnamefont
  {Nelson}}, \ and\ \bibinfo {author} {\bibfnamefont {A.~W.}\ \bibnamefont
  {Murray}},\ }\href@noop {} {\bibfield  {journal} {\bibinfo  {journal}
  {eLife}\ }\textbf {\bibinfo {volume} {10}},\ \bibinfo {pages} {e62932}
  (\bibinfo {year} {2021})}\BibitemShut {NoStop}%
\bibitem [{\citenamefont {Berry}\ \emph {et~al.}(2018)\citenamefont {Berry},
  \citenamefont {Brangwynne},\ and\ \citenamefont {Haataja}}]{Berry_2018}%
  \BibitemOpen
  \bibfield  {author} {\bibinfo {author} {\bibfnamefont {J.}~\bibnamefont
  {Berry}}, \bibinfo {author} {\bibfnamefont {C.~P.}\ \bibnamefont
  {Brangwynne}}, \ and\ \bibinfo {author} {\bibfnamefont {M.}~\bibnamefont
  {Haataja}},\ }\href@noop {} {\bibfield  {journal} {\bibinfo  {journal}
  {Reports on Progress in Physics}\ }\textbf {\bibinfo {volume} {81}},\
  \bibinfo {pages} {046601} (\bibinfo {year} {2018})}\BibitemShut {NoStop}%
\bibitem [{\citenamefont {Weber}\ \emph {et~al.}(2019)\citenamefont {Weber},
  \citenamefont {Zwicker}, \citenamefont {J{\"u}licher},\ and\ \citenamefont
  {Lee}}]{weber2019physics}%
  \BibitemOpen
  \bibfield  {author} {\bibinfo {author} {\bibfnamefont {C.~A.}\ \bibnamefont
  {Weber}}, \bibinfo {author} {\bibfnamefont {D.}~\bibnamefont {Zwicker}},
  \bibinfo {author} {\bibfnamefont {F.}~\bibnamefont {J{\"u}licher}}, \ and\
  \bibinfo {author} {\bibfnamefont {C.~F.}\ \bibnamefont {Lee}},\ }\href@noop
  {} {\bibfield  {journal} {\bibinfo  {journal} {Reports on Progress in
  Physics}\ }\textbf {\bibinfo {volume} {82}},\ \bibinfo {pages} {064601}
  (\bibinfo {year} {2019})}\BibitemShut {NoStop}%
\bibitem [{\citenamefont {Perlekar}\ \emph {et~al.}(2014)\citenamefont
  {Perlekar}, \citenamefont {Benzi}, \citenamefont {Clercx}, \citenamefont
  {Nelson},\ and\ \citenamefont {Toschi}}]{perlekar2014spinodal}%
  \BibitemOpen
  \bibfield  {author} {\bibinfo {author} {\bibfnamefont {P.}~\bibnamefont
  {Perlekar}}, \bibinfo {author} {\bibfnamefont {R.}~\bibnamefont {Benzi}},
  \bibinfo {author} {\bibfnamefont {H.~J.}\ \bibnamefont {Clercx}}, \bibinfo
  {author} {\bibfnamefont {D.~R.}\ \bibnamefont {Nelson}}, \ and\ \bibinfo
  {author} {\bibfnamefont {F.}~\bibnamefont {Toschi}},\ }\href@noop {}
  {\bibfield  {journal} {\bibinfo  {journal} {Physical review letters}\
  }\textbf {\bibinfo {volume} {112}},\ \bibinfo {pages} {014502} (\bibinfo
  {year} {2014})}\BibitemShut {NoStop}%
\bibitem [{\citenamefont {Berti}\ \emph {et~al.}(2005)\citenamefont {Berti},
  \citenamefont {Boffetta}, \citenamefont {Cencini},\ and\ \citenamefont
  {Vulpiani}}]{berti2005turbulence}%
  \BibitemOpen
  \bibfield  {author} {\bibinfo {author} {\bibfnamefont {S.}~\bibnamefont
  {Berti}}, \bibinfo {author} {\bibfnamefont {G.}~\bibnamefont {Boffetta}},
  \bibinfo {author} {\bibfnamefont {M.}~\bibnamefont {Cencini}}, \ and\
  \bibinfo {author} {\bibfnamefont {A.}~\bibnamefont {Vulpiani}},\ }\href@noop
  {} {\bibfield  {journal} {\bibinfo  {journal} {Physical review letters}\
  }\textbf {\bibinfo {volume} {95}},\ \bibinfo {pages} {224501} (\bibinfo
  {year} {2005})}\BibitemShut {NoStop}%
\bibitem [{\citenamefont {Zwicker}\ \emph {et~al.}(2017)\citenamefont
  {Zwicker}, \citenamefont {Seyboldt}, \citenamefont {Weber}, \citenamefont
  {Hyman},\ and\ \citenamefont {J{\"u}licher}}]{zwicker2017growth}%
  \BibitemOpen
  \bibfield  {author} {\bibinfo {author} {\bibfnamefont {D.}~\bibnamefont
  {Zwicker}}, \bibinfo {author} {\bibfnamefont {R.}~\bibnamefont {Seyboldt}},
  \bibinfo {author} {\bibfnamefont {C.~A.}\ \bibnamefont {Weber}}, \bibinfo
  {author} {\bibfnamefont {A.~A.}\ \bibnamefont {Hyman}}, \ and\ \bibinfo
  {author} {\bibfnamefont {F.}~\bibnamefont {J{\"u}licher}},\ }\href@noop {}
  {\bibfield  {journal} {\bibinfo  {journal} {Nature Physics}\ }\textbf
  {\bibinfo {volume} {13}},\ \bibinfo {pages} {408} (\bibinfo {year}
  {2017})}\BibitemShut {NoStop}%
\bibitem [{\citenamefont {Singh}\ and\ \citenamefont
  {Cates}(2019)}]{singh2019hydrodynamically}%
  \BibitemOpen
  \bibfield  {author} {\bibinfo {author} {\bibfnamefont {R.}~\bibnamefont
  {Singh}}\ and\ \bibinfo {author} {\bibfnamefont {M.}~\bibnamefont {Cates}},\
  }\href@noop {} {\bibfield  {journal} {\bibinfo  {journal} {Physical review
  letters}\ }\textbf {\bibinfo {volume} {123}},\ \bibinfo {pages} {148005}
  (\bibinfo {year} {2019})}\BibitemShut {NoStop}%
\bibitem [{\citenamefont
  {Chandrasekhar}(1981)}]{chandrasekhar1981hydrodynamic}%
  \BibitemOpen
  \bibfield  {author} {\bibinfo {author} {\bibfnamefont {S.}~\bibnamefont
  {Chandrasekhar}},\ }\href@noop {} {\emph {\bibinfo {title} {Hydrodynamic and
  Hydromagnetic Stability}}},\ Dover Books on Physics Series\ (\bibinfo
  {publisher} {Dover Publications},\ \bibinfo {year} {1981})\BibitemShut
  {NoStop}%
\bibitem [{Note1()}]{Note1}%
  \BibitemOpen
  \bibinfo {note} {Note that the conventional Rayleigh-Plateau instability is
  driven by surface tension and requires a $3d$ fluid column \cite
  {chandrasekhar1981hydrodynamic}, unlike the flow driven pinch-off of a $1d$
  strip obtained here.}\BibitemShut {Stop}%
\bibitem [{\citenamefont {Guccione}\ \emph {et~al.}(2019)\citenamefont
  {Guccione}, \citenamefont {Benzi}, \citenamefont {Plummer},\ and\
  \citenamefont {Toschi}}]{guccione2019}%
  \BibitemOpen
  \bibfield  {author} {\bibinfo {author} {\bibfnamefont {G.}~\bibnamefont
  {Guccione}}, \bibinfo {author} {\bibfnamefont {R.}~\bibnamefont {Benzi}},
  \bibinfo {author} {\bibfnamefont {A.}~\bibnamefont {Plummer}}, \ and\
  \bibinfo {author} {\bibfnamefont {F.}~\bibnamefont {Toschi}},\ }\href@noop {}
  {\bibfield  {journal} {\bibinfo  {journal} {Physical Review E}\ }\textbf
  {\bibinfo {volume} {100}},\ \bibinfo {pages} {062105} (\bibinfo {year}
  {2019})}\BibitemShut {NoStop}%
\end{thebibliography}

%
\end{document}